\documentclass[11pt]{article}%
\usepackage{amssymb}
\usepackage{amsfonts}%
\usepackage{amsmath}%
\usepackage{hep}
\usepackage{graphicx}
\usepackage[margin=2.5cm]{geometry}
\providecommand{\U}[1]{\protect\rule{.1in}{.1in}}
\begin{document}
\title{\begin{flushright}
\vspace{-2cm} 
\small{CERN-PH-TH/2009-185}\\
\small{NSF-KITP-09-193}\\
\end{flushright}
\vspace{1cm}
Deep Inelastic Scattering  in Conformal QCD}
\author{Lorenzo Cornalba$^{a}$, Miguel S. Costa$^{b,c}$, Jo\~ao Penedones$^{d}$ 
%\footnote{{\small {\texttt{Lorenzo.Cornalba@mib.infn.it, miguelc@fc.up.pt, penedon@kitp.ucsb.edu}}}}
\medskip\\$^{a}$  Centro Studi e Ricerche E. Fermi, Compendio Viminale, I-00184, Roma\\  Universit\`{a} di Milano-Bicocca and INFN, sezione di Milano-Bicocca\\   Piazza della Scienza 3, I--20126 Milano, Italy 
\medskip\\$^{b}$   Departamento de F\'{\i}sica e Centro de F\'{\i}sica do Porto,\\   Faculdade de Ci\^{e}ncias da Universidade do Porto,\\  Rua do Campo Alegre 687, 4169--007 Porto, Portugal 
\medskip\\$^{c}$    Theory Group, Physics Department, CERN\\   CH-1211 Geneva 23, Switzerland
\medskip\\$^{d}$    Kavli Institute for Theoretical Physics\\ University of California, Santa Barbara, CA 93106-4030, USA\medskip\\
{   {\texttt{Lorenzo.Cornalba@mib.infn.it, miguelc@fc.up.pt, penedon@kitp.ucsb.edu}}}
}
\date{}
\maketitle

\begin{abstract}
We consider the Regge limit of a CFT correlation function of two vector and two scalar operators, 
as appropriate to study small-x deep inelastic scattering in $\mathcal{N}=4$ SYM or in QCD assuming approximate conformal symmetry.
After clarifying the nature of the Regge limit for a CFT correlator, we use its conformal partial wave expansion to obtain
an impact parameter representation encoding the exchange of a spin $j$ Reggeon for any value of the coupling constant. The CFT impact parameter space is the 
three-dimensional hyperbolic space $H_3$, which is the impact parameter space for high energy scattering in the dual  AdS  space.
We determine the small-x structure functions associated to the exchange of a Reggeon.
We discuss unitarization from the point of view of scattering in  AdS  and comment on the validity of the eikonal approximation.

We then focus on  the weak coupling limit of the theory where the amplitude is dominated by the exchange of  the BFKL pomeron. 
Conformal invariance fixes the form of the vector impact factor and its decomposition in transverse spin 0 and spin 2 components. 
Our formalism reproduces exactly the general results predict by the Regge theory, both for a scalar target and for $\gamma^*-\gamma^*$
scattering. We compute current impact factors
for the specific examples of  $\mathcal{N}=4$ SYM and QCD, obtaining very simple results.  In the case of the R-current of $\mathcal{N}=4$ SYM, we show that the 
transverse spin 2 component vanishes. We conjecture that the impact factors of all chiral primary operators of $\mathcal{N}=4$ SYM only 
have components with 0 transverse spin.
\end{abstract}

\newpage
\tableofcontents

%%%%%%%%%%%%%%%%%%%%%%%%%%%%%%%%%%%%%%
\section{Introduction}
%%%%%%%%%%%%%%%%%%%%%%%%%%%%%%%%%%%%%%

The QCD Regge limit of high center of mass energy, with other
kinematical invariants kept fixed and larger than $\Lambda_{QCD}$, is greatly simplified
by the smallness of the coupling and the approximate conformal invariance of the
theory. This is the case in deep inelastic scattering (DIS) experiments
in the limit of vanishing Bjorken ${\rm x}$, at
fixed and large photon virtuality $Q^2$ with $Q\gg \Lambda_{QCD}$.
In this case the important observable that contains information regarding the scattering process 
is the correlation function
\begin{equation}
A^{ab}(y_i)= \left\langle j^a(y_1) {\cal O}(y_2) j^b(y_3) {\cal O}(y_4) \right\rangle\,, 
\label{4ptcor}
\end{equation}
where $j^a$ is a vector operator of dimension $\xi$, given in  QCD by the quark electromagnetic current operator
$j^a=\bar{\psi}\gamma^a \psi$, of dimension $\xi=3$.  
The scalar operator  ${\cal O}$ of dimension $\Delta$ creates a state
that represents the target hadron.  

Although $\alpha_s$ is small for large photon virtualities, in low ${\rm x}$ DIS one still needs to resum many diagrams because of the kinematical
enhancement in $\ln(1/{\rm x})$, so that, in this sense, the dynamics is still strongly coupled. For instance, 
the power like growth in $1/{\rm x}$  of the cross section is determined by the exchange of a pomeron \cite{BFKL,KLF,BL} between 
the quark dipole created by the photon \cite{dipole} and the target hadron, 
which resums diagrams of the order of $(\alpha_s\ln(1/{\rm x}))^n$. This growth breaks unitarity and leads to gluon saturation in the target hadron structure functions 
\cite{satLine,Mueller,Iancu,MuellerTrian,MunierPesch1,MunierPesch2,MunierPesch3}, which 
for large $Q$ still occurs for small coupling $\alpha_s$ \cite{LipatovSmallx}. 
In this kinematical regime of approximate conformal symmetry, a proposal for restoring unitary of the amplitude was given in \cite{Saturation}, based on a conformal phase shift
derived from the dual strong coupling picture  as a scattering process in  AdS  space \cite{EikonalShock,EikonalPartial,AdSeikonal,EikonalBrower1,EikonalBrower2}. 
At weak coupling,
the growth of the imaginary part of the  conformal phase shift leads in general to saturation, giving a prediction for the data inside the saturation 
region, where at present the resumation of Feynman diagrams is not under control, as shown in detail in \cite{Saturation} for external scalar operators. 
This paper develops the necessary techniques to include photon polarization in this analysis.

We will assume throughout conformal invariance and
we will write the general form of the correlator (\ref{4ptcor}) in the planar limit, exploring its CFT Regge limit, 
as dictated by conformal symmetry and {\em for any value of the 't Hooft coupling}.
To define an effective hadron in the conformal theory, we shall then consider the above correlation function in momentum space with a 
spacelike momentum for the operator ${\cal O}$ with virtuality $\bar{Q}^2$ set by $\Lambda_{QCD}$. 
To make conformal invariance manifest, it is useful to apply different conformal transformations, $y_i\rightarrow x_i$, 
to each of the external points in the correlator \cite{Lorenzo}.
Then the CFT Regge limit corresponds simply to taking small $x_i$'s. These new coordinates can be understood by realizing the
conformal symmetry on a dual  AdS  space, where  they correspond to parametrizing each external point close to the
origin of different Poincar\'e patches. Then, defining  $x \approx x_1-x_3$ and $\bar{x} \approx x_2-x_4$, 
we will show that the contribution of a Regge pole to the amplitude has the form
\begin{align*}
A^{mn}(x,\bar{x})  \approx- 2\pi i  \sum_{k=0}^4 \int 
d\nu~ \frac{\alpha_k (  \nu) \, \mathcal{D}_k^{mn} \Omega_{i\nu} ( x,\bar{x})}{
( x^2)^{\xi +\frac{j(\nu)-1}{2}}  ( \bar{x}^2)^{\Delta+\frac{j(\nu)-1}{2}}     }~,
\end{align*}
where  $j(\nu)$ is the spin of the Reggeon 
%as a function of the dual  AdS  momentum transfer 
and    $\alpha_k (  \nu)$ are the corresponding residues. 
The tensor  structure of the amplitude is fixed by conformal 
invariance and  is best expressed in terms of four differential operators $\mathcal{D}_k^{mn} $ of the variable $x$ acting on a function $\Omega_{i\nu} ( x,\bar{x})$, as we shall explain.
We will then be able to write the contribution of a Regge pole to  the structure functions of the target hadron in terms of the spin $j(\nu)$ and residues $\alpha_k (  \nu)$.
This analysis is carried in section 2, while the more technical derivation of the general form of the correlator (\ref{4ptcor}) in a CFT is presented in section 3.

The general form of  the amplitude (\ref{4ptcor}) in a CFT for any value of the coupling constant is also relevant to the strong coupling regime
of DIS, to the extent that one considers graviton exchange in the dual  AdS  space and breaks conformal invariance by simulating confinement with the
hadron wave function \cite{DISStrongCouplingWaveFunction,Hatta,Levin,Beuf,Albacete,LevinPotas,shockDIS}.
Other ways of simulating confinement at strong coupling in DIS include the hard and soft wall models \cite{DISStrongCouplingWall,Bayona1,Bayona2,Bayona3}.

At high energies the kinematics of the external operators breaks the full conformal group to the residual transverse conformal group $SO(3,1)$ that acts on
the two-dimensional space $\mathbb{R}^2$ transverse to the scattering plane.
We explore this residual symmetry in the weak coupling regime, where the amplitude is dominated by the exchange of a hard pomeron of spin 1.  In this limit the amplitude factorizes as the product of 
the BFKL propagator $F$ \cite{BFKL,KLF,BL} times impact factors $V^{mn}$ and $\bar{V}$, all integrated over   transverse space. 
More precisely, defining a reduced amplitude $\mathcal{A}^{mn} = (x^{2\xi}\bar{x}^{2\Delta})A^{mn}$, we will see that the 
amplitude has the following form
\begin{align*}
\mathcal{A}^{mn} (x,\bar{x})
\approx&-\frac{1}{N^2}\,\int_{\mathbb{R}^{2}} \frac{dz_{1\perp} dz_{3\perp}}{\left(z_{1\perp}-z_{3\perp}\right)^4}\,
\frac{dz_{2\perp} dz_{4\perp}}{\left(z_{2\perp}-z_{4\perp}\right)^4}\,
 \\&  V^{mn}(  x,z_{1\perp},z_{3\perp})  \; F(z_{1\perp},z_{3\perp},z_{2\perp},z_{4\perp}) \; \bar{V}(\bar{x},z_{2\perp},z_{4\perp})  \,,
\end{align*}
generalising previous results for external scalar operators \cite{BFKLpaper}. 
We shall then focus on the vector impact factor,  
which can in general be written as a linear combination of five given conformal tensor structures ${\cal I}_k^{mn}$,
with coefficients depending on a single cross ratio $u$. The most natural basis, given by a decomposition in  conformal transverse spin, has  
four components with transverse spin 0 and one with transverse spin 2.
With this decomposition, and with the decomposition of the BFKL propagator in transverse conformal blocks of definite spin \cite{Lipatov}, it is simple to relate the above BFKL and Regge forms 
of the amplitude, therefore determining the residues $\alpha_k (  \nu)$. In this case only the four transverse spin 0 components of the vector  impact factor survive, because the spin 2
component has no overlap with the impact factor of the scalar operator ${\cal O}$ associated with the target. This analysis is done in section 4.

We also extend the above results to the four-point function of the current operator $j^a$, relevant to the study
of $\gamma^*-\gamma^*$ scattering in the Regge limit \cite{gamma-gamma, gamma-gammaBrodsky}. In this case one does not need to introduce any scale near $\Lambda_{QCD}$,
since we are probing the QCD vacuum with point like sources.
The spin 2 component of each impact factor  contributes to this amplitude
and we derive the corresponding contribution to the Regge form of the amplitude, valid for any value of the coupling constant. 
The strong coupling limit, in the case of the R-current operator of ${\cal  N}=4$ SYM, was studied in \cite{Bartels} using the dual gravity description.
 
Finally, in section 5 we show how to compute impact factors in leading order in perturbation theory. In particular we compute the current impact factors for QCD with a massless quark.
Conformal invariance restricts considerably the form of this impact factor to a very simple form given in equation (\ref{FermionIF}) bellow.

Another motivation for our work is to consider ${\cal  N}=4$ SYM and the AdS/CFT duality \cite{AdSCFT}. In this context conformal invariance, and therefore the general form 
derived for the correlation function (\ref{4ptcor}), as well as its form in the Regge limit, are exact. The spin of the Regge trajectory varies from 1 to 2, corresponding to the pomeron
trajectory at weak coupling and to the  AdS  graviton trajectory at strong coupling \cite{Brower1}.  More precisely, it is the transverse spin 0 component of the pomeron
that becomes the graviton trajectory at strong coupling. By computing, at weak coupling, the impact factor for the R-current operator \cite{Bartels:2008zy}, 
we show that its transverse spin 2 component vanishes. The computation leading to this result is quite elaborate. We expect that this cancelation
is related to supersymmetry. In fact, since at strong coupling the graviton trajectory only contains  the
transverse spin 0 component of the pomeron trajectory, it is natural to expect that this holds for the all the operators that are dual to the
supergravity multiplet. This result leads us to conjecture that 
{\it half-BPS single-trace operators in ${\cal N}=4$ SYM have impact factors without transverse conformal spin}.
%{\it the transverse spin components greater than 0 of the impact factor of ${\cal N}=4$  chiral primary operators vanishes}. 
We shall comment on this conjecture and give some open questions in the concluding  section 6.

%%%%%%%%%%%%%%%%%%%%%%%%%%%%%%%%%%%%%%
\section{Regge kinematics in CFTs \label{Regge}}
%%%%%%%%%%%%%%%%%%%%%%%%%%%%%%%%%%%%%%

\begin{figure}[t]
\begin{center}
\includegraphics[
height=6cm]{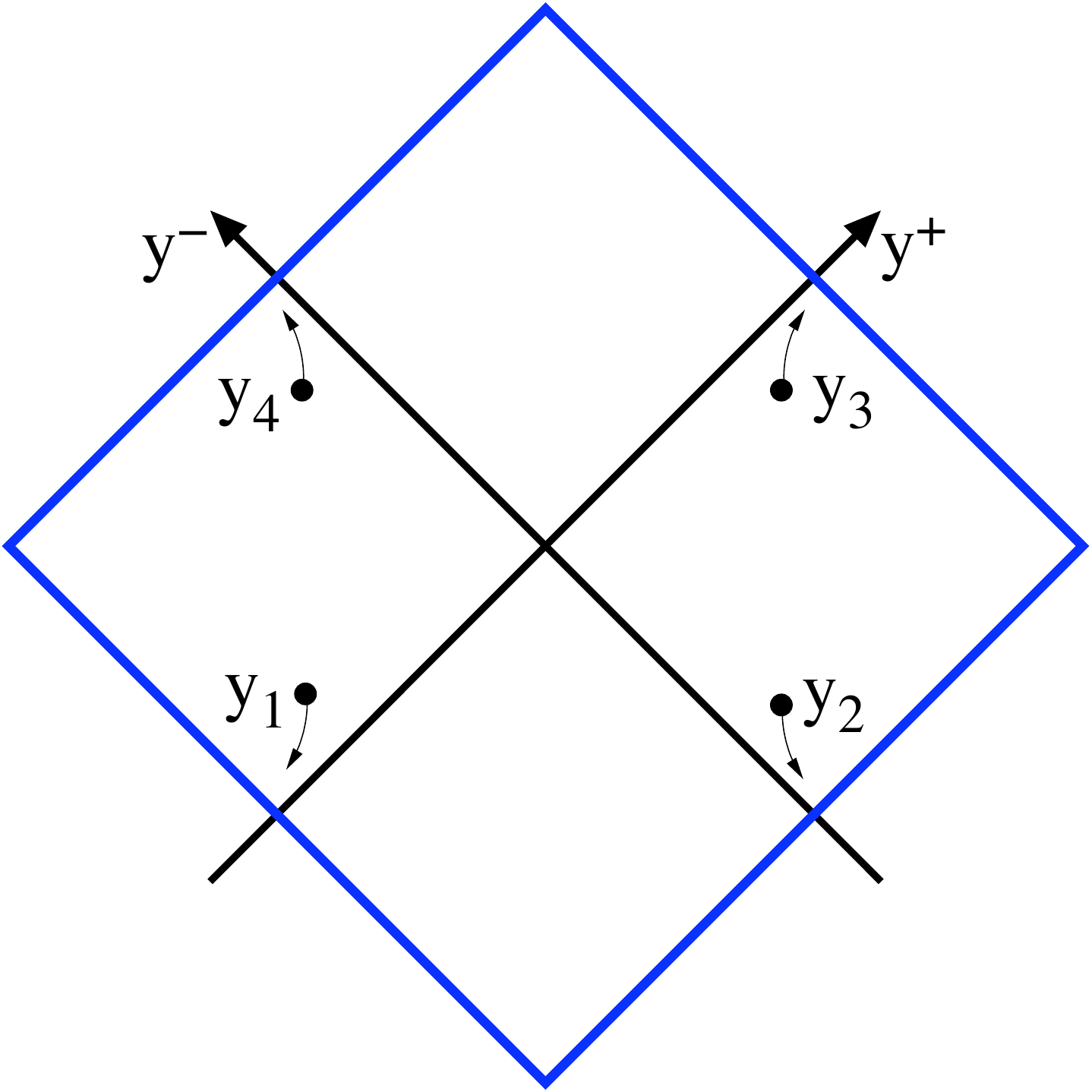}
\end{center}
\caption{ Conformal compactification of the $(y^+,y^-)$   Minkowski plane. The Regge kinematics corresponds to having all points
close to null infinity. }
\label{kinematics}
\end{figure}

Consider the correlation function (\ref{4ptcor}) in momentum space with external momenta $k_i$. The  Regge limit is usually described as the limit 
$$
s=-(k_1+k_2)^2 \to \infty \,,
$$
with $t=-(k_1+k_3)^2$ and $k_i^2$ fixed.
However, to make the conformal symmetry manifest,  it is more convenient to define the Regge limit in position space.
In the case of  the four point function (\ref{4ptcor}), the Regge limit corresponds to the
limit where the points $y_i$ are sent to null infinity, as described in \cite{Lorenzo} for scalar operators. 
Defining light-cone coordinates in four dimensional Minkowski space  $\mathbb{M}^4$ by
\[
y=(y^+,y^-, y_\perp)\,,
\]
where $y_\perp$ is a point in transverse space $\mathbb{R}^2$,
this limit is attained by sending
\[
y_1^+\rightarrow -\infty\,,\ \ \ \ \ \ 
y_3^+\rightarrow +\infty\,,\ \ \ \ \ \ 
y_2^-\rightarrow -\infty\,,\ \ \ \ \ \ 
y_4^-\rightarrow +\infty\,,\ \ \ \ \ \ 
\]
while keeping $y_i^2$ and $y_{i\perp}$ fixed,
as represented in figure \ref{kinematics}.

\begin{figure}[pt]
\begin{center}
\includegraphics[
height=6cm]{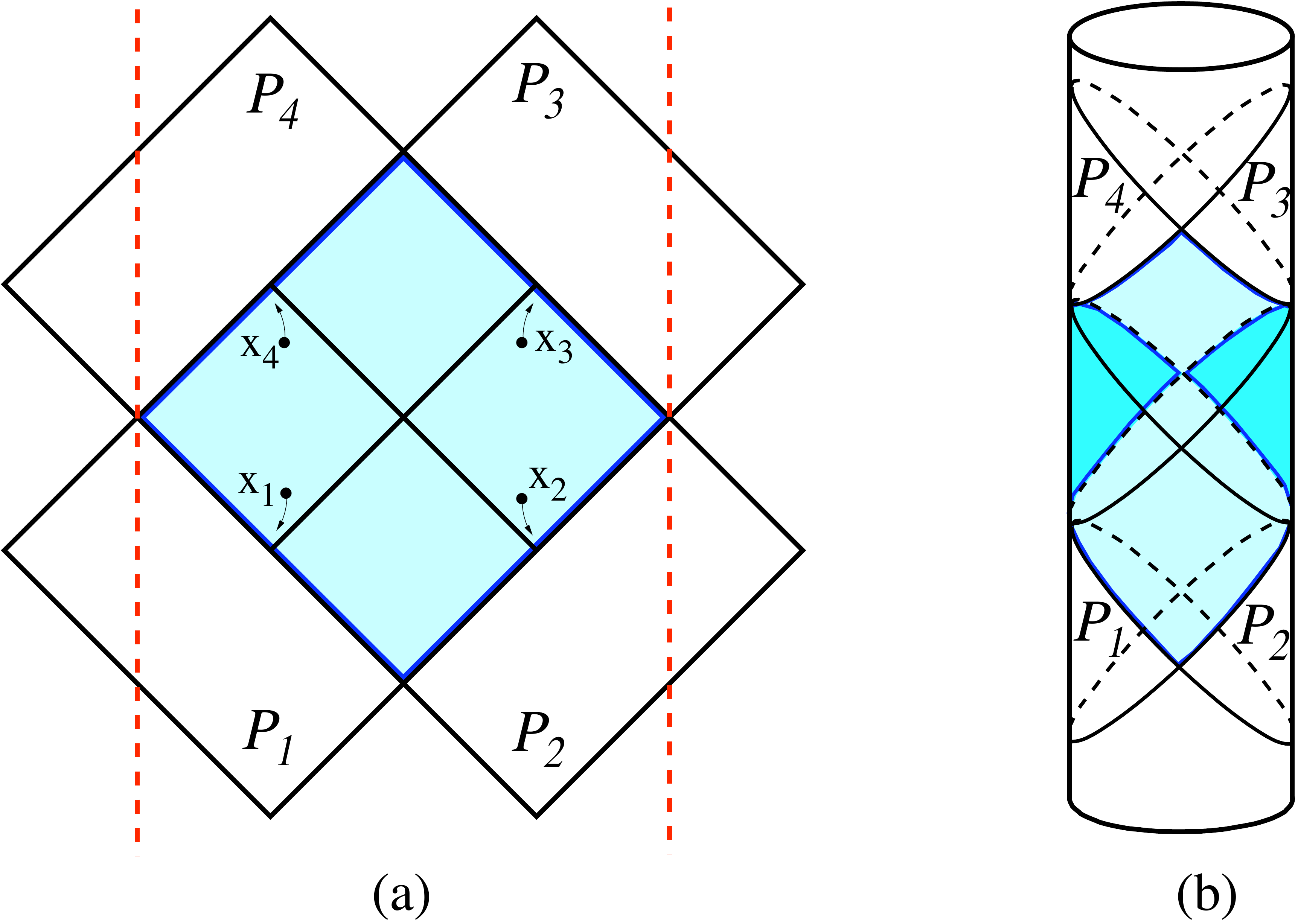}
\end{center}
\caption{(a)  Conformal compactification of the $(y^+,y^-)$   Minkowski plane and 
its relation to the Poincar\'e patches $\mathcal{P}_i$.  In the Regge limit the insertion points $x_i$ approach the origin of the Poincar\'e patches $\mathcal{P}_i$.
The vertical dashed lines are identified in the boundary of  AdS. 
(b) The $y_i$ Minkowski plane is shown as the central Poincar\'e patch on the boundary of global AdS. The other patches are also represented.}
\label{kinematicsAdS}
\end{figure}

It is instructive to visualize the operator insertions at the positions $y_i$ on the boundary of global  AdS, where by the AdS/CFT duality they act as sources for an  AdS  scattering process.
We take the original $\mathbb{M}^4$ as the central Poincar\'e patch in figure
  \ref{kinematicsAdS}(b).
Then, in the Regge limit, each $y_i$ tends to the origin of a different Poincar\'e patch in the boundary of global  AdS ,
which overlaps with the central patch as shown in figure
  \ref{kinematicsAdS}(a).
The Regge limit of the correlation function (\ref{4ptcor}) is best described 
parametrizing the position of each operator using different Poincar\'e patches,
so that the operators are always close to the origin of their patch \cite{Lorenzo}.
Using several coordinate patches is a natural procedure in any CFT where, to define local operators, 
one needs a local coordinate system with a metric, but this choice of coordinate system may be different for each operator.
This picture works both for Euclidean and Lorentzian signatures.

To find a new parametrization of the correlation function (\ref{4ptcor}) we consider the conformal 
transformation\footnote{There is an implicit length scale in this transformation that makes $x_i$ dimensionless.}
\begin{equation}
x_i= (x_i^+,x_i^-, x_{i\perp})  = - \frac{1}{y_i^+} \left(1, y_i^2, y_{i\perp} \right)\,,\ \ \ \ \ \ \ \ \ \ \ \ \ i=1,3\,,
\label{ConformalTrans1}
\end{equation}
that maps the points $y_1$ and $y_3$ into the Poincar\'e patches $\mathcal{P}_1$ and 
$\mathcal{P}_3$, respectively.
As intended, in the Regge limit the points $x_1$ and $x_3$ approach the origin of $\mathcal{P}_1$ and  $\mathcal{P}_3$.
We remark that the conformal transformation (\ref{ConformalTrans1}) is discontinuous, mapping points with negative
$y^+$ to points with positive $x^+$ in $\mathcal{P}_1$, and points with positive $y^+$ to points
 with negative $x^+$ in $\mathcal{P}_3$. 
The metric transforms  according to
\[
-dy^+dy^- + dy_\perp^2 = \frac{1}{(x^+)^2} \left(-dx^+dx^- + dx_\perp^2\right)\,.
\]
For the points  $y_2$ and $y_4$ we consider instead the conformal transformation
\begin{equation}
x_i= (x_i^+,x_i^-, x_{i\perp})  = - \frac{1}{y_i^-} \left( 1,y_i^2, y_{i\perp} \right)\,,\ \ \ \ \ \ \ \ \ \ \ \ \ i=2,4\,.
\label{ConformalTrans2}
\end{equation}
This transformation maps $y_2$ and $y_4$ into the Poincar\'e patches $\mathcal{P}_2$ and 
$\mathcal{P}_4$, respectively.
Again, the points $x_2$ and $x_4$ approach the origin in the Regge limit.

We shall study the limit $x_i \to 0$ of the correlation function 
\begin{equation*}
A^{mn}(x_i)= \langle j^m(x_1) {\cal O}(x_2) j^n(x_3) {\cal O}(x_4) \rangle\, ,
\end{equation*}
where the coordinates $x_i$ parametrize the Poincar\'e patches $\mathcal{P}_i$.
The original correlator  $A^{ab}(y_i)$ can be obtained by a simple conformal transformation. 
This requires taking  into account the transformation rule for spin 0 and spin 1 primary operators,
\begin{equation*}
\mathcal{O}(y)= \left| \frac{\partial x}{\partial y}\right|^{\frac{\Delta}{d}} \mathcal{O}(x)\ ,\ \ \ \ \ \ \ \ \ \ \ \  
j^a(y) =  \left| \frac{\partial x}{\partial y}\right|^{\frac{\xi+1}{d}} \frac{\partial y^a}{\partial x^m}\, j^m(x)\, ,
\end{equation*}
where $d$ is the spacetime dimension.
The Jacobian of the transformation (\ref{ConformalTrans1}) is $|y^+|^{-d}$, while the transformation
(\ref{ConformalTrans2}) has Jacobian  $|y^-|^{-d}$. Thus,  
\begin{equation}
A^{ab}(y_i) = (-y_1^+ y_3^+)^{-1-\xi}  (-y_2^- y_4^-)^{-\Delta} \,\frac{\partial y_1^a}{\partial x_1^m}\,\frac{\partial y_3^b}{\partial x_3^n}\, A^{mn}(x_i)\,.
\label{AyAx}
\end{equation}

The $x_i$ coordinate systems are very convenient to explore the  Regge limit of the correlator, while keeping the conformal symmetry manifest.
Let us consider the action of a translation in the Poincar\'e patch $\mathcal{P}_1$ 
(or $\mathcal{P}_3$),
\begin{align*}
x_1\rightarrow   x_1+a\, , \ \ \ \ \ \ \  x_3\rightarrow   x_3+a\, .
\end{align*}
It corresponds to a special conformal transformation in the Poincar\'e patch $\mathcal{P}_2$
(or $\mathcal{P}_4$), 
\begin{align*}
x_2\rightarrow \frac{x_2+x_2^2\, a}{1+2x_2\cdot a +a^2x_2^2}  \, , \ \ \ \ \ \ \ 
x_4\rightarrow  \frac{x_4+x_4^2 \,a}{1+2x_4\cdot a +a^2x_4^2} \, .
\end{align*}
In order to preserve the Regge limit, the translation parameter $a$ must be small. 
More precisely, we must consider $a$ of the same order as $x_i$. 
Therefore, points $x_2$ and $x_4$ are left invariant to leading order in the Regge limit. We conclude that
the  action of a small translation in the Poincar\'e patch $\mathcal{P}_1$ is given by
\begin{align}
x_1\rightarrow   x_1+a\, , \ \ \ \ \ \ \  x_3\rightarrow   x_3+a\, , \ \ \ \ \ \ \  
x_2\rightarrow   x_2\, , \ \ \ \ \ \ \  x_4\rightarrow   x_4\ . \label{translation}
\end{align}
Similarly, a small special conformal transformation in the Poincar\'e patch $\mathcal{P}_1$ yields
\begin{align}
x_1\rightarrow   x_1 \, , \ \ \ \ \ \ \  x_3\rightarrow   x_3 \, , \ \ \ \ \ \ \  
x_2\rightarrow   x_2+b\, , \ \ \ \ \ \ \  x_4\rightarrow   x_4+b\ . \label{special}
\end{align}
Lorentz transformations $SO(3,1)$ of the Poincar\'e patch $\mathcal{P}_1$ act as Lorentz transformations in all the 
four patches $\mathcal{P}_i$,
\begin{align}
x_i \rightarrow  \Lambda x_i\, . \label{Lorentz}
\end{align}
This $SO(3,1)$ acts on the  $y_i$ transverse space $\mathbb{R}^2$, with coordinates  $y_{i\perp}$, as 
the conformal group\footnote{This $SO(3,1)$ is also the isometry group of the three-dimensional hyperbolic space $H_3$, which is the transverse space of the 
dual AdS scattering process.}.
Finally, dilations in the Poincar\'e patch $\mathcal{P}_1$ give
\begin{align}
x_1\rightarrow \lambda x_1\, , \ \ \ \ \ \ \  x_3\rightarrow \lambda x_3\, , \ \ \ \ \ \ \
x_2\rightarrow\frac{x_2}{\lambda}\, , \ \ \ \ \ \ \ x_4\rightarrow\frac{x_4}{\lambda}\, . \label{dilatations}
\end{align}
This transformation is a $SO(1,1)$ boost in the $(y^+,y^-)$ plane.
We now wish to construct invariants under the conformal transformations (\ref{translation})-(\ref{dilatations}).
The transformations (\ref{translation}) and (\ref{special}) show that the correlator $A^{mn}(x_i)$
only depends on the combinations
\begin{equation}
x\approx x_1 - x_3\,,\ \ \ \ \ \ \ \ \ \ \ \ 
\bar{x}\approx x_2 - x_4\,.
\label{xxbar}
\end{equation}
Moreover, we can write the only two independent invariants as 
\begin{equation}
\sigma^2 = x^2 \bar{x}^2 \,,\ \ \ \ \ \ \ \ \  \cosh\rho = -\frac{x\cdot\bar{x}}{|x||\bar{x}|} \,,
\label{SigmaRho}
\end{equation}
so that the Regge limit corresponds to sending $\sigma\rightarrow 0$ with $\rho$ fixed.

An alternative but less intuitive way to derive this result is to consider the standard cross ratios
of the original correlator,
\begin{equation*}
z\bar{z} = \frac{y_{13}y_{24}}{y_{12}y_{34}}\,,\ \ \ \ \ \ \ \ \ \
(1-z)(1-\bar{z}) = \frac{y_{14}y_{23}}{y_{12}y_{34}}\,,
\end{equation*}
where $y_{ij} = (y_i - y_j)^2$. 
We may now perform the conformal transformations on the  $y_i$ to express the cross ratios in terms of  the
$x_i$. In the above Regge limit, a simple computation shows that
\begin{equation*}
z\bar{z} = x^2 \bar{x}^2\,,\ \ \ \ \ \ \ \ \ z+\bar{z} = -2x\cdot\bar{x}\, ,
\end{equation*}
with $x$ and $\bar{x}$ given by (\ref{xxbar}). We conclude that $\sigma$ and $\rho$ are just
a convenient parametrization of the two conformal invariants $z$ and $\bar{z}$. 
 
To simplify our exposition we shall often assume that $y_{13}>0$ and $y_{24}>0$.
This implies that both $x$ and $\bar{x}$ are future directed timelike vectors.
These conditions are not essential to our discussion and could be dropped.

%%%%%%%%%%%%%%%%%%%%%%%%%%%%%%%%%%%%%%
\subsection{Relation to scattering amplitude \label{relscat}}
%%%%%%%%%%%%%%%%%%%%%%%%%%%%%%%%%%%%%%

The usual discussions of the Regge limit consider the
  scattering amplitude
\begin{align}
(2\pi)^d\, \delta\left( \sum k_j \right) i\,T^{ab}(k_j) = \int \prod_{j=1}^4 dy_j \,e^{i  k_j \cdot y_j} \, A^{ab}(y_j)\ .
\label{Tamp}
\end{align}
We now wish to relate the Regge behavior of the correlation function $A^{mn}(x,\bar{x})$ with 
this more common approach.
To that end, it is convenient to introduce the Fourier transform
\begin{align}
A^{mn}(x,\bar{x})=  \int dp\,  d\bar{p} \,  e^{ -2 i p\cdot x -2 i \bar{p} \cdot \bar{x}}
B^{mn}(p,\bar{p})\ .\label{AB}
\end{align}
We shall see in section \ref{ReggeTheory}  that the contribution of a Regge pole $j(\nu)$ to the four point function can be written as
\begin{align}
A^{mn}(x,\bar{x})  \approx - 2\pi i  \int 
  d\nu~ \frac{\sum_{k=0}^4\alpha_k (  \nu) \,\mathcal{D}_k^{mn} \,\Omega_{i\nu} ( \rho)}{
( x^2-i\epsilon_{x})^{\xi +\frac{j(\nu)-1}{2}}  ( \bar{x}^2-i\epsilon_{\bar{x}})^{\Delta+\frac{j(\nu)-1}{2}}     }~,
\label{ReggeA}
\end{align}
where  the differential operators $ {\cal D}^{mn}_{k}$ depend only on $x$ and are given by
\begin{eqnarray}
&&\mathcal{D}_{1}^{mn}  =  \eta^{mn}-\frac{x^{m}x^{n}}{x^2}\,,\nonumber\\
&&\mathcal{D}_{2}^{mn}  =  \frac{x^m x^n}{x^2}\,,
\label{DTensors}\\
&&\mathcal{D}_{3}^{mn}  =  x^m\partial^n + x^n\partial^m\,,\nonumber\\
&&\mathcal{D}_{4}^{mn}  =  x^2 \partial^m\partial^n + \left(x^m\partial^n + x^n\partial^m\right)-\frac{1}{3}\left(\eta^{mn}-\frac{x^m x^n}{x^2}\right)x^2\square_{x}\,.\nonumber
\end{eqnarray}
with $\displaystyle{\partial_m=\frac{\partial\ }{\partial x^m}}$.
The functions $\Omega_{i\nu} ( \rho)$ are a basis of harmonic functions on the unit 3-dimensional hyperbolic space $H_3$.
The coefficients $\alpha_k(\nu)$ and the spin $j(\nu)$ encode the dynamical information of the correlation function.
The $i\epsilon$-prescription, $ \epsilon_x =  \epsilon \,{\rm sgn}\, x^0 $, is the appropriate one \cite{iepsilon,Mythesis}
for the propagator between $x_1$ in the Poincar\'e patches $\mathcal{P}_1$ and $x_3$ in $\mathcal{P}_3$ 
(and similarly between  points in $\mathcal{P}_2$ and $\mathcal{P}_4$).

We compute the Fourier transform $B^{mn}(p,\bar{p})$ of (\ref{ReggeA}) in appendix \ref{Ftrans}.
The $i\epsilon$-prescription in (\ref{ReggeA}) implies that $B^{mn}(p,\bar{p})$ only has support   on the 
future light-cones.
%$$
% \left\{ p  \in \mathbb{M}^4: p^2 <0  \wedge  p^0 >0\right\} \cap
%\left\{ \bar{p} \in \mathbb{M}^4:  \bar{p}^2<0  \wedge   \bar{p}^0>0 \right\} \ .
%$$
The result for future directed timelike vectors $p$ and $\bar{p}$ can be written as
\begin{align*}
B^{mn}(p,\bar{p})  = \frac{\mathcal{B}^{mn}(p,\bar{p}) }{
( -p^2 )^{\frac{d}{2}-\xi  }  ( -\bar{p}^2 )^{\frac{d}{2}-\Delta }     }~,
\end{align*}
with $\mathcal{B}^{mn}$ given by
\begin{align}
\mathcal{B}^{mn}(p,\bar{p})  \approx 2\pi i  \int  
  d\nu~ S^{j(\nu)-1}\, \sum_{k=0}^4\beta_k (  \nu)\,\hat{\mathcal{D}}_k^{mn}\, \Omega_{i\nu} \left(L \right) ~, \label{calB}
\end{align}
where  
\begin{align}
S=4|p||\bar{p}| \ ,\ \ \ \ \ \ \ \ \ \ \ \cosh L= -\frac{p\cdot \bar{p}}{|p||\bar{p}|}\ .
\label{SL}
\end{align}
The differential operators $ \hat{{\cal D}}^{mn}_{k}$ have the same form as in (\ref{DTensors}) but now depend on $p$,
\begin{eqnarray*}
&&\hat{\mathcal{D}}_{1}^{mn}  =  \eta^{mn}-\frac{p^{m}p^{n}}{p^2}\,,\nonumber\\
&&\hat{\mathcal{D}}_{2}^{mn}  =  \frac{p^m p^n}{p^2}\,,
\\
&&\hat{\mathcal{D}}_{3}^{mn}  =  p^m\hat{\partial}^n + p^n\hat{\partial}^m\,,\nonumber\\
&&\hat{\mathcal{D}}_{4}^{mn}  =  p^2 \hat{\partial}^m\hat{\partial}^n + \left(p^m\hat{\partial}^n + p^n\hat{\partial}^m\right)-\frac{1}{3}\left(\eta^{mn}-\frac{p^m p^n}{p^2}\right)p^2\square_{p}\,.\nonumber
\end{eqnarray*}
with $\displaystyle{\hat{\partial}_m=\frac{\partial\ }{\partial p^m}}$.
The coefficients $\beta_k(\nu)$ are   linear combinations of the $\alpha_k(\nu)$, as given explicitly in appendix \ref{Ftrans}.

Using (\ref{AyAx}) and (\ref{AB}) we can  write the scattering amplitude (\ref{Tamp}) as follows
\begin{align}
&(2\pi)^d \,\delta\left( \sum k_j \right) i\,T^{ab}(k_j) = 
 \int dp \,d\bar{p}  \,B^{mn}(p,\bar{p}) \label{ScatAmp}\\
 &
  \int \prod_{j=1}^4 dy_j\, e^{i  k_j \cdot y_j} \, (-y_1^+ y_3^+)^{-1-\xi}  (-y_2^- y_4^-)^{-\Delta} \,
  \frac{\partial y_1^a}{\partial x_1^m}\,\frac{\partial y_3^b}{\partial x_3^n}\, e^{ - 2i p\cdot x - 2i \bar{p} \cdot \bar{x}}\,.
\nonumber
\end{align}
To compute the integrals in the second line we consider the Regge kinematics in momentum space,
\begin{align*}
k_1\cdot y_1&= -\omega_1 y_1^- + \frac{k_1^2}{4 \omega_1} \,y_1^+\ , \ \ \ \ \ 
&k_3\cdot y_3= \omega_3 y_3^- - \frac{k_3^2-q_\perp^2}{4 \omega_3} \,y_3^+ -q_\perp \cdot y_{3\perp} \ ,\\
k_2\cdot y_2&= -\omega_2 y_2^+ + \frac{k_2^2}{4 \omega_2} \,y_2^-\ ,\ \ \ \ \ 
&k_4\cdot y_4= \omega_4 y_4^+ - \frac{k_4^2-\bar{q}_\perp^2}{4 \omega_4} \,y_4^-+\bar{q}_\perp \cdot y_{4\perp}\ ,
\end{align*}
with $\omega_i \to \infty$. 
With this choice  the momentum conserving delta-function becomes
$$
\delta\left( \sum k_j \right)
\approx
\frac{1}{2}\,\delta(\omega_1-\omega_3) \,\delta(\omega_2-\omega_4) \,
\delta(q_\perp-\bar{q}_\perp)  \ ,
$$
and the Mandelstam invariants are given by $s\approx4\omega_1\omega_2$ and $t=-q_\perp^2$.
In this limit, the integral is dominated by the 
position space Regge kinematics of small $x_i$, which corresponds to  large and positive 
$-y_1^+$, $y_3^+$, $-y_2^-$ and $y_4^-$ with fixed $y_i^2$ and $y_{i\perp}^2$.
Therefore, we can use equation (\ref{xxbar}) to factorize the integrals in the second line of (\ref{ScatAmp}).
Moreover, we can restrict the $y$ integrals to positive 
$-y_1^+$, $y_3^+$, $-y_2^-$ and $y_4^-$.

It is also convenient to parametrize  $p=E\,e$ and $\bar{p}=\bar{E}\,\bar{e}$ with $E$ and $\bar{E}$ positive
and $e$ and $\bar{e}$ points in 3-dimensional hyperbolic space,
\begin{equation}
e= \frac{1}{r}\left(1,r^2+e_\perp^2,e_\perp\right) \ ,\ \ \ \ \ \ \ 
\bar{e}=\frac{1}{\bar{r}}\left(1,\bar{r}^2+\bar{e}_\perp^2,\bar{e}_\perp\right)\  .
\label{e}
\end{equation}
In the new coordinates $p^\mu=(E,r,e_\perp)$ the reduced amplitude  $ \mathcal{B}$ has components
\begin{align*}
 \mathcal{B}^{\mu}_{\  \tau}  =  \frac{\partial p^\mu }{\partial p^m}\frac{\partial p_\tau}{\partial p^n}\,\mathcal{B}^{mn}\ .
 \end{align*}
Then a simple computation shows that the amplitude $T^{ab}(k_j)$ in (\ref{ScatAmp}) has the form
\begin{align}
 i\,T^{ab}(k_j) \approx 2s\int dl_\perp
 e^{iq_\perp\cdot l_\perp}
 \int \frac{dr}{r^3}\, \frac{d\bar{r}}{\bar{r}^3} \,
F_{1\,\mu}^{\,a}(r)\,F_{3}^{\,b\tau}(r)\,F_2(\bar{r})\,F_4(\bar{r})  \,
\mathcal{B}^{\mu}_{\ \tau} (S,L)  \,,
\label{finalScat}
\end{align}
where
$$
S=r\bar{r} s \ ,\ \ \ \ \ \ \ \ \ \ 
\cosh L=\frac{r^2+\bar{r}^2+l_\perp^2}{2r\bar{r}}
 $$
and  $l_\perp=e_\perp-\bar{e}_\perp$ is the standard impact parameter in the  field theory.

In the representation (\ref{finalScat}), the scattering
amplitude $T^{ab}(k_j)$ is written in terms of the CFT reduced amplitude $\mathcal{B}^{\mu}_{\ \tau} (S,L)$,
of scalar functions $F_2$ and $F_4$ associated with the operator ${\cal O}$ and of 
tensor functions $F_{1\,\mu}^{\,a}$ and $F_{3\, \tau}^{\,b}$ associated with the vector operator $j^a$.
This conformal representation is quite natural from the dual AdS scattering process point of view, with
transverse space given by the hyperbolic space $H_3$, whose boundary is the field theory transverse 
space $\mathbb{R}^2$.  In the AdS  picture the  functions $F_i$ are the radial
part of the wave functions of the dual fields and the quantities $S$ and $L$ are, respectively, the
AdS generalization of the energy squared and impact parameter of the associated geodesics. The functions $F_4$  and $F_3$ are defined respectively by \footnote{More precisely, we restrict the integration region to $y_4^- , y_3^+>0$ because this is the dominant region in the Regge limit.} 
$$
 \int  dy_4  \,   ( y_4^-  )^{-\Delta}  \, e^{ i  k_4 \cdot y_4+2i \bar{p} \cdot  x_4}=   2\sqrt{\pi}  \,
 \delta( \bar{r}\omega_4-\bar{E}) \,
 \bar{E}^{1-\Delta}  e^{i\bar{q}_\perp\cdot \bar{e}_\perp} F_4(\bar{r})\ ,
$$  
and 
\begin{align*}
\frac{\partial p^n}{\partial p^ \tau}  \int  dy_3 \, ( y_3^+)^{-1-\xi}  
  \,\frac{\partial y_3^b}{\partial x_3^n}\, e^{i  k_3 \cdot y_3+  2i p\cdot x_3  } = 
   2\sqrt{\pi}   \,\delta(  r \omega_3- E )   \,
E^{1-\xi} e^{-i q_\perp \cdot e_\perp}  F_{3\, \tau}^{\,b}(r)\ ,
\end{align*}
with similar expressions for $F_2$ and $F_1$. They can be expressed explicitly  in terms of Bessel functions as shown 
in appendix \ref{Ftrans1}.
 
 %%%%%%%%%%%%%%%%%%%%%%%%%%%%%%%%%%%%%%
 \subsection{Structure functions}
%%%%%%%%%%%%%%%%%%%%%%%%%%%%%%%%%%%%%%

We now consider the special case of a conserved current  with $\xi=3$ in $d=4$. 
Conservation means that the Fourier transformed amplitude $\mathcal{B}^{mn}$ satisfies
$$
p_m\mathcal{B}^{mn}(p,\bar{p}) =p_n\mathcal{B}^{mn}(p,\bar{p}) =0\ .
$$
From (\ref{calB}) we see that this condition implies $\beta_2(\nu)=\beta_3(\nu)=0$.
In the coordinate system $p^\mu=(E,r,e_\perp)$, conservation gives simply $\mathcal{B}^{E\mu}=0$.
It is then natural to use the indices $\hat{\mu}$ and $\hat{ \tau}$ to denote only the directions tangent to the hyperboloid parametrized by 
$e$, as given in (\ref{e}). Using this notation, the Regge form of the reduced amplitude (\ref{calB}), can be written
in the geometrical form
\begin{align}
 \mathcal{B}^{\hat{\mu}}_{\   \hat{\tau}}    \approx   2\pi i  \int  
  d\nu~ S^{j(\nu)-1}\,  
  \left[ \beta_1 (  \nu) \delta^{\hat{\mu}}_{\ \hat{\tau}} +\beta_4(\nu) \left(\nabla^{\hat{\mu}} \nabla_{\hat{\tau}} -\frac{1}{3}\,\delta^{\hat{\mu}}_{\ \hat{\tau}} \right)
  \right] \Omega_{i\nu} \left(L \right) \,,\label{calBcurrent}
\end{align}
with $\nabla_{\hat{\mu}}$ the Levi-Civita connection on the hyperboloid and hated indices are raised and lowered with the $H_3$ metric.

The structure functions  are directly related to the forward scattering amplitude, with kinematics $k_1=-k_3$ and $k_2=-k_4$.
 We also take $k_1^2=k_3^2=Q^2$ to be the photon virtuality and $k_2^2=k_4^2=\bar{Q}^2>0$ to simulate confinement, as explained in the introduction.
 The Bjorken variable is
 $$
{\rm x}=-\frac{Q^2}{2 k_1\cdot k_2} \approx 
 \frac{Q^2}{s}\,.
$$
As usual, Lorentz invariance and conservation restricts  $T^{ab}$ to the form
$$
T^{ab}= \left( \frac{k_1^a k_1^b}{k_1^2}-\eta^{ab}\right)\Pi_1 
-\frac{2{\rm x}}{Q^2}  \left( k_2^a+\frac{k_1^a  }{2{\rm x}}\right)\left( k_2^b+\frac{k_1^b }{2{\rm x}}\right) \Pi_2 \,.
$$
 In the Regge limit explained above (${\rm x} \ll 1$) we have
 \begin{align*}
T^{ij}= -\delta^{ij}\Pi_1 \ ,\ \ \ \ \ \ \ \ \ \ \ \ 
T^{++}\approx \frac{(2\omega_1)^2}{Q^2}\left( \Pi_1 -\frac{1}{2{\rm x}}\,\Pi_2  \right)  \,,
\end{align*}  
where the Latin indices $i$ and $j$ run over the transverse space $\mathbb{R}^2$ directions.

To determine the structure functions from expression (\ref{finalScat}) we need the following 
integrals\footnote{The last integral is subtle, but it can be defined as the zero momentum limit of the two dimensional Fourier transform of $ \Omega_{i\nu}$.} 
\begin{align*}
  \int dl_\perp  \left(\nabla^i\nabla_j-\frac{1}{3}\delta^i_{\,j} \right)   \Omega_{i\nu} (L )   
  &= - \delta^i_{\,j}  \left(\frac{1}{3} +r  \partial_r    \right)   \int dl_\perp \,   \Omega_{i\nu} (L) \,,\\
  \int dl_\perp  \left(\nabla^r\nabla_r-\frac{1}{3}  \right)   \Omega_{i\nu} (L)   
  &=    \left((r  \partial_r )^2-\frac{1}{3}    \right)   \int dl_\perp  \,  \Omega_{i\nu} (L)  \,,\\
     \int dl_\perp \,   \Omega_{i\nu}(L )&= \frac{1}{4\pi} \left( r^{1-i\nu}\bar{r}^{1+i\nu} + r^{1+i\nu}\bar{r}^{1-i\nu}\right) \,.
      \end{align*}   
We can then write
 \begin{align}
 T^{ij}  \approx&\,2\pi    \int d\nu  \, 2s 
 \int \frac{dr}{r^3}\, \frac{d\bar{r}}{\bar{r}^3} \,
 F_2(\bar{r})\,F_4(\bar{r})\,  F_{1\, k}^{\,i}(r)\, F_{3\, l}^{\,j}(r) \,\delta^{kl}r^2 \nonumber\\&
 (r\bar{r}s)^{j(\nu)-1}  \left[ \beta_1(\nu) +  \left( i\nu -\frac{4}{3} \right)  \beta_4(\nu)\right]
  \frac{1}{2\pi} \,r^{1-i\nu}\bar{r}^{1+i\nu}  \ ,     \label{Piintdr}\\
T^{++} \approx&\, 2\pi    \int d\nu  \, 2s 
 \int \frac{dr}{r^3}\, \frac{d\bar{r}}{\bar{r}^3} \,
 F_2(\bar{r})\,F_4(\bar{r})\,   F_{1\ r}^{\,+}(r) \,F_{3\ r}^{\,+}(r)\,r^2\nonumber
 \\&
 (r\bar{r}s)^{j(\nu)-1} \left[ \beta_1(\nu) + \left( \frac{2}{3}-2i\nu-\nu^2    \right) \beta_4(\nu)\right]
  \frac{1}{2\pi}\, r^{1-i\nu}\bar{r}^{1+i\nu}  \nonumber \,,
 \end{align}
where we used the invariance $\beta_k(-\nu)=\beta_k(\nu)$ and $j(-\nu)=j(\nu)$.
Finally, using the explicit expressions for the functions $F_i$ given in appendix \ref{Ftrans2}, we can perform the integrals over $r$ and $\bar{r}$ to obtain
the Regge representation of the structure functions
\begin{align}
 \Pi_1 &\approx\bar{Q}^{2\Delta-6}
   \int d\nu \,  \gamma_1(\nu) \, {\rm x}^{-j(\nu)}  \left(\frac{Q}{\bar{Q}}\right)^{i\nu+j(\nu)}\ ,\nonumber\\
    \Pi_1 -\frac{1}{2{\rm x}}\,\Pi_2  &\approx\bar{Q}^{2\Delta-6}
   \int d\nu \,  \gamma_2(\nu) \, {\rm x}^{-j(\nu)}  \left(\frac{Q}{\bar{Q}}\right)^{i\nu+j(\nu)}   \,,
\label{StructFunc}
 \end{align}   
where   $\gamma_1(\nu)$ and  $\gamma_2(\nu)$ are given explicitly in appendix \ref{Ftrans2} in terms of $\beta_1$ and $\beta_4$.
The structure functions are then given by the imaginary part of $\Pi_1$ and $\Pi_2$.

%%%%%%%%%%%%%%%%%%%%%%%%%%%%%%%%%%%%%%
\subsection{Unitarization}\label{Unitarization}
%%%%%%%%%%%%%%%%%%%%%%%%%%%%%%%%%%%%%%

When $j(\nu)$ is greater than 1 the scattering amplitude (\ref{finalScat}) grows too quickly with energy and violates unitarity. 
We shall address this issue in ${\cal N}=4$ SYM where conformal invariance is exact and we have a good strong coupling description using AdS/CFT. 
At strong coupling and to leading order in the planar expansion, the high energy behavior of the scattering amplitude is dominated by one graviton exchange, which gives $j(\nu)=2$ and strongly violates unitarity. 
In this regime, unitarity is recovered by including multiple graviton exchanges in the eikonal approximation \cite{AdSeikonal,EikonalBrower1}.

Let us focus  on the more relevant case of conserved currents ($\xi=d-1$).
In the bulk, we are considering elastic scattering of a scalar particle and a gauge boson. The physical polarizations of the gauge boson, which we label with $\hat{\mu},\hat{\tau}$,  
are the directions along the $(d-1)$-dimensional hyperboloid   transverse to the scattering plane.
 Then,  the eikonal approximation in AdS gives rise to 
(\ref{finalScat}) with $\mathcal{B}^{\hat{\mu}}_{\ \hat{\tau}}(S,L) $ replaced by 
\begin{align*}
  \mathcal{N}\, \left[e^{i\chi(S,L)}\right]^{\hat{\mu}}_{\ \hat{\tau}}\ ,
\end{align*} 
where $\chi^{\hat{\mu}}_{\ \hat{\tau}}(S,L)$ is the phase shift matrix for the scattering process in AdS and $\mathcal{N}$ is a normalization constant that we shall fix below. 
Then  the  correlation function has the form
\begin{align}
A^{mn}(x,\bar{x})=  \mathcal{N} \int_{\rm M} dp\,  d\bar{p} \,  \frac{e^{ -2 i p\cdot x -2 i \bar{p} \cdot \bar{x}} }{
( -p^2 )^{1-\frac{d}{2}   }  ( -\bar{p}^2 )^{\frac{d}{2}-\Delta }     }
\frac{\partial p^m}{\partial p^{\hat{\mu}}} \frac{\partial p^n}{\partial p_{\hat{\tau}}} 
\left[e^{i\chi(S,L)}\right]^{\hat{\mu}}_{\ \hat{\tau}} \ , \label{Aphaseshift}
\end{align} 
where  the integration is over future directed vectors (Milne wedge). 
We recall that in the AdS process the quantities $S$ and $L$ given in (\ref{SL}) are, respectively, 
the generalization of the energy squared and impact parameter of the associated geodesics. 
The constant $\mathcal{N}$ can be fixed by matching  the disconnected correlator with zero phase shift,
\begin{align*}
 \langle j^m(x_1) j^n(x_3)  \rangle\,  \langle{\cal O}(x_2) {\cal O}(x_4) \rangle=  \mathcal{N} \int_{\rm M} dp\,  d\bar{p} \,  \frac{e^{ -2 i p\cdot x -2 i \bar{p} \cdot \bar{x}} }{
( -p^2 )^{1-\frac{d}{2}  }  ( -\bar{p}^2 )^{\frac{d}{2}-\Delta }     }
\frac{\partial p^m}{\partial p^{\hat{\mu}}} \frac{\partial p^n}{\partial p_{\hat{\mu}}} 
  \ .
\end{align*} 
The integrals over $p$ and $\bar{p}$ factorize and can be done explicitly (see appendix B of \cite{AdSeikonal}).
Using
%\footnote{We recall that the $\hat{\mu}$ index only runs over the tangent directions to hyperboloid $p^2=-1$.}
\begin{equation*}
\frac{\partial p^m}{\partial p^{\hat{\mu}}} \frac{\partial p^n}{\partial p_{\hat{\mu}}} =
\eta^{mn}-\frac{p^m p^n}{p^2}\,,
\end{equation*}
and the normalization of the external operators 
\begin{equation}
\langle {\cal O}(x_2) {\cal O}(x_4) \rangle = \frac{1}{
\big(  \bar{x}^{2} - i\epsilon_{\bar{x}}\big)^{\Delta}}\ ,\ \ \ \ \ \ 
\langle j^m(x_1) j^n(x_3) \rangle = \frac{x^2 \eta^{mn} -2  x^m x^n  }{\big(  x^{2} - i\epsilon_x\big)^{d}}\,,\label{normop}
\end{equation}
we obtain
 \begin{equation*}
 \mathcal{N} =
 \frac{4 \pi ^{2-d}(d-2)}{ \Gamma \left(d\right)\Gamma \left(\frac{d}{2}\right) \Gamma \left(\Delta \right)
   \Gamma \left(\Delta -\frac{d}{2}+1\right)}\ .
\end{equation*}

We can also determine the phase shift associated to the exchange of a  Reggeon by matching equation (\ref{calBcurrent}) to the second term in the expansion of the  exponential $e^{i\chi}$, 
with the result
\begin{align*}
\chi^{\hat{\mu}}_{\ \hat{\tau}}(S,L) \approx  \frac{ 2\pi}{\mathcal{N}}  \int  
  d\nu~ S^{j(\nu)-1}\,  
  \left[ \beta_1 \left(  \nu\right) \delta^{\hat{\mu}}_{\ \hat{\tau}} +\beta_4(\nu) \left(\nabla^{\hat{\mu}} \nabla_{\hat{\tau}} -\frac{1}{3}\,\delta^{\hat{\mu}}_{\ \hat{\tau}} \right)
  \right] \Omega_{i\nu} \left(L \right) \ .
\end{align*}
This is a phase shift of order $1/N^2$ in  't Hooft's expansion and therefore corresponds to a tree level process in the dual string theory. In other words, the exponentiation of this phase shift corresponds to the eikonal approximation in AdS where the external states scatter elastically by the exchange of multiple soft Reggeons (the graviton at strong coupling and the hard pomeron at weak coupling). Let us now comment on the  validity of this approximation.

We shall analyze this issue by considering large and fixed $S$ and decreasing the impact parameter $L$. The spirit will be similar to the discussion \cite{GGM} of high energy scattering in flat space.
In units where the AdS radius is 1, and for AdS impact parameter $L \gg 1$, the Reggeon phase shift is approximately given by\begin{align}
\chi  \sim \frac{b}{N^2}    \,S^{j_0-1} \, e^{-a L}    \label{chiapprox}
\end{align}
where $a,\ b$ and $j_0>1$ are functions of the 't Hooft coupling
$g^2$. At weak coupling, this generic form can be obtained by a saddle point approximation to the integral over $\nu$. At strong coupling, the same generic form describes the contribution of the gravi-reggeon dominant pole ($j_0=2-O(1/g)$) at large impact parameter.
The parameter $a$ is of order 1 both at weak and strong coupling.
We also assume that $S$ is large enough so that the phase shift can become large for some $L\gg 1$.
We have dropped the index structure and powers of $L$ and $\log S$ because they are not important for the argument we want to make.

At very large impact parameter $L$, the phase shift is very small and the scattering amplitude is dominated by single Reggeon exchange.
Then, as we decrease the impact parameter, there are three important transitions. The first transition corresponds to the eikonal exchange of multiple Reggeons. The   impact parameter where this process becomes important can be estimated by
\begin{align*}
\chi(S,L_{eik}) \sim 1  \ .
\end{align*}
The second transition corresponds to tidal excitation of the scattering strings.  When the tidal forces induced by one string on the other become stronger than the string tension, the string can change their internal state.
This  crossover can be estimated by the condition
\begin{align*}
\frac{\partial^2  \chi}{\partial L^2} (S,L_{tid})   \sim {\rm string\  tension} \sim g \ .
\end{align*}
The third transition corresponds to the breakdown of the eikonal approximation when the momentum transfer $\sim\partial_L \chi$ is of order or greater than the center-of-mass energy $\sim \sqrt{S}$,
\begin{align*}
\frac{\partial   \chi}{\partial L}(S,L_{bh}) \sim  \sqrt{S}  \ .
\end{align*}
 In flat space, this condition corresponds to black hole formation for impact parameters smaller that the Schwarzschild radius associated with the total energy of the scattering process.
The eikonal approximation breaks down for $L<L_{tid}$ or $L<L_{bh}$.

Using the approximation  (\ref{chiapprox}) we obtain
\begin{align*}
  L_{eik}  \sim\frac{1}{a} \log  \frac{b\,S^{j_0-1}}{N^2}     \ ,\ \ \ \ \ \ 
L_{eik}- L_{tid}  \sim \frac{1}{a} \log  g\ .
\end{align*} 
Notice that the difference $L_{eik}-L_{tid}$ is independent of $S$. 
This is in sharp contrast with what happens in flat space.
The source of this qualitatively different behavior is that in AdS, the phase shift decays exponentially with the impact parameter $L$ (for $L\gg 1$) and in flat space it decays with a power law. The fact that in AdS these two effects appear at the same impact parameter, parametrically in $S$, was noticed before in \cite{DISStrongCouplingWaveFunction}.
This limits the validity of the eikonal approximation in AdS to a rather small interval of impact parameters $L_{tid}<L<L_{eik}$. The situation is even worst at small $g$ where $L_{tid} >L_{eik}$ and it seems that the eikonal approximation is never useful. 
Nevertheless, we can still use the representation  (\ref{Aphaseshift}) of the correlator, which only relies on conformal symmetry. At weak coupling one does not expect the phase 
shift $\chi$ to be dictated by the tree level result, but one still expects that a black disk in a conformal field theory (dual to an AdS black disk) corresponds to a large imaginary phase
shift. This basic input was used in \cite{Saturation} to  successfully fit available experimental data for the proton structure function at small-x inside the saturation region.
 
Finally, let us consider the critical impact parameter for black hole formation
\begin{align*}
  L_{bh}  \sim\frac{1}{a} \log  \frac{b\,S^{j_0-\frac{3}{2}}}{N^2}     \ .
\end{align*} 
When $j_0<\frac{3}{2}$ increasing the energy does not increase $  L_{bh}$ as expected from the gravitational intuition.
Indeed, for $j_0<\frac{3}{2}$ black hole formation should be irrelevant for all impact parameters \cite{Brower1}.
Furthermore, $j_0$ is bounded from above by the intercept $j(0)$.
As the 't Hooft coupling $g^2$ varies from 0 to $\infty$ and the intercept $j(0)$ goes from 1 to 2, there must be a critical value $g_c$ for which $j(0)=\frac{3}{2}$.
Therefore, black hole production should be absent in high energy scattering at weak coupling $g <g_c$ \cite{Shiraz}.
On the other hand, for $g> g_c$ there is black hole formation in the scattering process. 

%As in flat space, $L_{eik}-L_{bh}$ grows with $S$ and leaves ample room for using the eikonal approximation at high energies for intermediate values of the impact parameter.

%%%%%%%%%%%%%%%%%%%%%%%%%%%%%%%%%%%%%%
\subsection{Example}
%%%%%%%%%%%%%%%%%%%%%%%%%%%%%%%%%%%%%%
  
We shall now illustrate the use of the previous formulas in a particular example in $\mathcal{N}=4$ SYM.
We consider the R-current as the vector operator and Tr$Z^2$ as our scalar operator with $\Delta=2$. 
Using the results of \cite{BFKLpaper} and of sections \ref{SectionBFKL} and \ref{ImpactFactors} of this paper, for operators normalized as in (\ref{normop}) we found the weak coupling expressions,
 \begin{align*}
 \alpha_1(\nu)&=-\frac{i g^4 \left(\nu ^2+19\right)
  \tanh
   \left(\frac{\pi  \nu }{2}\right)}{192 N^2 \nu   \cosh^2\left(\frac{\pi  \nu }{2}\right)}\ ,\ \ \ \ \ \ \ \ \
    \alpha_2(\nu)= -\frac{i g^4 \left(\nu ^2-11\right)
    \tanh
   \left(\frac{\pi  \nu }{2}\right)}{128 N^2 \nu   \cosh^2\left(\frac{\pi  \nu }{2}\right)}\ ,\\
      \alpha_3(\nu)&=-\frac{5 i g^4 \tanh \left(\frac{\pi  \nu }{2}\right)}{128 N^2
   \nu   \cosh^2\left(\frac{\pi  \nu }{2}\right)} \ ,\ \ \ \ \ \ \ \ \ \ \ \ \ \ \ 
  \alpha_4(\nu)=\frac{i g^4   \tanh \left(\frac{\pi  \nu }{2}\right)}{128 N^2
   \nu   \cosh^2\left(\frac{\pi  \nu }{2}\right)} \ ,
  \end{align*} 
where $g^2=g_{\rm YM}^2N$ is the 't Hooft coupling. The Reggeon spin is the well known $n=0$ component of the BFKL spin
$$
j(\nu)=1+\frac{g^2}{4\pi^2}\left(2\Psi(1)-\Psi\left(\frac{1+i\nu}{2}\right)
-\Psi\left(\frac{1-i\nu}{2}\right)\right)+\cdots\ .
$$
   
Using the formulas in appendix \ref{Ftrans}, we can determine the functions $\beta_k(\nu)$. This gives $\beta_2=\beta_3=0$ as expected for a conserved current. However, we also get $\beta_4=0$ which is not required by conservation. Indeed, the only non-zero $\beta_k$ is 
 \begin{align*}
 \beta_1(\nu)=\frac{2 i g^4 \tanh \left(\frac{\pi  \nu
   }{2}\right)}{N^2 \pi ^4 \nu  \left(\nu ^2+1\right)^2}\ .
 \end{align*} 
This gives rise to a purely diagonal phase shift
\begin{align*}
\chi^{\hat{\mu}}_{\ \hat{\tau}}(S,L) \approx  \delta^{\hat{\mu}}_{\ \hat{\tau}} \, \frac{6 i g^4 }{\pi N^2 } \int  
  d\nu~ S^{j(\nu)-1}\,  \frac{\tanh \left(\frac{\pi  \nu
   }{2}\right)}{  \nu  \left(\nu ^2+1\right)^2} \, \Omega_{i\nu} (L) \,.
\end{align*}
It is interesting to compare this result with the result at strong coupling
\begin{align*}
\chi^{\hat{\mu}}_{\ \hat{\tau}}(S,L) \propto \delta^{\hat{\mu}}_{\ \hat{\tau}} \, \frac{S}{N^2  } \int  
  d\nu  
\frac{1}{   \nu ^2+4}  \,\Omega_{i\nu} (L) \,,
\end{align*}
computed in \cite{shockDIS} by studying the propagation of a gauge boson across a gravitational shockwave produced by the very energetic scalar particle. 
At strong coupling, the Reggeon is just the graviton with spin $j(\nu)=2$ and the dependence on the impact parameter is given by the scalar propagator in $H_3$.
The diagonal nature of the phase shift at strong coupling is easy to understand from the form of the gauge field energy-momentum tensor that couples to the graviton. The fact that the phase shift is also diagonal at weak coupling suggests that this property holds for all  values of the coupling.

One can also compute 
\begin{align*}
\gamma_1(\nu)=-\frac{3-i\nu}{1-i\nu} \gamma_2(\nu)\ ,\ \ \ \ \ \  
   \gamma_2(\nu)= \frac{i \pi ^5 g^4 \left(\nu ^2+1\right) \sinh^2\left(\frac{\pi  \nu }{2}\right) }{128 N^2 \nu ^2\left(\nu ^2+4\right)\cosh^4\left(\frac{\pi  \nu }{2}\right)}\ ,
   \end{align*}
to leading order in $g^2$, therefore determining the Regge representation (\ref{StructFunc}) of the structure functions.

%%%%%%%%%%%%%%%%%%%%%%%%%%%%%%%%%%%%%%
\section{CFT   amplitude}\label{Amplitude}
%%%%%%%%%%%%%%%%%%%%%%%%%%%%%%%%%%%%%%

In this section we shall analyze the general form of the  four-point correlation function $A^{mn}(x_i)$,
where we recall that the points $x_i$ are taken on different Poincar\'e patches.
The vector can be a conserved current only for $\xi=d-1$, where $d$ is the space-time dimension. For now we 
leave $d$ as arbitrary and specify to $d=4$ later on.
The general form of $A^{mn}(x_i)$ here derived is actually exact for any CFT, before the Regge limit.
Then we will consider the correlation function in the CFT Regge limit.

This section is rather technical and uses the embedding formalism \cite{embeddingformalismDirac,embeddingformalism} that we develop in a separate publication \cite{embedding}.  Here
the goal is to justify the Regge form of the amplitude already written in  (\ref{ReggeA}). The reader may wish to skip the technicalities,
jumping to section \ref{ReggeTheory} bellow.

We shall now  introduce very briefly the necessary notation used in the embedding formalism.
Let $P\in \mathbb{R}^{2,d}$ be a point in the embedding space. 
Points in physical space-time are identified with null vectors in  $\mathbb{R}^{2,d}$ up to the re-scallings
\[
P^2=0 \,,\ \ \ \ \ \ \ \ P\sim \lambda P\ \ \ (\lambda>0)\,.
\]
In the following we shall use light-cone coordinates
\[
P^M=(P^+,P^-,P^m)\,,
\]
with $P^m\in \mathbb{M}^d$. In these coordinates the metric takes the form
\[
P\cdot P = \eta_{MN} P^M P^N = -P^+P^- + \eta_{mn} P^m P^n\,.
\] 
We shall consider the projection of embedding points to different Poincar\'e patches, for example, to the Poincar\'e patch
\[
P^M=(1,x^2, x^m)\,,
\]
where $x\in \mathbb{M}^d$. Given two points in the same Poincar\'e patch, we have that
\[
P_{ij}\equiv -2P_i\cdot P_j  =  (x_i-x_j)^2
\]
is the Lorentzian distance in the physical space $\mathbb{M}^d$.

In the embedding formalism  we start by defining the embedding amplitude
\begin{equation}
A_{MN}(P_i)  =  \frac{{\cal A}_{MN}(P_i)}{(P_{13})^{\xi}(P_{24})^{\Delta}}\,.
\label{EmbeddingAmp}
\end{equation}
After some specific choices of light cone sections for the external points, the physical amplitude is
given by the projection
\[
A_{mn} (x_i)= \frac{\partial P_1^M}{\partial x_1^m} \,\frac{\partial P_3^N}{\partial x_3^n} \, A_{MN}(P_i)\,.
\]
In (\ref{EmbeddingAmp}) the embedding reduced amplitude  ${\cal A}_{MN} $
depends on all the external points $P_i$, has weight zero under any re-scalling $P_i\rightarrow \lambda P_i$ of a point $P_i$,  satisfies the orthogonality conditions
\begin{equation}
P_1^M  {\cal A}_{MN} = 0 \,,\ \ \ \ \ \ \  P_3^M {\cal A}_{MN}= 0\,,
\label{Transverse}
\end{equation}
and is defined up to the equivalence
\begin{equation}
{\cal A}_{MN}  \sim {\cal A}_{MN} + p_{1M} R_N + p_{3N} S_M\,,
\label{Equiv}
\end{equation}
since both tensors have the same projection to the light-cone sections.

The problem of finding the generic form of the amplitude reduces to finding  the most general embedding tensor ${\cal A}_{MN}$, which will have the form
\[
{\cal A}_{MN} = \sum_k f_k(z,\bar{z}) \,T_{MN}^{k}\,,
\]
where the $f_k(z,\bar{z}) $ are functions of the cross ratios
\begin{eqnarray}
z\bar{z} = \frac{P_{13}P_{24}}{P_{12}P_{34}}\,,
\ \ \ \ \ \ \ \ \  \  \ \ \ \  \ 
(1-z)(1-\bar{z})= \frac{P_{14}P_{23}}{P_{12}P_{34}} \,.
\nonumber
\end{eqnarray}
Thus, after finding  all possible tensor structures  $T_{MN}^{k}$, all the dynamical information regarding the amplitude is contained in the functions $f_k(z,\bar{z}) $.
The physical amplitude is simply given by
\begin{equation}
A_{mn} (x_i) = \frac{{\cal A}_{mn}(x_i)}{(P_{13})^{\xi}(P_{24})^{\Delta}}\,, \label{genAPi}
\end{equation}
with reduced amplitude
\[
{\cal A}_{mn}(x_i) = \sum_k f_k(z,\bar{z}) \,t_{mn}^{k}(x_i)\,,
\]
where
\[
t_{mn}^{k}(x_i) = \frac{\partial P_1^M}{\partial x_1^m} \,\frac{\partial P_3^N}{\partial x_3^n} \, T_{MN}^{k}(P_i)\,.
\]

There are five independent weight zero  tensor structures $T_{MN}^{k}$ satisfying conditions (\ref{Transverse}) and (\ref{Equiv}). 
However, we must impose the additional symmetry constraint 
\[
{\cal A}_{MN}(P_1,P_2,P_3,P_4)={\cal A}_{NM}(P_3,P_2,P_1,P_4)={\cal A}_{MN}(P_1,P_4,P_3,P_2)\,,
\]
coming from invariance under exchange of points 
$P_1$ and $P_3$, and also under exchange of points $P_2$ and $P_4$.
For both  transformations the action on the cross ratios is
\[
z\rightarrow \frac{z}{z-1}\,,\ \ \ \ \ \ \bar{z}\rightarrow \frac{\bar{z}}{\bar{z}-1}\,.
\]
These symmetries reduce the number of independent  tensor structures to four and impose constrains on the corresponding
functions  $f_k(z,\bar{z}) $. To write these tensors it is convenient to define
the basic weight zero building block
\[
(ij)^{MN} = \frac{P_{13}} {P_{1i} \, P_{3j}}\, P_i^M P_j^N\,,\ \ \ \ \ \ \ \ \ \ \ \ (i\ne 1,\, j\ne 3).
\]
Then,  the four independent tensors $T_{MN}^{k}$ can be chosen to be
\begin{align}
T_{MN}^{1} &= \eta_{MN} +2 (31)_{MN} \,,
\nonumber\\
T_{MN}^{2}&= (22)_{MN} + (44)_{MN} - (24)_{MN} - (42)_{MN}  \,,
\nonumber\\
\frac{1}{2}\left( T_{MN}^{3} + T_{MN}^{4}\right) &= (21)_{MN} + (34)_{MN} - (31)_{MN} - (24)_{MN}  \,,
\label{TTensors}\\
\frac{1}{2}\left( T_{MN}^{3} - T_{MN}^{4}\right) &= (41)_{MN} + (32)_{MN} - (31)_{MN} - (42)_{MN}  \,.
\nonumber
\end{align}
Under the exchange of points $P_1$ and $P_3$, or of points  $P_2$ and $P_4$, we have
\begin{eqnarray}
&&T_{MN}^{k} \rightarrow T_{MN}^{k}\,,\ \ \ \ \ \ \ \ \ \ \ (k=1,2,3)
\nonumber\\
&&T_{MN}^{4} \rightarrow -\,T_{MN}^{4}\,.
\nonumber
\end{eqnarray}
Thus, exchange symmetry implies that
\begin{eqnarray}
&&
f_k(z,\bar{z}) = f_k\left(\frac{z}{z-1},\frac{\bar{z}}{\bar{z}-1}\right)\,,\ \ \ \ \ \ \ \ \ \ \ (k=1,2,3)
\nonumber\\
&&f_4(z,\bar{z}) =- f_4\left(\frac{z}{z-1},\frac{\bar{z}}{\bar{z}-1}\right)\,.
\label{fsymmetry}
\end{eqnarray}
We conclude that the general form of the amplitude is determined by these four functions, with the embedding reduced amplitude given by
\begin{align}
{\cal A}_{MN}=& \ f_1(z,\bar{z})  \,\Big[ \eta_{MN} +2 (31)_{MN} \Big] +
%\nonumber\\&
 f_2(z,\bar{z}) \,\Big[ (22)_{MN} + (44)_{MN} - (24)_{MN} - (42)_{MN} \Big]  
\nonumber\\
&  + \big(f_3(z,\bar{z})+ f_4(z,\bar{z})\big) \, \Big[ (21)_{MN} + (34)_{MN} - (31)_{MN} - (24)_{MN} \Big]  
\nonumber\\
&  + \big(f_3(z,\bar{z})- f_4(z,\bar{z})\big)  \, \Big[  (41)_{MN} + (32)_{MN} - (31)_{MN} - (42)_{MN} \Big]   \,.
\nonumber
\end{align}

In \cite{embedding} we show that, when the vector primary is a conserved current of dimension $\xi=d-1$,
the projection  to the light cone sections of the embedding conservation equation
\begin{equation}
\partial_M A^{MN}= 0\,,
\label{Conservation0}
\end{equation}
gives precisely  the usual  Ward identity $\partial_m A^{mn} = 0$. Equation (\ref{Conservation0})
gives three differential equations involving the functions of the cross ratios $f_i(z,\bar{z})$, arising from the coefficients multiplying $P_1^N$, $P_2^N$ and $P_4^N$,
since the term proportional to $P_3^N$ is pure gauge.
It turns out that only two of these equations are linearly independent, so that we have two differential equations implied by current conservation
\begin{align}
&(d-1) \Big( (1 + (1 - z) (1 - \bar{z})) f_3 - (z (1 - \bar{z}) + \bar{z}) f_4\Big) 
\nonumber\\&
+2\,\frac{(1-z) (1-{\bar{z}})z^2 }{z-{\bar{z}}}\,\partial f_1
- (1 - z) (2 - \bar{z})z \,\partial f_3
+ (1 - z) z \bar{z} \,\partial f_4
\nonumber\\&
+2\,\frac{(1-z) (1-{\bar{z}})\bar{z}^2 }{{\bar{z}}-z}\,\bar{\partial}f_1
 - (1 - \bar{z})(2 - z) \bar{z}\,\bar{\partial}f_3
 + (1 - \bar{z}) z\bar{z}\,\bar{\partial}f_4=0\,,
\label{Conservation} \\
&(d-2) (z  + \bar{z} - z\bar{z}) f_2 
-\big(f_3- f_4\big)
- (d-2) (1 - z) (1 - \bar{z})\big( f_3 +  f_4\big)
\nonumber\\
&-2\,\frac{(1-z) (1-{\bar{z}})z }{z-{\bar{z}}}\,\partial f_1
-(1-z) z {\bar{z}} \,\partial f_2
 + (1 - z)(1 - \bar{z}) z \big( \partial f_3 + \partial f_4\big)
\nonumber\\&
-2\,\frac{(1-z) (1-{\bar{z}})\bar{z} }{{\bar{z}}-z}\,\bar{\partial}f_1
-(1-{\bar{z}})z {\bar{z}} \,\bar{\partial} f_2
 + (1 - z)(1 - \bar{z}) \bar{z} \big( \bar{\partial} f_3 + \bar{\partial} f_4\big)=0\,.
\nonumber
\end{align}

%%%%%%%%%%%%%%%%%%%%%%%%%%%%%%%%%%%%%%
\subsection{Kinematics}
%%%%%%%%%%%%%%%%%%%%%%%%%%%%%%%%%%%%%%

Now we define  the kinematics of the embedding external points, as appropriate to study the CFT Regge limit of section \ref{Regge}. 
First we parametrize  the external points $P_i$ in the the central Poincar\'e patch with the embedding coordinates
\begin{equation}
P^A = (y^+,y^-,1,y^2,y_\perp)\,,
\label{CentralPatch}
\end{equation}
where the  coordinates of the external points in the physical Minkowski space  $\mathbb{M}^d$  are given by  $y^a=(y^+,y^-,y_\perp)$.
However, we are interested in parameterizing these points in different  Poincar\'e patches, with coordinates $x_i$, such that each point sits close to the origin of the corresponding
patch, as explained in section \ref{Regge}. In the embedding formalism this corresponds to the choice
\begin{eqnarray}
&&P_1^M =\left(-1,-x_1^2,x_1^m\right)\,, 
\ \ \ \ \ \ \ \ \ \ \ \ \ \ \ 
P_2^M = \left(-x_2^2,-1,x_2^m\right)\,,
\nonumber\\
&&P_3^M = \left(1,x_3^2,-x_3^m\right) \,,
\label{xPatches}\ \ \ \ \ \ \ \ \ \ \ \ \ \ \ \ \ 
P_4^M =  \left(x_4^2,1, -x_4^m\right)\,,
\end{eqnarray}
where $x_i^m=(x_i^+,x_i^-,x_{i\perp})$. An easy and elegant way to derive the conformal transformations between the $y_i$ and the $x_i$ given in (\ref{ConformalTrans1})
and (\ref{ConformalTrans2}) is to equate (\ref{CentralPatch}) and (\ref{xPatches}) for each point, and then use the identification $P\sim \lambda P\ (\lambda>0)$.

Let us now comment on the exactness of the general form of the amplitude that will be presented bellow.
We can use conformal invariance to fix the two external points $x_1$ and $x_4$ to the origin, and then define $x=-x_3$ and $\bar{x}=x_2$, so that
\begin{eqnarray}
&&P_1^M =(-1,0,0)\,,
\ \ \ \ \ \ \ \ \ \ \ \ \ \ \ \ \ \  P_2^M = \left(-\bar{x}^2,-1,\bar{x}^m\right)\,,
\nonumber\\
&&P_3^M = \left(1,x^2,x^m\right)\,,
\label{ExactxPatches}\ \ \ \ \ \ \ \ \ \ \ \ \ \ \ P_4^M  = (0,1,0)\,.
\end{eqnarray}
After projecting the embedding amplitude to these light-cone sections, we will obtain an exact expression for the general form of the amplitude, which
is manifestly invariant under the residual transverse conformal group $SO(1,1)\times SO(3,1)$. Hence the 
general form of the amplitude presented bellow is exact. 
When taking the Regge limit, we may then choose instead the more symmetric choice  with $x\simeq x_1-x_3$ and $\bar{x}\simeq x_2-x_4$, which corresponds to
the choice of light-cone sections  (\ref{xPatches}) given in the previous paragraph.

To  project the different tensor structures $T_{MN}$ given in (\ref{TTensors}) to the light-cone sections (\ref{ExactxPatches})
we need to compute
\begin{eqnarray}
&&
\frac{\partial P_1^A}{\partial x_1^m} =  \left(0,-2x_{1m},\delta^a_m\right) = \left(0,0,\delta^a_m\right)\,,
\nonumber\\
&&
\frac{\partial P_3^B}{\partial (-x_3^n)} =   \left(0,-2x_{3n},\delta^b_n\right) = \left(0,-2x_n,\delta^b_n\right)\,.
\nonumber
\end{eqnarray}
Then the projection $t_{mn}$ to the above  light-cone sections simplifies to
\[
t_{mn} = - 2x_n T_{m-} + T_{mn}\,.
\]
Noting that  (\ref{ExactxPatches}) gives
\begin{eqnarray}
&P_{12}=1\,,\ \ \ \ \ \ \ 
P_{13}=-x^2\,,\ \ \ \ \ \ \ 
P_{14}=-1\,,&
\nonumber\\
&
P_{23}=-1-2x\cdot \bar{x}-x^2 \bar{x}^2\,,\ \ \ \ \ \ \ 
P_{24}=-\bar{x}^2\,,\ \ \ \ \ \ \ 
P_{34}=1\,,&
\nonumber
\end{eqnarray}
a simple computation yields
\begin{align}
t_{mn}^{1}&= \eta_{mn} - 2\,\frac{x_m x_n}{x^2}  \,,
\nonumber\\
t_{mn}^{2}&= \frac{x^2 \bar{x}_m \bar{x}_n}{1+2 x\cdot\bar{x}+x^2 \bar{x}^2}
+\frac{x^2 \bar{x}^2  \bar{x}_m x_n}{1+2 x\cdot\bar{x}+x^2 \bar{x}^2} \,,
\nonumber\\
\frac{1}{2} \left(t_{mn}^{3} + t_{mn}^{4}\right)&= \frac{x_m x_n}{x^2}+x_n\bar{x}_m\,,
\nonumber\\
\frac{1}{2} \left(t_{mn}^{3} - t_{mn}^{4}\right)&=\frac{x_m x_n}{x^2}
-\frac{ x_m \bar{x}_n}{1+2 x\cdot\bar{x}+x^2 \bar{x}^2}  
- \frac{\bar{x}^2 x_m x_n}{1+2 x\cdot\bar{x}+x^2 \bar{x}^2}\,.
\nonumber
\end{align}
All tensors are invariant under the residual scale invariance  and transform as tensors under the residual $SO(3,1)$.

We arrive at the final result for the amplitude in the Lorentzian kinematical setting of interest
\begin{align}
A_{mn} = \frac{{\cal A}_{mn}(x,\bar{x})}{(x^2-i\epsilon_x)^{\xi}
(\bar{x}^2-i\epsilon_{\bar{x}})^{\Delta}}\,, \label{genAxxbar}
\end{align}
with 
\[
{\cal A}_{mn} = \sum_{k=1}^4 f_k(\sigma,\rho) \,t_{mn}^{k}(x,\bar{x})\,.
\]
where $\sigma$ and $\rho$ are the $SO(1,1)\times SO(3,1)$ cross ratios defined in (\ref{SigmaRho}). 
In (\ref{genAPi}) we have been careless about the choice of the $i\epsilon$-prescription determining the branch cuts in the denominator. 
The correct $i\epsilon$-prescription is the one of the two-point function and was studied in detail in \cite{iepsilon,Mythesis}. 
In (\ref{genAxxbar}) we have inserted the appropriate $i\epsilon$ for the kinematical choice (\ref{ExactxPatches}), as already done in (\ref{ReggeA}) of section
\ref{relscat}. This is the general form of the four-point correlation
function as dictated by conformal symmetry. 

%%%%%%%%%%%%%%%%%%%%%%%%%%%%%%%%%%%%%%
\subsection{Regge limit}
%%%%%%%%%%%%%%%%%%%%%%%%%%%%%%%%%%%%%%

As explained in section \ref{Regge}, the high energy CFT Regge limit is defined by $x_i\to 0$ (or $\sigma\rightarrow 0$ with $\rho$
fixed). In this limit the above $t_{mn}$ tensor structures simplify to
\begin{align}
t_{mn}^{1} &= \eta_{mn} - 2\,\frac{x_m x_n}{x^2}  \,,
\nonumber\\
t_{mn}^{2} &=\sigma^2\,\frac{\bar{x}_m \bar{x}_n}{\bar{x}^2} + O(\sigma^3) \,,
\nonumber\\
t_{mn}^{3} &= 2\,\frac{x_m x_n}{x^2}+O(\sigma)   \,,
\nonumber\\
t_{mn}^{4} &= \sigma\, \frac{x_m \bar{x}_n + \bar{x}_m x_n}{|x||\bar{x}|}  + O(\sigma^2) \,.
\nonumber
\end{align}
It is now clear that the behaviour of the amplitude will depend on the expansion in
powers of $\sigma$ of the functions $f_k(\sigma,\rho)$, as these contain all the dynamical information.
In particular, we are interested in the leading behaviour of $f_k(\sigma,\rho)$ when a particle of spin 
$J$ is exchanged in the t-channel. Since the amplitude can be expanded in conformal partial waves of
definite spin   and conformal dimension, one way to derive the behaviour of the functions $f_k(\sigma,\rho)$ is 
to study the Regge limit of the t-channel conformal partial waves of spin $J$. The general construction of  the conformal partial wave expansion using the embedding formalism is given in \cite{embedding}. 
For the present purposes all we need to know is that, in the limit of $\sigma\rightarrow 0$, the functions
$f_k(\sigma,\rho)$ associated to an exchange of a spin $J$ state have the expansion
\begin{eqnarray}
&&f_1(\sigma,\rho) = E(\rho) \,\sigma^{1-J} + O(\sigma^{2-J}) \,,
\nonumber\\
&&f_2(\sigma,\rho) = G(\rho)\,\sigma^{-1-J}+ O(\sigma^{-J}) \label{CPWRegge}\,,
 \\
&&f_3(\sigma,\rho) =  \left(E(\rho) + \frac{F(\rho)}{2}\right)\,\sigma^{1-J} + O(\sigma^{2-J}) \,,
\nonumber\\
&&f_4(\sigma,\rho) =  H(\rho)\,\sigma^{-J} + O(\sigma^{1-J})\,,
\nonumber
\end{eqnarray}
for some functions $E(\rho)$, $F(\rho)$, $G(\rho)$ and $H(\rho)$, which  depend on the conformal dimension  of the exchanged state and whose
explicit form is not important for the present argument.
We refer the reader to appendix \ref{CPW} for a proof of this result.

Thus, we finally arrive at a very simple form of the reduced amplitude ${\cal A}_{mn}$  for a spin $J$ conformal partial wave in the Regge limit,
\begin{equation}
{\cal A}_{mn} \approx 
\sigma^{1-J} \left[ E(\rho)  \,\eta_{mn}
+F(\rho)\,\frac{x_m x_n}{x^2}
+ G(\rho) \, \frac{\bar{x}_m \bar{x}_n}{\bar{x}^2} 
+H(\rho) \,\frac{x_m \bar{x}_n+\bar{x}_m x_n }{|x||\bar{x}|}\right]\,.
\label{AmplitudeRegge}
\end{equation}
To this leading order the amplitude is determined by four unknown functions of the cross ratio $\rho$.
One can also replace the expansion in powers of $\sigma$ of the functions $f_k(\sigma,\rho)$ in the conservation equations (\ref{Conservation}) to obtain 
\begin{eqnarray}
(2 d-3+J)\, E(\rho)
+(d-2+J) \,F(\rho) 
-(2 d-2+J) \cosh\rho \,H(\rho) 
\nonumber\\
+ \coth\rho\, E'(\rho)-\sinh\rho \,H'(\rho)=0\,,
\nonumber\\
(2 d-3+J) \sinh\rho \cosh\rho\, G(\rho) - (d-3+J) \sinh\rho\, H(\rho)
\nonumber\\
-E'(\rho) +(\sinh\rho)^2 G'(\rho)=0\,,
\nonumber
\end{eqnarray}
where $'$ stands for the $\rho$ derivative. We conclude that, in the  Regge limit, the amplitude with a conserved current operator  
depends on two unknown functions of the  transverse conformal group cross ratio $\rho$.

As a final remark we note that, in the Regge limit, under the exchange of points $P_1$ and $P_3$, we have $x\rightarrow - x$, while $\bar{x}$ is unchanged.
From the symmetry properties of the functions $f_k(z,\bar{z})$ given in (\ref{fsymmetry}), we see that the functions  $E(\rho)$, $F(\rho)$ and $G(\rho)$ are even under this transformation, while
the function $H(\rho)$ is odd. It is then clear that the amplitude (\ref{AmplitudeRegge}) remains invariant under this transformation.

%%%%%%%%%%%%%%%%%%%%%%%%%%%%%%%%%%%%%%
\subsection{Regge theory}\label{ReggeTheory}
%%%%%%%%%%%%%%%%%%%%%%%%%%%%%%%%%%%%%%
 
Let us now consider an amplitude which includes contributions of conformal  partial waves of all spins.
The conformal partial wave expansion can be written in the form \cite{classic,classicPolyakov,CPWSofia,Lorenzo}
\begin{align*}
 \frac{{\cal A}^{MN}(P_i)}{(P_{13})^{\xi}(P_{24})^{\Delta}}= \sum_J \int d\nu\, a(\nu,J) \,G_{\nu,J}^{MN} (P_i)
\end{align*}
where $G_{\nu,J}^{MN}$ is the conformal partial wave of spin $J$ and dimension $\frac{d}{2}+i\nu$. 
In this case one must resum all contributions, in order to determine the
small $\sigma$ behaviour of the amplitude. 
Following \cite{Lorenzo}, we use the Sommerfeld-Watson transform to write the sum over $J$ as a contour integral. 
From (\ref{AmplitudeRegge}) we conclude that, for each $\nu$, the Regge limit of the integral over $J$ is dominated by the right most non-analyticity  in the complex $J$ plane. 
Since we are considering the planar amplitude, dual to tree-level string theory, we expect this non-analyticity to be a simple Regge pole at $J=j(\nu)$.

The analysis is entirely analogue to the case of the amplitude for scalar operators, which has the form \cite{Lorenzo}
\[
{\cal A}= 2\pi i \int d\nu \,(-)^{j(\nu)} \sigma^{1-j(\nu)} \alpha(\nu)\,\Omega_{i\nu}(\rho)\,,
\]
where $J=j(\nu)$ is the leading Regge pole. The function $\Omega_{i\nu}(\rho) $ is just the harmonic scalar function on the conformal transverse space $H_3$, 
satisfying
\[
\Omega_{i\nu}(\rho) = \frac{\nu}{4\pi^2}\,\frac{\sin{\nu\rho}}{\sinh\rho}\,,\ \ \ \ \ \ \ \ \left(\square_{H_3} + \nu^2 +1\right)\,\Omega_{i\nu}(\rho) =0\,.
\]

In the present case of the four-point function of two vector and two scalar operators, the Regge form of the reduced amplitude is
\begin{equation}
{\cal A}^{mn} = 2\pi i \sum_{k=1}^4\int d\nu \,(-)^{j(\nu)} \sigma^{1-j(\nu)} \alpha_k(\nu)\left(\Omega_{i\nu}^{k}(\rho)\right)^{mn}\,,
\label{ReggeAmplitude}
\end{equation}
where now the functions $\Omega_{i\nu}^{k}(\rho)$ are a basis of tensor functions.
To derive their explicitly form, we first observe that in the Regge limit $x\approx x_3-x_1$, where the $x$'s are the coordinates  of points 1 and 3 in 
the corresponding  Poincar\'e patches, as explained in section \ref{Regge}. 
Since the vector operator is inserted at points 1 and 3, and the other points contain only scalar operators, one expects
the tensor structures of the $\Omega$ functions to be local in $x$, i.e. to be constructed from the metric and from  $x$ and its derivatives acting on a scalar function. This expectation follows from
the factorization of the Regge theory amplitude. Indeed there 
are precisely four such structures that can be constructed, correctly matching the counting of (\ref{AmplitudeRegge}). 
We shall introduce the following convenient basis of tensor functions
\[
\left(\Omega_{i\nu}^{k}(\rho)\right)^{mn} = {\cal D}^{mn}_{k} \,\Omega_{i\nu}(\rho)\,,
\]
where the explicit expressions for the differential  operators $ {\cal D}^{mn}_{k}$ were anticipated in (\ref{DTensors}). 
The reason for this particular choice of the $ {\cal D}^{mn}_{k}$ will become clear in the next section. 
Note that these operators are independent of $\bar{x}$.

%%%%%%%%%%%%%%%%%%%%%%%%%%%%%%%%%%%%%%
\section{Hard Pomeron in conformal gauge theories}\label{SectionBFKL}
%%%%%%%%%%%%%%%%%%%%%%%%%%%%%%%%%%%%%%

The general form of the four-point function $A^{mn}(x,\bar{x})$ derived in the previous section relies only on conformal symmetry 
and then on standard arguments in Regge theory. In particular (\ref{ReggeAmplitude})  applies to
any CFT at any value of the coupling constant, whenever the amplitude is dominated by a Regge pole. In this section we focus on the case of gauge theories in the weak coupling regime. We shall focus on an amplitude dominated by the planar diagram associated to the exchange of a hard BFKL
 pomeron, whose spin is approximately $1$. As explained in the introduction, this is important in the Regge limit of low Bjorken ${\rm x}$ in DIS, where the
exchange of a hard pomeron sets the growth of the cross section leading to gluon saturation. In the context of the AdS/CFT duality, the Regge pole interpolates between
the Pomeron at weak coupling and a reggeized spin $2$ graviton in the bulk of AdS at strong coupling \cite{Brower1}.

In the limit of vanishing 't Hooft coupling $g\rightarrow 0$, the leading contribution to the pomeron comes from a pair of gluons in a color singlet state. The spin is strictly $1$,
and therefore in this limit  the amplitude will not depend on $\sigma=|x| |\bar{x}|$. It will depend on the single cross ratio $\rho$, given by
$\cosh\rho = - x\cdot \bar{x} /(|x||\bar{x}|)$,
where we recall that we chose both $x$ and $\bar{x}$ are in the future light-cone of four-dimensional Minkowski space $\mathbb{M}^{4}$.
The cross ratio $\rho$ is the geodesic distance between $x/|x|$ and $\bar{x}/|\bar{x}|$ on the unit  three-dimensional hyperboloid $H_{3}$.
The amplitude takes then the BFKL form in position space, similar to the case of scalar operators given in   \cite{BFKLpaper},
\begin{align}
\mathcal{A}^{mn} (x,\bar{x})
  \simeq&-\frac{1}{N^2}\,\int_{\partial H_{3}} \frac{dz_1 dz_3}{\left(z_{13}\right)^2}\,
   \frac{dz_2 dz_4}{\left(z_{24}\right)^2}~
    V^{mn}(  x,z_1,z_3)  \; F(z_1,z_3,z_2,z_4) \; \bar{V}(\bar{x},z_2,z_4)  \,,
     \label{BFKLposition} 
\end{align}
where $F$ is the BFKL kernel and $V^{mn}$ and $ \bar{V}$ are the impact factors, respectively describing the coupling of the current and scalar operators to the pomeron.

In (\ref{BFKLposition})  the action of the transverse conformal group $SO(3,1)$ is made manifest by working again in the
embedding space, which in this case is the four dimensional Minkowski space $\mathbb{M}^{4}$.
Transverse space is then recovered by taking an arbitrary slice of the future 
light-cone, choosing a specific representative for each ray.
We shall denote with $\partial H_{3}$ any given choice of such slice, since it can be identified 
with the conformal boundary of the unit three-dimensional hyperboloid $H_{3}$.
This is the notation followed in  (\ref {BFKLposition}) where we replaced the integrals over the transverse space 
$\mathbb{R}^{2}$ with integrals over an arbitrary section $\partial H_{3}$ of the future light--cone. 
The standard transverse space $\mathbb{R}^{2}$ is recovered with the usual Poincar\'{e}
choice 
\begin{equation*}
z^a= (z^+,z^-,z_\perp) = \left(  1, z_\perp^2, z_\perp \right) \,,
\end{equation*}
where  $z_\perp\in\mathbb{R}^{2}$.
The inner product between two point $z_i$ and $z_j$ of this form computes the 
Euclidean distance in $\mathbb{R}^{2}$,
\[
z_{ij} \equiv -2 z_i\cdot z_j = \left(  z_{i\perp}- z_{j\perp}\right)  ^{2}\,.
\]
We refer the reader to \cite{BFKLpaper} for a more thoroughly discussion of the 
realization of the transverse conformal symmetry in the context of (\ref {BFKLposition}).\footnote{To clear our presentation we have changed  notation
with respect to \cite{BFKLpaper}. In this paper we call the embedding points $z=\left(  1, z_\perp^2, z_\perp \right)$, while in \cite{BFKLpaper} we used instead
${\bf z}=\left(  1, z^2,z\right)$. The present notation is less heavy because we mostly use embedding points.}

%%%%%%%%%%%%%%%%%%%%%%%%%%%%%%%%%%%%%%
\subsection{BFKL propagator}\label{BFKLkernel}
%%%%%%%%%%%%%%%%%%%%%%%%%%%%%%%%%%%%%%

%In the Born approximation, 
The leading contribution at high energies to the pomeron propagator 
comes from the exchange of a pair of gluons in a color singlet state, with transverse propagator
\begin{equation}
F(z_1,z_3,z_2,z_4) = 2\ln(z_{1\perp} - z_{2\perp})^2\ln(z_{3\perp}-z_{4\perp})^2 = 2\ln z_{12} \ln z_{34}\,.
\label{TwoGluon}
\end{equation}
When the scattering states are colorless the impact factors satisfy the infrared finiteness condition
\begin{equation}
\int_{\partial H_{3}} \frac{dz_1}{\left(z_{13}\right)^2}\,V^{mn}(  x,z_1,z_3)  =0\,,
\label{IRfinite}
\end{equation}
and similarly for $\bar{V}(\bar{x},z_2,z_4) $. The leading BFKL propagator may then be replaced with
the equivalent conformally invariant scalar function of dimension zero,
\[
F(z_1,z_3,z_2,z_4) = \ln\frac{z_{13}z_{24}}{z_{12}z_{34}}\,\ln\frac{z_{14}z_{23}}{z_{12}z_{34}}\,.
\]

\begin{figure}[t]
\begin{center}
\includegraphics[height=3.5cm]{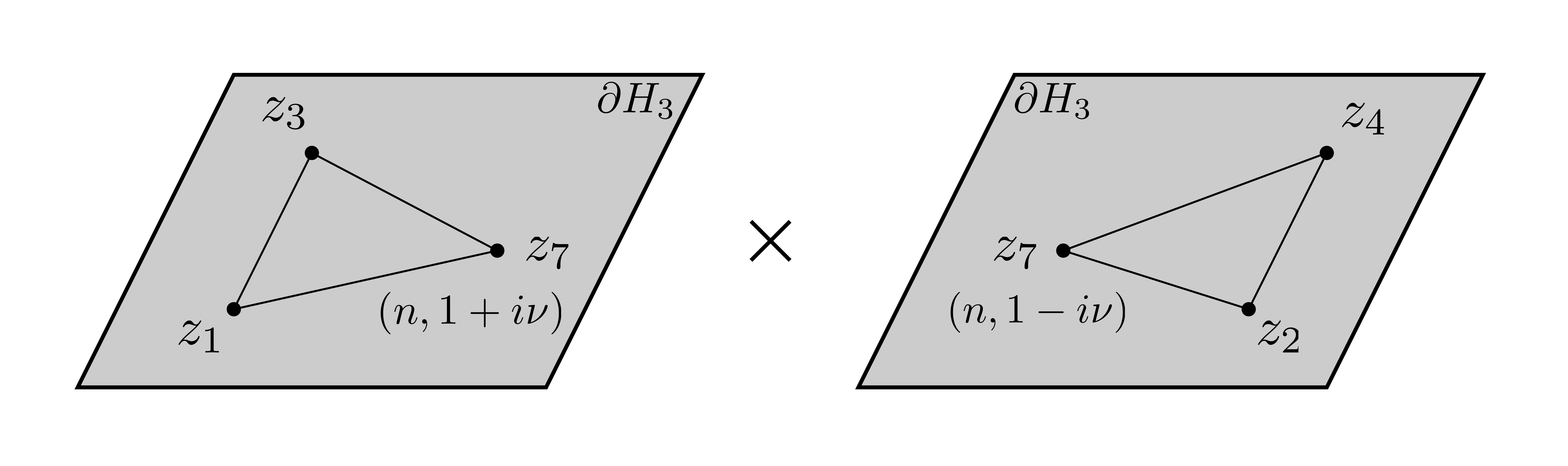}
\caption{The conformal partial waves used in the decomposition  of the BFKL propagator are obtained from the integral of the 
product of two 3-point functions. 
One 3-point function has scalars of dimension zero at $z_1$ and $z_3$ and a spin $n$ operator of dimension $1+i\nu$ at $z_7$,
while the other has scalars of dimension zero at $z_2$ and $z_4$ and a spin $n$ operator of dimension $1-i\nu$ at $z_7$.
\label{Lipatovkernel}}
\end{center}
\end{figure}

Lipatov \cite{Lipatov} found a beautiful decomposition of the leading BFKL propagator in  conformal partial waves
of transverse spin $n\ge 0$ and dimension $1+(n \pm 1)$, given by
\begin{align}
F(z_1,z_2,z_3,z_4) = \frac{4}{\pi^2} \,\sum_{n=0}^\infty\,2^n \int d\nu \,\frac{\nu^2+n^2}{\big( \nu^2 + (n-1)^2\big)\big( \nu^2 + (n+1)^2\big) }&
\nonumber\\
\int_{\partial H_3}
d z_7 \,E^{a_1\cdots a_n}_{\nu}(z_1,z_3,z_7) \,E^{b_1\cdots b_{n}}_{-\nu\ \ }(z_2,z_4,z_7)\;\eta_{a_1b_1}\cdots\,\eta_{a_nb_n}&\,,
\label{Kernel}
\end{align}
where we note that the poles at $i\nu=n \pm 1$ determine the dimension of the partial waves. The  tensor functions
$E_{\nu}(z_i,z_j,z_k)$ are conformal 
3-point functions on the transverse space of two scalar fields of zero dimension at  $z_i$ and $z_j$
and one symmetric and traceless spin $n$ field of dimension $1+i\nu$ at $z_k$. 
We shall see in the next section that  these conformal 3-point functions are essential for the construction of an impact factor basis with definite transverse spin. 
In appendix \ref{Efunctions} the explicit form of $E_{\nu}$ in the embedding formalism is given, as well as their projection to  
complex coordinates on $\mathbb{R}^{2}$  used in \cite{Lipatov}. The above representation of the pomeron propagator is pictured in figure \ref{Lipatovkernel}.

The 3-point functions $E_{\nu}$ satisfy the important orthogonality relation \cite{Lipatov}
\begin{align}
&\int_{\partial H_3} \frac{dz_1 dz_3 }{(z_{13})^2}\, E^{a_1\cdots a_{n'}}_{-\mu}(z_1,z_3,z_5) \,E^{b_1\cdots b_{n}}_\nu(z_1,z_3,z_7) = 
\label{Orthogonality} \\
&\delta_{n,n'}\,\delta(\nu-\mu) \,a_{n}(\nu)\, \delta(z_5,z_7)\, U^{a_1\cdots a_nb_1\cdots b_{n}}+ 
\delta_{n,n'}\,\delta(\nu+\mu)\,b_{n}(\nu) \,\frac{1}{(z_{57})^{1+i\nu}} \, V^{a_1\cdots a_nb_1\cdots b_{n}}(z_5,z_7) \,,\nonumber
\end{align}
where
\begin{eqnarray}
&&
a_{n}(\nu) = \frac{4\pi^4}{\nu^2+n^2}\,,
\ \ \ \ \ \ \ \ \ \ \ \ 
b_{n}(\nu) = 4\pi^3\,\frac{2^{2i\nu}}{n-i\nu}\,\frac{\Gamma\Big( \frac{n+1-i\nu}{2}\Big)\Gamma\Big( \frac{n+i\nu}{2}\Big)}{\Gamma\Big( \frac{n+1+i\nu}{2}\Big)\Gamma\Big( \frac{n-i\nu}{2}\Big)}\,.
\nonumber
\end{eqnarray}
and the tensors $U$ and $V$ are  symmetric and traceless in both $a_i$ and $b_i$ indices.
For the sake of clarity  in the exposition we only give the explicit expressions for $U$ and $V$ in the  appendix \ref{Efunctions}.

%%%%%%%%%%%%%%%%%%%%%%%%%%%%%%%%%%%%%%
\subsection{Impact factor}\label{ImpactFactors1}
%%%%%%%%%%%%%%%%%%%%%%%%%%%%%%%%%%%%%%

We shall now consider the general form of the vector current impact factor $V^{mn}(x,z_1,z_3)$. By conformal invariance 
it can only depend on functions of the single cross ratio
\[
u=\frac{(-x^2) z_{13}}{(-2x\cdot z_1)(-2x\cdot z_3)}\,,
\]
it must be constructed from tensor structures of weight zero in $x$, $z_1$ and $z_3$, and it must be symmetric under $z_1\leftrightarrow z_3$.
A simple counting shows that there are only five allowed tensor structures. Let us then introduce the following tensor basis for the impact factor
\begin{align}
&{\cal I}_{1}^{mn}  =  \eta^{mn}\:,\ \ \ \ \ \ \ \ \ \ \ \ 
{\cal I}_{2}^{mn}  =  \frac{x^{m}x^{n}}{x^{2}}\:,\ \ \ \ \ \ \ \ \ \ \ \ 
{\cal I}_{3}^{mn}  =  \frac{x^{m} z_1^{n}+x^{n} z_1^{m}}{-2 x\cdot z_1} + \frac{x^{m} z_3^{n}+x^{n} z_3^{m}}{-2 x\cdot z_3}\:,\nonumber
\\
&{\cal I}_{4}^{mn}  =  \frac{z_1^{m}z_1^{n}(-x^2)}{(-2x\cdot z_1)^2} + \frac{z_3^{m}z_3^n(-x^2)}{(-2x\cdot z_3)^2}\:,\ \ \ \ \ \ \ \ \ \ \ \ 
{\cal I}_{5}^{mn}  =  \frac{z_1^mz_3^n + z_3^mz_1^n}{z_{13}}\:.\ \ \ \ \ \ \ \ \ \label{ITensors}
\end{align}
A general impact factor will  be a linear combination of the form
\begin{equation}
V^{mn}=\sum_{k=1}^{5}\: h_{k}(u)\: {\cal I}_{k}^{mn}\:,
\label{IF}
\end{equation}
for general functions $h_{k}(u)$ of the cross ratio $u$. 

Although the general form of the four-point amplitude (\ref{AmplitudeRegge}) contains only four tensor structures, we see that the impact factor for the
vector operator contains five tensor structures. The reason for this apparent mismatch is clear. The impact factor for the scalar operator has a  single
scalar structure, which only overlaps with the spin 0 component of the BFKL propagator expansion (\ref{Kernel}). Consequently, the spin 2 component of the 
vector impact factor will not contribute to a 4-point amplitude of the form (\ref{AmplitudeRegge}), as it is clear from  (\ref{BFKLposition}). This leaves only four 
tensor structures in the vector impact factor, which have, as we shall see, transverse spin 0. 

In the following we shall construct a basis for both the spin 0 and spin 2 
components of the vector impact factor. We shall then explain how to decompose a general impact factor of the form (\ref{IF}) in its spin 0 and spin 2 components.

%%%%%%%%%%%%%%%%%%%%%%%%%%%%%%%%%%%%%%
\subsubsection{Transverse spin 0}
%%%%%%%%%%%%%%%%%%%%%%%%%%%%%%%%%%%%%%

\begin{figure} 
\begin{center}
\includegraphics[height=6cm]{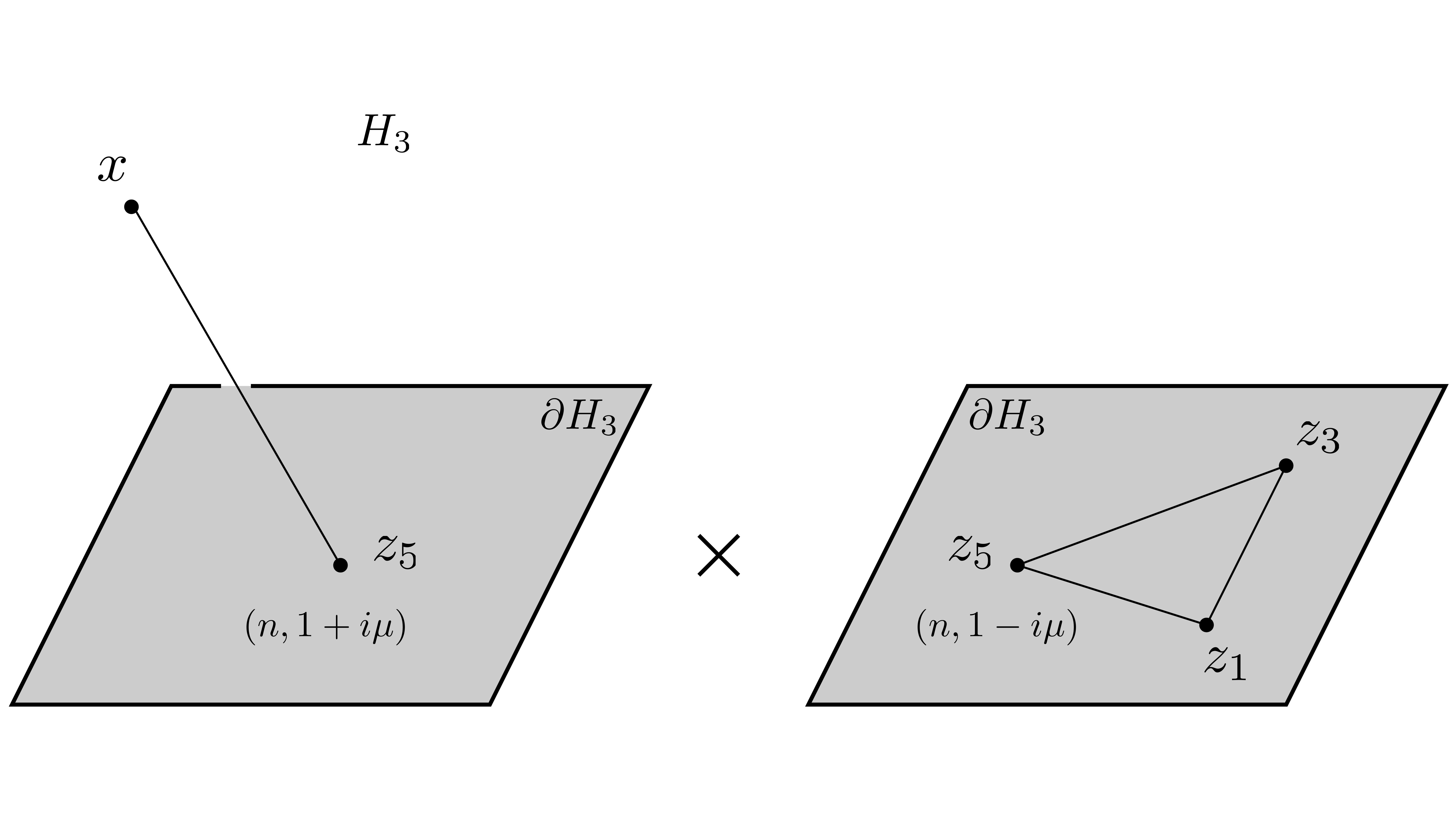}
\caption{\label{IFspin0} The functions $\chi_{\mu}(u)$ and $\psi_{\mu}^{mn}(u)$, respectively 
used as a complete basis for the spin $n=0$ and $n=2$ components of the impact factor $V^{mn}(u)$.
These functions are given by the integral over $z_5$ of the spin $n$ bulk to boundary propagator from $x$ to $z_5$ with dimension $1+i\mu$, multiplied by the 3-point
function of scalars with zero dimension at $z_1$ and $z_3$ and a spin $n$ operator of dimension $1-i\mu$  at $z_5$.}
\end{center}
\end{figure}

The scalar part of the impact factor $V_0^{mn}$ can be constructed simply by acting with the differential operators ${\cal D}_k^{mn}$ given in (\ref{DTensors}) on scalar functions 
of the cross ratio $u$. Hence we define the following basis for the scalar components of the impact factor
\[
V^{mn}_0 (x,z_1,z_3)= \sum_{k=1}^4 {\cal D}_k^{mn} (x) \,S_k(u)\,,
\]
with
\begin{equation}
S_k(u) = \int d\mu \,S_k(\mu) \,\chi_\mu(u)\,.
\label{Spin0Expansion}
\end{equation}
We are expanding the scalar functions $S_k(u)$ in the basis introduced in \cite{BFKLpaper}, defined by 
\footnote{In the notation of \cite{BFKLpaper},  $\chi_\mu(u)$ and $c_0(\mu)$ are given by  $\phi_\mu(u) + \phi_{-\mu}(u)$ and $\mu^2c(\mu)$, respectively.}
\begin{equation}
\chi_\mu (u) = c_0(\mu)  \int_{\partial H_{3}}dz_5\,
\Pi_\mu(x,z_5) \,E_{-\mu}(z_1,z_3,z_5)
\label{Spin0Basis}
\end{equation}
where 
\[
\Pi_\mu(x,z_5) = \left(\frac{|x|}{-2x\cdot z_5}\right)^{1+i\mu}
\]
is the scalar bulk to boundary propagator of weight $1+i\mu$, 
\[
E_{-\mu}(z_1,z_3,z_5) = \left(  \frac{z_{13}}{z_{15}z_{35}}\right)  ^{\frac{1-i\mu}{2}}
\]
is the scalar 3-point function with zero weight at $z_1$ and $z_3$ and with weight $1-i\mu$  at $z_5$, as described in section \ref{BFKLkernel}, and the constant
\[
c_0(\mu)  =\frac{\mu^{2}(1+\mu^{2})}{64\pi^{5}}~\frac{\Gamma^{2}\left(
\frac{1-i\mu}{2}\right)  }{\Gamma\left(  1-i\mu\right)  }~.
\]
The functions $\chi_\mu (u)$ can be expressed in terms of hypergeometric functions.
In figure \ref{IFspin0} we represent schematically the basis for the spin 0 components of the impact factor.

%%%%%%%%%%%%%%%%%%%%%%%%%%%%%%%%%%%%%%
\subsubsection{Transverse spin 2}
%%%%%%%%%%%%%%%%%%%%%%%%%%%%%%%%%%%%%%

We now seek, analogously to the scalar case, for a Fourier representation of 
the spin 2 part of the impact factor $V_2^{mn}$. The natural generalisation of the spin 0 case is
to consider the  spin 2  bulk to boundary propagator from $x$ to $z_5$ of weight $1+i\mu$  and then
the 3-point coupling $E_{-\mu}(z_1,z_3,z_5)$ with a spin 2 state of weight $1-i\mu$ at $z_5$, as also
represented in figure \ref{IFspin0}.

Let us start by defining the metric on the space 
orthogonal to a given bulk vector $x$ and a boundary vector $z$,  satisfying $z^2 = 0$, given by 
\[
\pi^{mn} (x,z)= \eta^{mn} - \frac{x^mz^n}{x\cdot z} -\frac{z^m x^n}{x\cdot z} + \frac{x^2z^mz^n}{(x\cdot z)^2}\,.
\]
By construction it is orthogonal
\[
x_m \pi^{mn} = 0 \,,\ \ \ \ \ \ \ \ \ z_m \pi^{mn} = 0\,,
\]
and it satisfies
\[
\pi^{mn} \pi_{np} = \pi^{m}_{\ p}\,,\ \ \ \ \ \ \ \ \pi^m_{\ m}=2\,.
\]
The spin 2 bulk to boundary propagator of weight $1+i\mu$ is then given by
\begin{equation}
\Pi_\mu^{mnab} (x,z)= \left(\frac{|x|}{-2x\cdot z}\right)^{1+i\mu} \pi^{mnab}\,, 
\label{Pispin2}
\end{equation}
where $\pi^{mnab}$ is the weight zero tensor structure
\[
\pi^{mnab}  (x,z)=  \frac{1}{2}\,\left(  \pi^{ma}\pi^{nb} + \pi^{mb}\pi^{na} - \pi^{mn}\pi^{ab}\right)\,.
\]
This propagator is transverse to $x$ and $z$, and is traceless and symmetric in both pairs of indices $mn$ and $ab$. It is
important to note that the spin 2 propagator has zero divergence,
\[
\partial_m \Pi_\mu^{mnab} (x,z) = 0\,.
\]

The  3-point coupling $E_{-\mu}(z_1,z_3,z_5)$ with a spin 2 state at $z_5$ is given by 
\[
E^{ab}_{-\mu}(z_1,z_3,z_5) = \left( \frac{z_{13}}{z_{15}z_{35}}\right)^{\frac{1-i\mu}{2}} T^{ab}(z_1,z_3,z_5)\,,
\]
where the tensor $T^{ab}$ has weight zero, is symmetric and traceless, and is orthogonal to $z_5$. As explained in appendix \ref{Efunctions}, $T^{ab}$ can be written as
\[
T^{ab} =  T^a T^b -\frac{1}{2}\, G^{ab}\,,
\]
where 
\[
T^a (z_1,z_3,z_5)= \left( \frac{z_{15}z_{35}}{z_{13}}\right)^{\frac{1}{2}}\left( \frac{z_1^a}{z_{15}} - \frac{z_3^a}{z_{35}}\right)\,,
\]
and
\[
G^{ab} (z_1,z_3,z_5)= \eta^{ab} + \frac{z_{1}^a z_{5}^b + z_{1}^b z_{5}^a}{z_{15}} + \frac{z_{3}^a z_{5}^b + z_{3}^b z_{5}^a}{z_{35}} \,.
\]

We may now introduce a basis for the spin 2 component of the impact factor
\begin{equation}
V_2^{mn}(x,z_1,z_3) = \int d\mu \, T(\mu) \, \psi^{mn}_\mu (x,z_1,z_3)\,,
\label{Spin2Expansion}
\end{equation}
where we define the  Fourier tensor function by
\begin{equation}
\psi^{mn}_\mu (x,z_1,z_3) =   c_2(\mu)  \int_{\partial H_{3}}dz_5\,
\Pi^{mnab}_\mu(x,z_5) \,E_{-\mu}^{cd}(z_1,z_3,z_5)\, \eta_{ac}\eta_{bd}\,,
\label{Spin2Basis}
\end{equation}
with
\[
c_2(\mu) =  - \frac{1}{\pi^5}\,\frac{4+\mu^{2}}{1+\mu^2} ~
\frac{\Gamma^{2}\left(\frac{3-i\mu}{2}\right) }{\Gamma\left( 3-i\mu\right)  }~.
\]
The spin 2 component $V_2^{mn}$ inherits from the bulk to boundary spin 2 propagator the
basic conditions
\begin{equation}
x_mV_2^{mn}=0\,,\ \ \ \ \ \ \ \ 
\eta_{mn}V_2^{mn}=0\,,\ \ \ \ \ \ \ \ 
\partial_m V_2^{mn}=0\,.
\label{ConditionsSpin2}
\end{equation}

%%%%%%%%%%%%%%%%%%%%%%%%%%%%%%%%%%%%%%
\subsection{Back to Regge theory}
%%%%%%%%%%%%%%%%%%%%%%%%%%%%%%%%%%%%%%

\begin{figure}[tb]
\begin{center}
\includegraphics[width= 16cm]{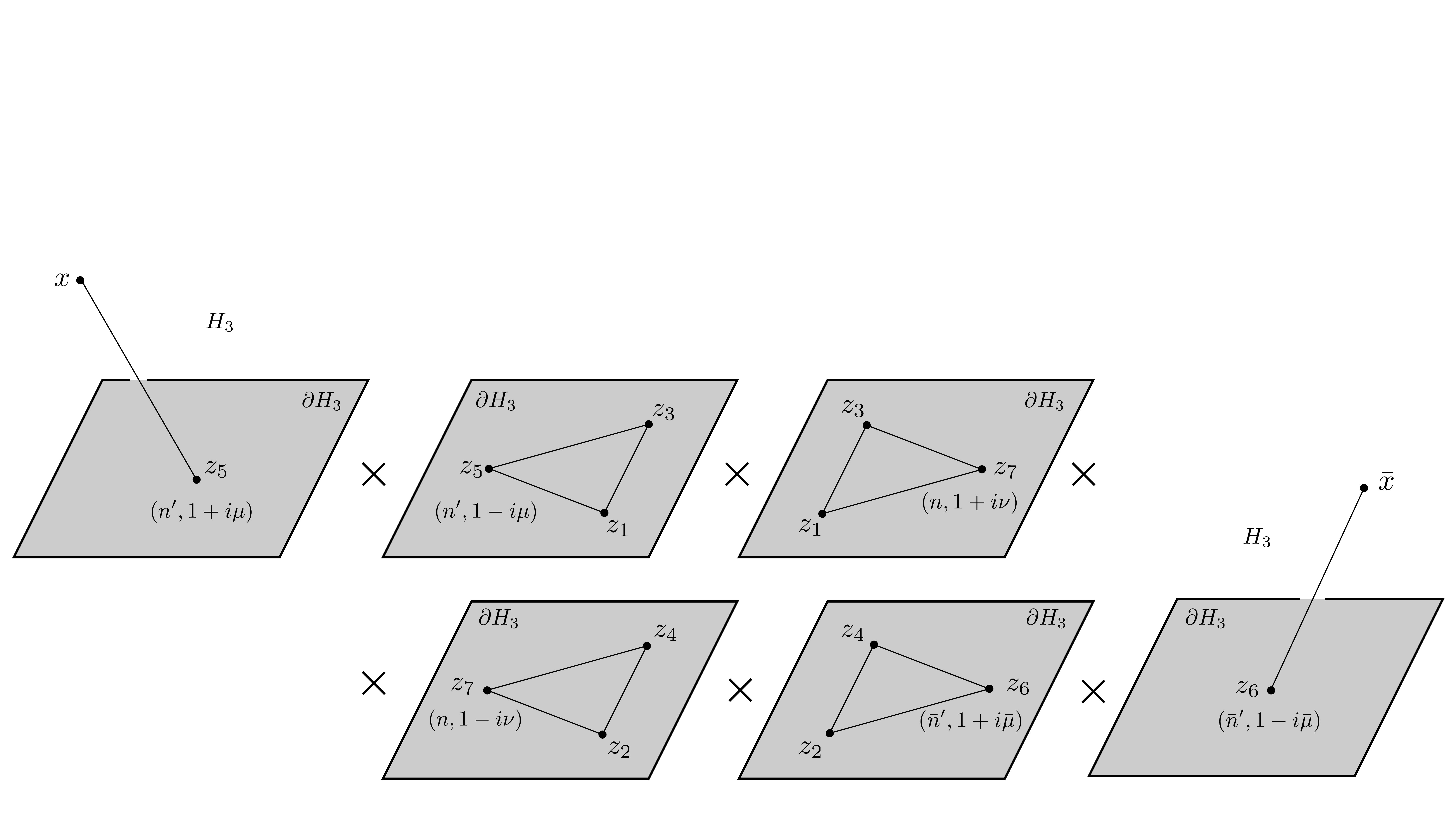}
\caption{Full BFKL amplitude, written as a product of the conformal basis for the left and right
impact factors and for the  BFKL propagator. After integrating over all $z$'s except $z_7$ one obtains the
integral representation of the $\Omega_{i\nu}$ function.
\label{FullBFKL}}
\end{center}
\end{figure}

It is now clear how to make contact with the general form  (\ref{ReggeAmplitude})  of the amplitude in Regge theory. We start with the 
BFKL   amplitude in position space (\ref{BFKLposition}), and then expand the BFKL propagator  and the impact factors in the
conformal basis given in sections \ref{BFKLkernel} and \ref{ImpactFactors1}, as represented by figure \ref{FullBFKL}. Then,
using the orthogonality conditions obeyed by the $E_\nu$ 3-point functions (\ref{Orthogonality}), 
one can do the integrals in $z_1$, $z_2$, $z_3$ and $z_4$.
In particular, since the impact factor $\bar{V}(\bar{x},z_2,z_4)$ will only overlap with the spin 0 component of the BFKL kernel, the spin 2
component of the impact factor $V_2^{mn}(x,z_1,z_3)$ will be projected out when integrating over $z_1$ and $z_3$. This explains
why the spin 2 component of the impact factor does not enter the four-point function 
of two vector and two scalar operators (\ref{BFKLposition}), correctly giving  the counting of four independent tensor structures in 
(\ref{ReggeAmplitude}).

The computation outlined in the previous paragraph was done in detail in \cite{BFKLpaper} for the case of scalar operators. In the present case the only 
difference is the operator ${\cal D}^{mn}_k$, which plays no role in the computation. We have therefore
\[
{\cal A}^{mn}(x,\bar{x}) = -\frac{1}{2N^2}  \sum_{k=1}^4\int d\nu~S_k(\nu)  \,\frac{\tanh\frac{\pi\nu}{2}}{\nu}\, \bar{V}(\nu) ~{\cal D}^{mn}_k \Omega_{i\nu}(x,\bar{x})  ~,
\]
where $\bar{V}(\nu) $ is the transform of the scalar impact factor $\bar{V}(\bar{u})$, with $\bar{u}$ the cross ratio constructed from $\bar{x}$, $z_2$ and $z_4$,
just as described in section \ref{ImpactFactors1} and in  \cite{BFKLpaper}. By doing this computation one discovers also an integral representation
for the harmonic functions $\Omega_{i\nu}(x,\bar{x})$, given by
\[
\Omega_{i\nu}(x,\bar{x}) = \frac{\nu^2}{4\pi^3} \int_{\partial \mathrm{H}_3} dz_7 \,\Pi_\nu(x,z_7)\, \Pi_{-\nu}(\bar{x},z_7)\,.
\]
This representation will be useful to generalise to the case of  transverse spin 2 explained below.
We conclude that the residues $\alpha_k(\nu)$ in the  Regge amplitude (\ref{ReggeAmplitude}) are given by
\[
\alpha_k(\nu) = -\frac{i}{4\pi N^2}~ S_k(\nu)  \,
\frac{\tanh\frac{\pi\nu}{2}}{\nu}\,\bar{V}(\nu) \,.
\]

%%%%%%%%%%%%%%%%%%%%%%%%%%%%%%%%%%%%%%
\subsubsection{Conserved current}
%%%%%%%%%%%%%%%%%%%%%%%%%%%%%%%%%%%%%%

We saw in section \ref{Amplitude} that when the vector operator is a conserved current, there are two differential equations that reduce to two the number 
of independent functions determining the amplitude. It is therefore of interest to understand the implications of conservation on the current impact factors
$S_k(u)$, as well as on its transform $S_k(\mu)$, and consequently on the Regge residue $\alpha_k(\mu)$ just discussed.

Starting from the BFKL amplitude in position space (\ref{BFKLposition}), the conservation equation 
$\partial_m A^{mn} =0$ becomes simply
\[
\partial_m \,\frac{V^{mn}}{(-x^2)^3} = 0 \ \ \Rightarrow\  \  \left( \partial_m -6\,\frac{x_m}{x^2} \right) V^{mn}= 0\,,
\]
which gives two independent differential equations from the  coefficient multiplying $x^n$ and $\bar{x}^n$, as expected.
Since the spin 2 component automatically satisfies the above condition (cf. (\ref{ConditionsSpin2})), the conservation equation only involves the  scalar functions $S_k(u)$.
This is best analyzed by considering
the integral representation of the various scalar functions given in (\ref{Spin0Expansion}) and (\ref{Spin0Basis}). In particular, we only need to keep track
of the pieces  involving the differential operators ${\cal D}_k^{mn} $ and the bulk to boundary propagator,
\[
\sum_{k=1}^{4}S_{k}(\mu)\,\mathcal{D}_{k}^{mn}\, \left(\frac{|x|}{-2x\cdot z_5}\right)^{1+i\mu}\:.
\]
A tedious computation shows that the conservation equation implies on the Fourier components $S_{k}(\mu)$ of the scalar impact factors
\begin{align}
S_{1}(\mu)-2S_{3}(\mu)+\frac{2}{3}\left(4+\mu^{2}\right)S_{4}(\mu)&  =  0\,,
\nonumber\\
3S_{1}(\mu)+3S_{2}(\mu)-\left(1+\mu^{2}\right)S_{3}(\mu) & =  0\,.
\label{nuconservation}
\end{align}
Recalling now that on the basis functions $\chi_{\mu}(u)$ the operator 
\begin{equation}
\nabla^2=4u^{2}\left(1-u\right)\frac{d^{2}}{du^{2}}-4u^{2}\frac{d}{du}\:,
\label{Delta}
\end{equation}
satisfies $\nabla^2=-\left(1+\mu^{2}\right)$,
we obtain the conditions on the functions $S_k(u)$,
\begin{align*}
S_{1}(u)-2S_{3}(u)-\frac{2}{3}\left(\nabla^2-3\right)S_{4}(u) = 0&\:,\\
3S_{1}(u)+3S_{2}(u)-\nabla^2 S_{3}(u) =  0&\:.
\end{align*}

Let us conclude this section by also writing the conservation equations in
terms of the functions $h_{k}(u)$ introduced in the expansion (\ref{IF}). A tedious calculation gives 
\begin{align}
2uh_{3}+2uh_{4}+3h_{5}-2u^{2}h_{1}^{\prime}-u^{2}\left(1-u\right)h_{4}^{\prime}-u\left(1-u\right)h_{5}^{\prime}  &= 0\,,
\nonumber
\\
6h_{1}+3h_{2}-6h_{3}-2uh_{1}^{\prime}+2u(1-u)h_{3}^{\prime}  &=  0 \,.
\label{hfunctionsdivergence}
\end{align}

%%%%%%%%%%%%%%%%%%%%%%%%%%%%%%%%%%%%%%
\subsection{$\gamma^*-\gamma^*$ scattering \label{gammagamma}}
%%%%%%%%%%%%%%%%%%%%%%%%%%%%%%%%%%%%%%

We have all the necessary ingredients to also understand the CFT Regge limit of  the four-point function of a vector operator $j^a$. This case is relevant to 
$\gamma^*-\gamma^*$ scattering in the conformal limit of QCD, so we shall spend this section analysing this amplitude.
In this case the  position space BFKL representation   takes the form
\begin{align}
\mathcal{A}^{mn\bar{m}\bar{n}} (x,\bar{x})
  \simeq&- \frac{1}{N^2}\int_{\partial H_{3}} \frac{dz_1 dz_3}{\left(z_{13}\right)^2}\,
   \frac{dz_2 dz_4}{\left(z_{24}\right)^2} 
  V^{mn}(  x,z_1,z_3)  \; F(z_1,z_3,z_2,z_4) \; \bar{V}^{\bar{m}\bar{n}}(\bar{x},z_2,z_4)  \,, \label{ggsca} 
\end{align}
As before, the impact factors can be written as
\begin{align}
V^{mn}(x,z_1,z_3) = \int d\mu \left[\sum_{k=0}^4 S_k(\mu) \mathcal{D}_k^{mn} 
\chi_\mu (x,z_1,z_3)+ T(\mu) \, \psi^{mn}_\mu (x,z_1,z_3)\right]\,, \label{fullIF}
\end{align}
with a similar expression for $\bar{V}^{\bar{m}\bar{n}}(\bar{x},z_2,z_4)$.
From the orthogonality relation (\ref{Orthogonality}) and the representations
 (\ref{Spin0Basis}) and (\ref{Spin2Basis}) of the functions $\chi_\mu$ and $\psi^{mn}_\mu$, it is clear that only the spin 0 and the spin 2
components of the  BFKL propagator (\ref{Kernel}) contribute to the correlator (\ref{ggsca}),
\begin{align*}
\mathcal{A} ^{mn\bar{m}\bar{n}} (x,\bar{x})=
\mathcal{A}_{(0)}^{mn\bar{m}\bar{n}} (x,\bar{x})+
\mathcal{A}_{(2)}^{mn\bar{m}\bar{n}} (x,\bar{x})\ .
\end{align*}
The spin 0 contribution is a direct generalization of the previous case,
\[
\mathcal{A}_{(0)}^{mn\bar{m}\bar{n}} (x,\bar{x}) = -\frac{1}{2N^2}  \sum_{k,\bar{k}=1}^4\int d\nu~S_k(\nu)  \,\frac{\tanh\frac{\pi\nu}{2}}{\nu}\, \bar{S}_{\bar{k}}(\nu) 
~{\cal D}^{mn}_k {\cal D}^{\bar{m}\bar{n}}_{\bar{k}} \Omega_{i\nu}(x,\bar{x})  ~.
\]
We shall therefore focus on the transverse spin 2 contribution to the correlator.
Using the representation (\ref{Kernel}) of the BFKL propagator we have 
\begin{align}
\mathcal{A}_{(2)}^{mn\bar{m}\bar{n}} (x,\bar{x}) 
=& -\frac{16}{\pi^2N^2}  \int d\nu \,\frac{\nu^2+4}{ ( \nu^2 + 1 ) ( \nu^2 + 9) }  \int_{\partial H_3}d z_7 
\,W^{mn a_1a_2}_{\nu}(x,z_7) \,\bar{W}^{\bar{m}\bar{n}b_1 b_2}_{-\nu \ \ }(\bar{x},z_7)\;\eta_{a_1b_1} \eta_{a_2b_2}\ ,\label{Aspin2}
\end{align}
where
\begin{align*}
W^{mn a_1a_2}_{\nu}(x,z_7) =
\int_{\partial H_{3}} \frac{dz_1 dz_3}{\left(z_{13}\right)^2}\,
V^{mn}(x,z_1,z_3) E_\nu^{a_1a_2}(z_1,z_3,z_7)\ .
\end{align*}
To determine $W$ we first write the impact factor as (\ref{fullIF}), and use the representation
(\ref{Spin2Basis}) and the orthogonality relation (\ref{Orthogonality}) to obtain
\begin{align*}
W^{mn a_1a_2}_{\nu}(x,z_7) &=
\frac{1}{2}\,T(\nu)c_2(\nu)a_2(\nu) \Pi^{mn a_1a_2}_{\nu}(x,z_7)\\& +
T(\nu)c_2(-\nu)b_2(\nu) 
\int_{\mathbb{R}^2} dz_5\,  \Pi_{-\nu}^{mn  b_1b_2} (x,z_5)\,
 \frac{V_{b_1b_2}^{\ \ \ \  a_1a_2}(z_5,z_7)}{(-2 z_5\cdot z_7)^{1+i\nu}}\ .
\end{align*}
The last integral is computed in appendix \ref{ConfInt} and gives  
\begin{align*}
\int_{\mathbb{R}^2} dz_5\,  \Pi_{-\nu}^{mn  b_1b_2} (x,z_5)
 \frac{V_{b_1b_2}^{\ \ \ \  a_1a_2}(z_5,z_7)}{(-2 z_5\cdot z_7)^{1+i\nu}} =
 \frac{i\pi}{2\nu} \Pi_{\nu}^{mn  a_1a_2} (x,z_7) +\cdots\ ,
\end{align*}
where the $\cdots$ are terms that drop out in (\ref{Aspin2}).
Therefore,
\begin{align*}
W^{mn a_1a_2}_{\nu}(x,z_7) &=
T(\nu)\left[ \frac{1}{2}\,c_2(\nu)a_2(\nu)+
\frac{i\pi}{2\nu}c_2(-\nu)b_2(\nu) 
\right] \Pi^{mn a_1a_2}_{\nu}(x,z_7)+\dots \\
&=
\frac{i +\nu}{\nu} \,c_2(\nu)a_2(\nu)T(\nu)  \Pi^{mn a_1a_2}_{\nu}(x,z_7)+\dots\,,
\end{align*}
and we conclude  that the transverse spin 2 contribution to the amplitude has the form
\begin{align*}
\mathcal{A}_{(2)}^{mn\bar{m}\bar{n}} (x,\bar{x}) 
= -  \frac{128}{N^2} \int d\nu \,\frac{ \tanh \left(\frac{\pi  \nu }{2}\right)T(\nu)\bar{T}(\nu)}{\nu ^3
   \left(\nu ^2+1\right) \left(\nu ^2+4\right) \left(\nu
   ^2+9\right)}\ 
   \Omega^{mn \bar{m}\bar{n}}_{i\nu} (x,\bar{x})\,,
   \end{align*}
where $\Omega^{mn \bar{m}\bar{n}}$ is a spin 2 harmonic function on $H_3$.
It is given by the natural generalization of the spin 0 case,
\begin{align*}
\Omega^{mn \bar{m}\bar{n}}_{i\nu} (x,\bar{x})= \frac{4+\nu^2}{4\pi^3}
\int_{\partial \mathrm{H}_3}d z_7 
\,\Pi^{mn a_1a_2}_{\nu}(x,z_7) \,\Pi^{\bar{m}\bar{n}b_1 b_2}_{-\nu \ \ }(\bar{x},z_7)\;\eta_{a_1b_1} \eta_{a_2b_2}\,.
\end{align*}

%%%%%%%%%%%%%%%%%%%%%%%%%%%%%%%%%%%%%%
\subsection{Disentangling spin 0 and spin 2 components}\label{Disentangling}
%%%%%%%%%%%%%%%%%%%%%%%%%%%%%%%%%%%%%%

As we shall see in section \ref{ImpactFactors},
when computing explicitly the impact factor $V^{mn}(x,z_1,z_3)$ using Feynman rules, one obtains a linear combination of the 
five tensor structures ${\cal I}^{mn}_k$, as written in (\ref{ITensors}) and (\ref{IF}). 
It is then desirable to split the spin 0 and spin 2 components of the impact factor. In particular, we wish to determine the functions 
$S_{k}(\mu)$ and $T(\mu)$ in the Fourier decomposition (\ref{Spin0Expansion})  and (\ref{Spin2Expansion}),  so that we can write amplitudes involving 
the impact factors in their Regge form.

Let us start by explaining how to extract the spin 0 components of a generic impact factor, written in a obvious notation as 
\begin{equation}
V^{mn} = V_0^{mn} + V_2^{mn}=\sum_{k=1}^4 \mathcal{D}_{k}^{mn} S_k(u) + V_2^{mn}\,.
\label{if}
\end{equation}
We note the following basic relations satisfied by the differential operators $\mathcal{D}_{k}^{mn}$
used to construct the tensor structures for the spin 0 components,
\begin{align*}
x_{m}\mathcal{D}_{1}^{mn}=x_{m}\mathcal{D}_{4}^{mn}  = 0&\:,\\
x_{m} x_{n}\mathcal{D}_{3}^{mn} = 0&\:,\\
\eta_{mn}\mathcal{D}_{3}^{mn}=\eta_{mn}\mathcal{D}_{4}^{mn} = 0&\:,
\end{align*}
while the spin $2$ component satisfies the conditions (\ref{ConditionsSpin2}).  
Given a generic impact factor written in the form (\ref{if}), we can determine the functions $S_{1}(u)$, $S_{2}(u)$ and $S_{3}(u)$ rather easily by noting that
\begin{align}
\frac{1}{3}\left(\eta_{mn}-\frac{x_m x_n}{x^2} \right)V^{mn} & =  S_{1}(u)\:,\nonumber\\
\frac{x_m x_n}{x^2}\,V^{mn} & = S_{2}(u)\:,\label{S1S2S3}\\
\left(\frac{x_m z_{1n}}{x\cdot z_1} - \frac{x_m x_n}{x^2}\right)V^{mn} & = 2u(1-u)\: S_{3}'(u)\:.\nonumber
\end{align}
Notice that a constant in $S_{3}$ is immaterial since
$\mathcal{D}_{3}^{mn}$ annihilates constant terms. 

Next let us consider the scalar function $S_{4}(u)$. After determining
$S_{1}$, $S_{2}$ and $S_{3}$, we consider the following transverse part of the impact
factor
\begin{equation*}
V_{\perp}^{mn}  =  V^{mn}-\sum_{i=1}^{3}\,\mathcal{D}_{i}^{mn}\, S_{i}(u)
 =  \mathcal{D}_{4}^{mn}\, S_{4}\left(u\right)+V_{2}^{mn}\,.
\end{equation*}
Since $V_{\perp}^{mn}$ is nothing but the part of $V^{mn}$ which
is traceless and transverse to $x^{m}$, it can also be computed
without reference to the functions $S_{1}$, $S_{2}$ and $S_{3}$ using
the explicit expression 
\begin{align}
V_{\perp}^{mn}  = \ & V^{mn}-\left(\frac{x^m x_pV^{pn}}{x^2}+\frac{x^n x_pV^{pm}}{x^2}\right)\label{VPerp}\\
 &  +\frac{1}{3}\,V_{pq} \,\frac{x^p x^q}{x^2}\left(\eta^{mn}+2\,\frac{x^m x^{n}}{x^2}\right) 
 -\frac{1}{3}\,V\left(\eta^{mn}-\frac{x^m x^n}{x^2}\right)\, , 
 \nonumber
\end{align}
where $V=V^{mn}\eta_{mn}$ is the trace of $V^{mn}$.
Since the spin 2 component of the impact factor has zero divergence, we can compute the divergence of $V_{\perp}^{mn} $ to isolate the
scalar function $S_4(u)$. A tedious but straight-forward computation shows that 
\begin{align}
\partial_{m} V_{\perp}^{mn} & = \partial_{m}\Big( \mathcal{D}_{4}^{mn}S_{4}(u) \Big)  =\left(\frac{2 x^n}{x^2}-\frac{z_1^n}{x\cdot z_1} - \frac{z_3^n}{x\cdot z_3} \right) \frac{2}{3}\,u\,\frac{d}{du}\left(3-\nabla^2\right)S_{4}(u) \:,
\label{VDivergence} 
\end{align}
where we recall that the operator $\nabla^2$ is defined in (\ref{Delta}).

We shall now describe how to extract the spin $2$ component of
a generic impact factor. We start by considering the transverse
traceless part $V_{\perp}^{mn}$. Since it is transverse to $x^m$,
it may be considered as a metric fluctuation on the hyperbolic space
$H_{3}$. Moreover, denoting with $\hat{\mu}, \hat{\tau}$ the coordinates on $H_{3}$,
we quickly discover that the scalar contribution $\mathcal{D}_{4}^{mn}S_{4}$
to $V_{\perp}^{mn}$ corresponds to
\[
\nabla^{\hat{\mu}}\nabla^{\hat{\tau}}S_{4}-\frac{1}{3}\,g^{\hat{\mu}\hat{\tau}}\nabla^{2}S_{4}\:.
\]
The above   corresponds to an infinitesimal diffeomorphism combined
with an infinitesimal conformal transformation. Since $H_{3}$ is a three-dimensional conformally flat space, we are led to computing
the linearized Cotton tensor corresponding to a metric fluctuation
$V_{\perp}^{mn}$. The spin $0$ contribution $S_{4}$ will not contribute
to the Cotton tensor, and we will be left uniquely with the spin $2$
contribution.

We first consider the linearized Ricci tensor and scalar for a metric  fluctuation
$V_{\perp}^{mn}$, which in the embedding space representation are given simply, recalling that
$V_{\perp}^{mn}$ is traceless, by
\begin{align*}
R^{mn} & =  \square V_{\perp}^{mn}-\partial^{m}\partial_{p}V_{\perp}^{pn}-\partial^{n}\partial_{p}V_{\perp}^{pm}\:,\ \ \ \ \ \ \ \ \ 
R   =  -2\partial_{p}\partial_{q}V_{\perp}^{pq}\:.
\end{align*}
Considering then the combination
\[
G_{mn}=R_{mn}-\frac{1}{4}\left(\eta_{mn}-\frac{x_m x_n}{x^2}\right)R\:,
\]
the Cotton tensor is defined by
\[
C_{abc}=\partial_{a}G_{bc}-\partial_{b}G_{ac}+\frac{x_a}{x^2}\,G_{bc}-\frac{x_b}{x^2}\,G_{ac}\:.
\]
Given that $C_{abc}$ is  traceless and transverse to $x^m$, and that it satisfies the symmetries
\begin{align*}
C_{abc}+C_{bac} &=  0\,\\
C_{abc}+C_{bca}+C_{cab} & =  0\,,
\end{align*}
we can construct a unique function of the cross ratio $u$ characterizing
$C_{abc}$. We shall call this function, explicitly given by
\[
C(u)=\frac{\left(-x^2\right)^{2}}{\left(-2z_{1}\cdot x\right) z_{13}} \,z_{1}^{a} z_{3}^{b} z_{1}^{c}\,C_{abc}\:,
\]
the Cotton function.

We may compute the Cotton function $C_\mu(u)$, corresponding the Fourier basis $\psi_\mu^{mn}(u)$ in the
expansion of the spin 2 component of the impact factor in (\ref{Spin2Expansion}). 
A long and mechanical computation,   shows the very simple result
\[
\left(u\,\frac{d}{du}-1\right) C_\mu(u) = \chi_{\mu}(u)\,,
\]
where we recall that $\chi_{\mu}(u)$ is the Fourier component for scalar impact factors (\ref{Spin0Basis}).
Given the Cotton function associated to the spin 2 component of a generic impact factor, we have therefore arrived at the following final representation
\begin{equation}
\left(u\frac{d}{du}-1\right)C(u)=\int d\mu\: T(\mu)\, \chi_{\mu}(u)\,.
\label{CottonTransform}
\end{equation}
This allows for the determination of $T(\mu)$ and hence of the spin 2 component of the impact factor via the transform (\ref{Spin2Expansion}). 

It is now just a mechanical computation to apply the general strategy
outlined in this section to determine  the spin 0 and spin 2 components of a generic impact factor given by a linear combination of the tensor
structures ${\cal I}^{mn}_k$ as written in (\ref{IF}). From (\ref{S1S2S3}) we deduce that
the functions $S_{1}(u)$,  $S_{2}(u)$ and $S_{3}(u)$ are given by
\begin{align}
&S_{1}=  h_{1}+\frac{1}{6}\,h_{4}+\frac{1-2u}{6}\,h_{5}\:,\nonumber\\
&S_{2}  =  h_{1}+h_{2}-2h_{3}-\frac{1}{2}\,h_{4}-\frac{1}{2u}\,h_{5}\:,\label{S1S2S3final}\\
&S_{3}=  \int\frac{du}{4u^{2}}\Big(2uh_{3}+uh_{4}+h_{5}\Big)\:.\nonumber
\end{align}
After computing the transverse and traceless part  (\ref{VPerp})  of the impact factor, one may compute its divergence and equate it to (\ref{VDivergence}), therefore  determining the function $S_{4}(u)$ through
\begin{equation}
\left(\nabla^2-3\right)S_{4}=\int\frac{du}{4u^{2}}\:\left(3uh_{4}+5h_{5}+u^{2}(3u-2)h_{4}^{\prime}+u(u-2)h_{5}^{\prime}\right)\:,
\label{S4}
\end{equation}
where $'$ denotes derivative with respect to $u$.
Finally, the Cotton function $C(u)$ for the transverse and traceless part (\ref{VPerp}) of the impact factor is explicitly given by 
\begin{align}
C = & \ (1-u)u^{2}\Big((3u-1)h_{4}^{\prime}+u(3u-2)h_{4}^{\prime\prime}+\frac{u^{2}}{2}\,(u-1)h_{4}^{\prime\prime\prime}\nonumber\\
 &  - h_{5}^{\prime}+(1-2u)h_{5}^{\prime\prime}-\frac{u}{2}\,(u-1)h_{5}^{\prime\prime\prime}\Big)\:,
 \label{Cotton}
\end{align}
which then allows for the determination of the spin 2 Fourier component $T(\mu)$ through (\ref{CottonTransform}).

%%%%%%%%%%%%%%%%%%%%%%%%%%%%%%%%%%%%%%
\section{Impact factors in QCD and SYM}\label{ImpactFactors}
%%%%%%%%%%%%%%%%%%%%%%%%%%%%%%%%%%%%%%

In this section we compute explicitly the impact factors for the electromagnetic
current operator in QCD with massless quarks and for the R-current operator in ${\cal N}=4$ SYM.
We consider first the impact factor for a Weyl fermion in an arbitrary representation of the gauge group, and then
for a complex scalar field. It is then trivial to apply the results to the cases of QCD and SYM. 

\begin{figure}[t]
\begin{center}
\includegraphics[width=13cm]{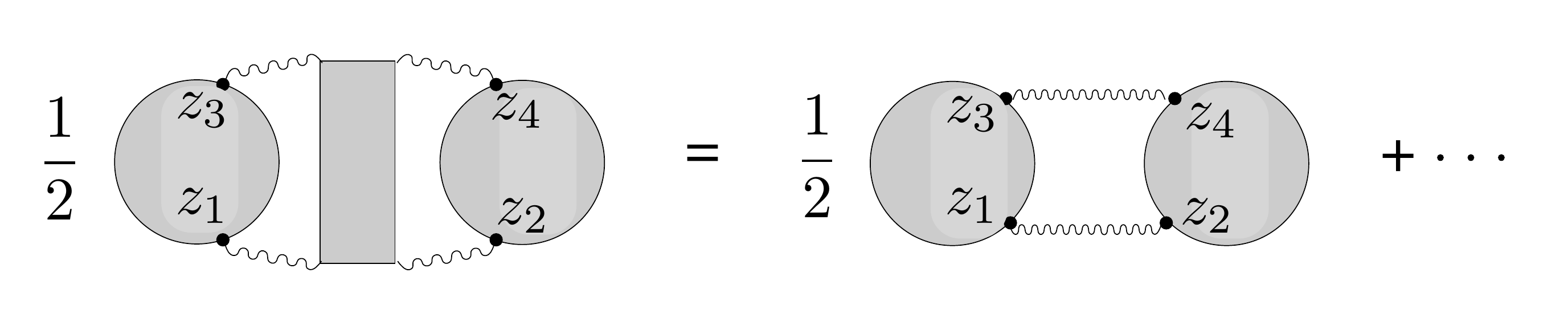}
\caption{Perturbative expansion of the BFKL propagator. The leading term
corresponds to the exchange of a pair of gluons in a color singlet state. The symmetry factor of 1/2 comes from 
the permutation of the the gluon lines.\label{pomeron}}
\end{center}
\end{figure}

To leading order in perturbation theory the BFKL kernel is given by the 
exchange of a pair of two gluons in a color singlet state, as represented
in figure \ref{pomeron}. The factor of $1/2$ is the symmetry factor of the diagram that comes from permuting
the gluon lines. 
Considering for now the diagram computed with standard Feynman rules
in the central Poincar\'e patch, we denote by $D^{c\bar{c}}_{A\bar{A}}(z_1,z_2)$ and by $D^{d\bar{d}}_{B\bar{B}}(z_3,z_4)$ the gluon propagators,
respectively between  $z_1$ and $z_2$ and between $z_3$ and $z_4$. The upper indices refer to the polarization and the 
lower indices to the color. In the CFT Regge limit gluons with polarizations $c=d=-$ and $\bar{c}=\bar{d}=+$, in a color
singlet $A=B$ and $\bar{A}=\bar{B}$, give the dominant contribution to the amplitude. 
This fact was shown in the computation done in \cite{BFKLpaper} and can also be shown in the present case.
Thus, in the Regge limit, the full diagram is given by
\begin{align}
A^{ab}(y_i) =\frac{i^4}{2} \int_{\mathbb{M}^4} &dz_1 dz_3 dz_2 dz_4  \big\langle j^a(y_1) j^b(y_3) g_-^A(z_1)g_-^A(z_3)\big\rangle
\label{AmplitudeYs}\\
&D^{-+}_{A\bar{A}}(z_1,z_2)D^{-+}_{A\bar{A}}(z_3,z_4)
\big\langle g_+^{\bar{A}}(z_2)g_+^{\bar{A}}(z_4){\cal O}(y_2) {\cal O}(y_4) \big\rangle\,,
\nonumber
\end{align}
where $g_c^A(z)$ is the gluon vector current operator,
%($\propto \bar{\psi}_i \gamma_c\psi_j T_{ij}^A$ in QCD)
 determining the coupling between the gluons and the other internal lines of the diagram. 
We already see that the amplitude will factorise as a product of the two impact factors and  the BFKL propagator, with the impact factors related to four-point functions involving the external operators and 
the gluon current operator $g_c^A(z)$.

To express the amplitude in the position space BFKL form  (\ref{BFKLposition}), 
and in particular to compute the impact factor $V^{mn}$ for vector current operators, we need to consider the Poincar\'e patches ${\cal P}_1$ and ${\cal P}_3$,
where the points $x_1$ and $x_3$ are close to the respective origin, as discussed early in section \ref{Regge}.   
We perform the conformal transformation  (\ref{ConformalTrans1}) from the external points  $y_1$ and $y_3$ to $x_1$ and $x_3$, and also on the
internal points $z_1$ and $z_3$, which become ($i=1,3$)
\begin{equation}
w_i= \left(w_i^+, w_i^-, w_{i\perp} \right) =- \frac{1}{z_i^+} \left(1, z_i^2, z_{i\perp} \right)  \,.
\label{ConformalTransS}
\end{equation}
The current operator $g^A_c(z)$ transforms as
\[
g^A_c(z) =  (w^+)^2 \,\frac{\partial w^p}{\partial z^c}\, g^A_p(w)\,,
\]
so that, since $\partial_{z^-}=\partial_{w^-}$, we have
\[
g^A_-(z) =  (w^+)^2 \, g^A_-(w)\,.
\]
Thus, dropping the factors from the transformation of the external vectors $j^a(y_1)$ and $j^b(y_3)$, which give the transformation rule between $A^{ab}(y_i)$ and $A^{mn}(x_i)$, 
the four-point function in (\ref{AmplitudeYs}) associated to the impact factor $V^{mn}$ becomes
\begin{equation*}
(w_1^+)^2 (w_3^+)^2\left\langle j^m(x_1) j^n(x_3) g_-^A(w_1)g_-^A(w_3)\right\rangle\,.
\end{equation*}
Next we need to take care of the integration in $w_1$ and $w_3$. We shall parameterize  $w_i$ using the components of 
$z_i$ in the central Poincar\' e patch, as given by (\ref{ConformalTransS}).
Thus, the integration  measure for each internal point is still given by
\[
\int \frac{dz_i^+}{2}\, dz_i^-\,dz_{i\perp}\,.
\]

Considering also the conformal transformation of the internal points $z_2$ and $z_4$ to the points $w_2$ and $w_4$ in patches ${\cal P}_2$ and ${\cal P}_4$, it is now 
clear, starting from (\ref{AmplitudeYs}), that the reduced amplitude ${\cal A}^{mn}(x,\bar{x})$ as the following form
\begin{align}
&{\cal A}^{mn}(x,\bar{x}) = \int  \frac{dz_{1\perp}\, dz_{3\perp}}{(z_{13})^2} \frac{dz_{2\perp}\, dz_{4\perp}}{(z_{24})^2} 
\nonumber
\\
&
\left(-x^2\right)^3 \left(z_{13}\right)^2 \int  \frac{dz_1^-}{(z_1^+)^2}\,\frac{dz_3^-}{(z_3^+)^2} 
\big\langle j^m(x_1) j^n(x_3) g_-^A(w_1)g_-^A(w_3)\big\rangle
\nonumber
\\
&
\left(-\bar{x}^2\right)^\Delta \left(z_{24}\right)^2 \int  \frac{dz_2^+}{(z_2^-)^2}\,\frac{dz_4^+}{(z_4^-)^2} 
\big\langle g_+^{\bar{A}}(w_2)g_+^{\bar{A}}(w_4){\cal O}(x_2) {\cal O}(x_4) \big\rangle
\nonumber
\\
&
\frac{1}{2}\,\int \frac{dz_1^+}{2}\frac{dz_2^-}{2}D^{-+}_{A\bar{A}}(z_1,z_2)\int
\frac{dz_3^+}{2}\frac{dz_4^-}{2}D^{-+}_{A\bar{A}}(z_3,z_4)\,,
\label{showingBFKL}
\end{align}
where $z_{ij}=(z_{i\perp} - z_{i\perp})^2$ and the $w_i$ are parameterized by the components of $z_i$. We moved the integration in 
$z_1^+$, $z_3^+$,  $z_2^-$ and  $z_4^-$ to the last line of this equation because, 
as we shall see in the explicit computations presented below, in the Regge limit it contains all the dependence in these variables. 
We also kept the expressions of the gluon propagators in terms of the $z_i$ in the central Poincar\'e patch. 

The last line of (\ref{showingBFKL}) gives the  leading order BFKL propagator, as represented in figure \ref{pomeron}.
This can be derived by first noting that, in the Regge limit, the integrals in the second and third line of (\ref{showingBFKL})
are dominated  by poles located at $z_1^-\simeq 0$, $z_3^-\simeq 0$, $z_2^+\simeq 0$ and $z_4^+\simeq 0$.
Physically this corresponds, for instance, to the on-shell propagation of the fields created at $x_1$ by the operator $j_m(x_1)$ and annihilated 
at $x_3$ by  $j_n(x_3)$, with the emission of two soft gluons at $z_1$ and $z_3$. Computing the gluon propagators at these poles 
gives  the 
two gluon transverse propagator \cite{BFKLpaper}
\begin{equation*}
-\frac{g_{\mathrm{YM}}^{4}}{\left(  8\pi\right)  ^{2}} \,D(A)
\, 2\ln z_{12} \ln z_{34}\,,
\end{equation*}
where $D(A)$ is the dimension of the adjoint representation of the gauge group and where we used the 
conventions introduced in the next section for the gluon propagator  in the Feynman gauge,
\begin{equation}
\langle A_{c}^{A}(  z)  A_{d}^{B} ( 0)  \rangle  =
\frac{g_{\mathrm{YM}}^{2}}{4\pi^{2}}\frac{\delta^{AB} \eta_{cd}}{ z^{2}+i\epsilon}\,.
\label{GluonPropagator}
\end{equation}
To match the convention (\ref{TwoGluon}) for the two--gluon leading
propagator, we shall multiply, at the end of the computation, the
expression  (\ref{ImpactFactor})  used to compute the impact factor by 
\begin{equation}
k= N\,\frac{g_{\mathrm{YM}}^{2}}{8\pi}\, \sqrt{D(A)} \,, 
\label{prefactor}
\end{equation}
where the extra factor of $N$ comes from our convention on planar amplitudes
(\ref{BFKLposition}) which explicitly shows an overall factor of $N^{-2}$.

The second and third line of (\ref{showingBFKL}) give, respectively, the impact factor for the current and scalar operators. Focusing on the current operator, 
we have
\begin{equation}
V^{mn}=  \left(-x^2\right)^3 \left(z_{13}\right)^2  \sigma_1^2 \sigma_3^2 \int  d\lambda_1d\lambda_3 
\big\langle j^m(x_1) j^n(x_3) g_-^A(w_1)g_-^A(w_3)\big\rangle\,,
\label{ImpactFactor}
\end{equation}
where we rewrote (\ref{ConformalTransS}) as 
\begin{equation}
w_i = \sigma_i \, z_i+ \lambda_i \, n \,,
\label{wParameterization}
\end{equation}
with
\begin{equation*}
z_i = \left(1, z_{i\perp}^2, z_{i\perp} \right) \,,\ \ \ \ \ \ n=(0,1,0)\,.
\end{equation*}
This is the main formula of this section that we will use below  to compute in a simple manner the impact factor for different theories. 
First let us warn the reader that we started in the beginning of this section by using $z_i$ to denote the space-time points in the central Poincar\'e patch where the 
gluons were emitted.  The light-cone components of these space-time points are now denoted by $\sigma_i=w_i^+$ and 
$\lambda_i=w_i^--w_{i\perp}^2/w_i^+$.
Unless otherwise stated, from now on
we redefine $z_i= \left(1, z_{i\perp}^2, z_{i\perp} \right)$ to be the above null vector, exactly as in section \ref{SectionBFKL}. 
This slight abuse of notation is justified by the fact that the relevant transverse parts coincide, 
as the final  expressions depend only on the gluon positions $z_{i\perp}$ in transverse
space $\mathbb{R}^{2}$. Using the canonical Poincar\'{e} slice of the light--cone given by the above form of $z_i$, we shall
see that our expressions for the impact factor are invariant under the rescallings $z_i\rightarrow \alpha z_i$, with $\alpha>0$,
therefore rendering manifest the  transverse conformal invariance.

\begin{figure}[t]
\begin{center}
\includegraphics[width=11.5cm]{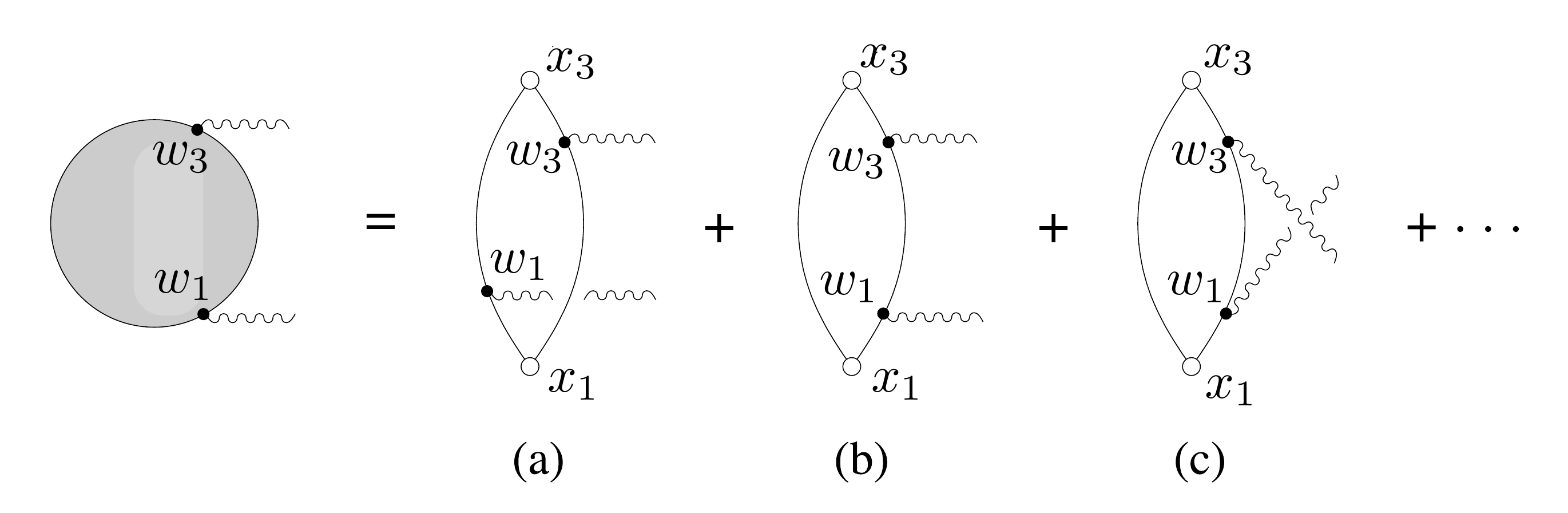}
\caption{Perturbative expansion of the four-point function necessary to compute the impact factor. Two gluons are emitted
in a color singlet at points $w_{1}$ and $w_{3}$, which may be in either of the Poincar\'e patches ${\cal P}_1$  and ${\cal P}_3$.\label{vertex}}
\end{center}
\end{figure}

Formula (\ref{ImpactFactor}) reduces the computation of the impact factor to the integral of a four-point function involving the external and the gluon current operators.
A word of caution is however necessary, since this four-point function includes operators placed at different Poincar\'e patches. 
We shall see in the next sections that the use of the $i\epsilon$-prescription for the propagators between points in the patches ${\cal P}_1$ and ${\cal P}_3$,
as given in the beginning of section \ref{Regge}, is crucial. This is already clear from figure \ref{vertex}, which shows the Feynman diagrams that contribute to the impact factor in leading order,
since the internal space-time points $w_i$  may be in either of the Poincar\'e patches ${\cal P}_1$  and ${\cal P}_3$. More precisely,
when $w_i^+>0$ ($z_i^+<0$) we have $w_i\in {\cal P}_1$, and when $w_i^+<0$ ($z_i^+>0$) we have $w_i\in {\cal P}_3$.
Thus, one needs to be careful when writing free propagators involving $x_1\in {\cal P}_1$, $x_3\in {\cal P}_3$ and
$w_i \in {\cal P}_1 \cup  {\cal P}_3$. 

Finally  we will normalize the current operators such that their two-point function, between $x_1\in {\cal P}_1$ and $x_3\in {\cal P}_3$, is given by
\begin{equation}
\langle j^m(x_1) j^n(x_3) \rangle = \frac{1}{\big(  x^{2} - i\epsilon_x\big)^{3}}\,
\left[ \eta^{mn} -2\,\frac{x^m x^n}{ x^{2}-i\epsilon_x} \right]\,, \label{CurrentNormalization}
\end{equation}
where $x=x_1-x_3$ and $\epsilon_x =\epsilon\, {\rm sgn}( x^0) $ gives the $i\epsilon$-prescription for propagators between points
in  ${\cal P}_1$ and ${\cal P}_3$.

%%%%%%%%%%%%%%%%%%%%%%%%%%%%%%%%%%%%%%
\subsection{Impact factor for a Weyl fermion}\label{Weyl}
%%%%%%%%%%%%%%%%%%%%%%%%%%%%%%%%%%%%%%

We start with the computation of  the impact factor for a Weyl fermion in an arbitrary representation $R$ of  a gauge group.
It will then be trivial to obtain the impact factor for a massless quark in QCD. The  Lagrangian is
\[
\frac{1}{g_{\mathrm{YM}}^{2}}\int d^{4}{y}
\left(  -\frac {1}{2}\,\mathrm{Tr} \left(F_{mn}F^{mn}\right)-
i\,\bar{\psi}_i\slashed{D}_{ij}\psi_j\right)  ~,
\]
with $F_{mn}=\partial_{m}A_{n}-\partial_{n}A_{m}-i\left[  A_{m},A_{n}\right] $ and 
$\slashed{D}_{ij}\psi_j = \gamma^m \left(  \delta_{ij}\partial_m- i T^A_{ij}A_m^A \right)\psi_j$. 
The gamma matrices obey the algebra $\{ \gamma_m,\gamma_n\} = 2\eta_{mn}$. 
We shall use $i,j,\cdots$ indices  for an arbitrary representation $R$ of the gauge group with dimension
$D(R)$, while  $A,B,\cdots$ are the  indices for the adjoint representation with dimension $D(A)$.
The gauge field $(A_m)_{ij}=A_m^{A}T^A_{ij}$ is in the adjoint representation, so that the $T^A_{ij}$  are the basis of $D(A)$ generators 
in the fundamental representation (here $i$ and
$j$ are fundamental and anti-fundamental color indices). For an arbitrary representation the generators are chosen with 
the normalisation
\[
\mathrm{Tr}\left( T^{A}T^{B} \right)  =C(R)\,\delta^{AB}~,
\]
where $C(R)$ is the first Casimir of that representation. We also have
\[
T^A_{ij} \,T^A_{jk} = C_2(R)\,\delta_{ik}\,,
\]
where 
\[
C_2(R)= \frac{D(A)}{D(R)}\,C(R)\,,
\]
is the second Casimir.
Finally the structure functions $f^{ABC}$ are defined as usual as
\[
\left[  T^{A},T^{B}\right]  =if^{ABC}T^{C}~,
\]
and satisfy
\[
f^{ACD}f^{BCD}    = C(A) \,\delta^{AB}~.
\]
In what follows we shall be interested in both the fundamental and the adjoint representation of the $SU(N)$ gauge group, for which we respectively have
\[
C(F) = \frac{1}{2}\,,\ \ \ \ \ \ \ \ C_2(F)=\frac{N^2-1}{2N}\,,\ \ \ \ \ \ \ \ D(F)=N\,,
\]
and
\[
C(A) =N\,,\ \ \ \ \ \ \ \ C_2(A)=N\,,\ \ \ \ \ \ \ \ D(A)=N^2-1\,.
\]

With the above conventions the free propagator for the gluon field is
given in (\ref{GluonPropagator}), while for the Weyl fermion it is
\begin{equation}
\langle \psi_i ( w)\, \bar{\psi}_j (  0) \rangle 
= \frac{g_{\mathrm{YM}}^{2}}{4\pi^{2}} \,\frac{1+\tilde{\gamma}}{2}  \,i\slashed{\partial}_w \,\frac{\delta_{ij}}{w^{2}+i\epsilon}~,
\label{FermionProp1}
\end{equation}
where $\tilde{\gamma} = i \gamma^0 \cdots \gamma^4$ is the chirality matrix. 
Notice that we wrote here the propagator 
between two points in the same patch. If $w$ is in the patch ${\cal P}_1$ and the other point is at the origin of ${\cal P}_3$, we have
\begin{equation}
\langle \psi_i ( w)\, \bar{\psi}_j (  0) \rangle 
= \frac{g_{\mathrm{YM}}^{2}}{4\pi^{2}} \,\frac{1+\tilde{\gamma}}{2}  \,i\slashed{\partial}_w \,\frac{\delta_{ij}}{w^{2}-i\epsilon_w}~,
\label{FermionProp2}
\end{equation}
where $\epsilon_w =\epsilon\, {\rm sgn}( w^0) $.
The  current operator, describing the coupling of the fermion to an external $U(1)$ gauge field, is
\begin{equation}
j_m({x}) = c\, \bar{\psi}_i ({x}) \gamma_m \psi_i({x}) \,,
\label{FermionCurrent}
\end{equation}
where the overall constant 
\begin{equation}
c=\frac{4\pi^2}{g^2_{\mathrm{YM}}}   \,\frac{1}{\sqrt{2D(R)}}
\label{2ptConstant}
\end{equation} 
is determined by the two-point function normalization (\ref{CurrentNormalization}).
The coupling to the gluon field $\eta^{pq}A_q^A$  reads 
\begin{equation}
g_p^A(w) = - \frac{1}{g_{\mathrm{YM}}^{2}}\,\bar{\psi}_i  \gamma_p \psi_j \,T^A_{ij} \,\,.
\label{FermionCoupling}
\end{equation}

We are now ready to compute the impact factor $V^{mn}$ using (\ref{ImpactFactor}). First we consider the contribution of the diagram in figure
\ref{vertex}a to the four-point function in (\ref{ImpactFactor}) with  the $w_i$ parameterized by (\ref{wParameterization}). 
We start with the case of $\sigma_1=w_1^+>0$ and $\sigma_3=w_3^+>0$, so that both points are in the patch 
${\cal P}_1$.  We may also set $x=x_1$ and  $x_3=0$. A simple computation gives 
\begin{align*}
&\langle j_m(x) j_n(0) g_-^A(w_1)g_-^B(w_3)\rangle = -  c^2 \left(  -\frac{1}{g_{\mathrm{YM}}^{2}}\right)^{2}  T^A_{ij} \,T^B_{ji}
\\
&
\mathrm{Tr}\,\Big\{ \gamma_m\, \langle \psi ({x})\, \bar{\psi} ({w}_1) \rangle 
\,\gamma_- \,\langle \psi ({w}_1)\, \bar{\psi} (0) \rangle 
\,\gamma_n\, \langle \psi (0)\, \bar{\psi} ({w}_3) \rangle
\,\gamma_- \,\langle \psi ({w}_3)\, \bar{\psi} ({x}) \rangle\Big\}  \,,
\nonumber
\end{align*}
where the trace in this expression acts on the Dirac indices and the overall minus sign comes from permuting the field $\bar{\psi} ({x})$ through all the 
other fields when applying Wick's theorem.  The fermionic propagators
have no color indices, which have already been taken care by the term $T^A_{ij} \,T^B_{ji}=C(R)\,\delta^{AB}$.
Dropping the colour delta function $\delta^{AB}$, which was already included in the computation of the  BFKL propagator,  the contribution of the diagram in figure \ref{vertex}a to the
impact factor is 
\begin{align}
&
-\frac{c^2}{g_{\mathrm{YM}}^{4}}\, C(R) \left(\frac{g_{\mathrm{YM}}^{2}}{4\pi^{2}}\right)^4
\left(-x^2\right)^3 \left(z_{13}\right)^2 \sigma_1^2\sigma_3^2
\int  d\lambda_1d\lambda_3
\nonumber\\
&
\mathrm{Tr}\left\{
\gamma_m\, \slashed{\partial}_{x} \frac{1}{(x-w_1)^2 +i\epsilon}
\,\gamma_-\,\slashed{\partial}_{{w}_1} \frac{1}{(w_1)^2 -i\epsilon_{w_1}}\right.
\label{CompFermIP}\\
&
\ \ \ \ \ \left.\,\gamma_n\left( - \slashed{\partial}_{{w}_{3}}\right) \,\frac{1}{({w}_{3})^{2}-i\epsilon_{w_3}}
\,\gamma_- \, \slashed{\partial}_{{w}_{3}}\,\frac{1}{\left({w}_{3}-{x}\right)  ^{2}+i\epsilon}
\frac{1-\tilde{\gamma}}{2}
\right\}\,,
\nonumber
\end{align}
where we recall that the $w_i$ are parametrised as in (\ref{wParameterization}).
Now we look at the integration of the second line of this equation
\[
\sigma_1^2 \int  d\lambda_1\, 
4\gamma_m \gamma_a \gamma_- \gamma_b
\frac{ (x-\sigma_1 z_1 - \lambda_1 n)^a   (\sigma_1 z_1 + \lambda_1 n)^b}
{\big(  (x-\sigma_1 z_1 - \lambda_1 n)^2 +i\epsilon \big)^2 \big(  (\sigma_1 z_1 + \lambda_1 n)^2 -i\epsilon_{w_1} \big)^2 }\,.
\]
This integral has two double poles located at
\begin{align*}
& \lambda_1\simeq  - 2 x\cdot z_1 + i\epsilon\,,
\\
 &\lambda_1= -i\epsilon_{w_1} =-  i \epsilon\,,
\end{align*}
where  we set  $\epsilon_{w_1}=\epsilon$ because at  $\lambda_1=0$ we have $w_1^-=w_{1\perp}^2/w^+>0$ and therefore $w_1^0>0$. 
We conclude that one pole is in the lower half plane and the other in the upper 
half plane. Thus,
deforming the $\lambda_1$ contour of integration in the lower half plane, we obtain
\[
(-2\pi i)\,
4\gamma_m \gamma_a \gamma_- \gamma_b\,
\partial_{\lambda_1} \left.\frac{ (x-\sigma_1 z_1 - \lambda_1 n)^a   (\sigma_1 z_1 + \lambda_1 n)^b}
{ \big((x-\sigma_1 z_1 - \lambda_1 n)^2\big)^2}\right|_{\lambda_1=0}\,.
\]
We note that the derivative of the numerator does not contribute because $(\gamma_-)^2=0$. A simple computation
shows that, in the Regge  limit of small $x$, we obtain
\[
%(-2\pi i)\,16\gamma_m \gamma_a \gamma_- \gamma_b\, \frac{ (x-\sigma_1 z_1)^a   (\sigma_1 z_1)^b(x-\sigma_1 z_1)\cdot n}{ \big( (x-\sigma_1 z_1)^2\big)^3} \simeq 
16\pi i\,\gamma_m \gamma_a \gamma_- \gamma_b\,
\frac{ z_1^a   z_1^b}{  (-2x\cdot z_1  )^3}
\,.
\]
At this point we realize that the result is independent of $w_1^+ = \beta_1$, as anticipated  in the previous section. Also, one could repeat the same computation with
$w_1^+<0$ and/or $w_3^+<0$, so that one point and/or both would be in the patch 
${\cal P}_3$, obtaining the same result.
Using the fact that $z_1$ is a null vector, the previous equation simplifies to
\[
32\pi i\, \gamma_m  \gamma_a \,\eta_{-b}  \,
\frac{ z_1^a   z_1^b}{  (-2x\cdot z_1 )^3} 
= -16\pi i \, \gamma_m  \gamma_a \, 
\frac{ z_1^a}{  (-2x\cdot z_1)^3} 
\,.
\]
Doing  in an entirely similar way the integration of the last line in (\ref{CompFermIP}), the contribution of the diagram in figure 
\ref{vertex}a to the impact factor is given by
\[
k\, c^2\, C(R) \,\frac{g^4_{\mathrm{YM}}}{\pi^6}
\left(-x^2\right)^3 \left(z_{13}\right)^2 \mathrm{Tr}\left\{ \gamma_m  \gamma_a\gamma_n  \gamma_b \, \frac{1-\tilde{\gamma}}{2}\right\}
\frac{ z_1^a z_3^b}{  (-2x\cdot z_1 )^3(-2x\cdot z_3)^3} \,,
\]
where we multiplied by the factor $k$ in (\ref{prefactor}) to match with our conventions for the BFKL propagator.
Using the  identity $\mathrm{Tr}\left\{\gamma_m\gamma_a \gamma_n \gamma_b \right\}=4(\eta_{ma}\eta_{nb} + \eta_{mb}\eta_{na} - \eta_{mn}\eta_{ab})$,
and noting that the term containing $\mathrm{Tr}\left\{\gamma_m\gamma_a \gamma_n \gamma_b \tilde{\gamma}\right\} = -4i \epsilon_{manb}$ will not contribute since
the final answer must be invariant under the exchange $x_1 \leftrightarrow x_3 $ and $m \leftrightarrow n$, the previous equation simplifies to
\[
 k\, c^2\, C(R) \,\frac{g^4_{\mathrm{YM}}}{\pi^6}\,
u^3 \left( \eta^{mn}  +  2\,\frac{ z_{1}^{m} z_{3}^{n} + z_{1}^{n} z_{3}^{m}}{z_{13}}\right)\,.
\]
Writing the result in terms of the tensor 
structures defined in (\ref{ITensors}), we conclude  that the diagram in figure 
\ref{vertex}a  contributes to the impact factor with
\[
k \,c^2\, C(R) \,\frac{g^4_{\mathrm{YM}}}{\pi^6}\, u^3 \Big( {\cal I}_1^{mn} + 2 {\cal I}_5^{mn}\Big)\,. 
\]
A simple check shows that this expression satisfies the current conservation conditions (\ref{hfunctionsdivergence}).

The contribution of the diagrams in figures  \ref{vertex}b and \ref{vertex}c to the  impact factor is proportional to a delta function $\delta(u)$. 
For the sake of clarity we compute these diagrams in appendix \ref{DeltaFunction}. Including these extra terms,
the final result for the impact factor of a Weyl fermion in a representation $R$ of  the colour  gauge group is
\begin{equation}
V^{mn}_{\rm fermion} (x,z_1,z_3)=  a
\left[
u^3 \Big( {\cal I}_1^{mn} + 2 {\cal I}_5^{mn}\Big)+
\frac{u^2\delta(u)}{2} \Big(  {\cal I}_1^{mn} + {\cal I}_3^{mn}\Big)
\right]\,,
\label{FermionIF}
\end{equation}
with
\begin{equation}
a= k \,c^2\, C(R) \,\frac{g^4_{\mathrm{YM}}}{\pi^6} = \frac{N}{\sqrt{D(A)}}\, \frac{C_2(R) g^2_{\mathrm{YM}}}{\pi^3}\simeq \frac{C_2(R) g^2_{\mathrm{YM}}}{\pi^3}\,,
\label{aConstant}
\end{equation}
where we took the large $N$ limit in the last step.

Let us remark that the role of the terms with a delta function is to enforce the IR finiteness condition (\ref{IRfinite}).
In fact, contracting the above impact factor with $\eta^{mn}$, $x^m x^n$, $x^m z_3^n$ and $z_3^m z_3^n$, it is simple to verify that this
condition is indeed satisfied.

%%%%%%%%%%%%%%%%%%%%%%%%%%%%%%%%%%%%%%
\subsection{Impact factor for a massless quark in QCD}
%%%%%%%%%%%%%%%%%%%%%%%%%%%%%%%%%%%%%%

From the result of the previous section 
it is trivial to compute the impact factor for the electromagnetic current operator in QCD $j_m=c \bar{\psi}\gamma_m\psi$, since a  massless quark is equivalent to
two Weyl fermions. The  constant $c$ computed from the normalisation of the current two-point function is now 
\[
c=\frac{2\pi^2}{g^2_{\mathrm{YM}}}   \,\frac{1}{\sqrt{D(F)}}\,.
\]
The impact factor is again given by (\ref{FermionIF}) with the overall constant $a$ given by 
\[
a= \frac{ g^2_{\mathrm{YM}} \sqrt{N^2-1}}{2\pi^3} \simeq \frac{g^2}{2\pi^3}\,,
\]
where $g^2=g^2_{\mathrm{YM}}N$ is the 't Hooft coupling.

%%%%%%%%%%%%%%%%%%%%%%%%%%%%%%%%%%%%%%
\subsection{Impact factor for a complex scalar field}\label{Scalar}
%%%%%%%%%%%%%%%%%%%%%%%%%%%%%%%%%%%%%%

Now we wish to compute the impact factor for a  charged complex scalar field in an arbitrary representation $R$ of the gauge group.
At the end of the computation, and also using the results of section \ref{Weyl},  it will be simple
to determine the impact factor for the $R$-current of ${\cal N}=4$  SYM theory. We compute the impact factor for a theory with Lagrangian 
\[
\frac{1}{g_{\mathrm{YM}}^{2}}\int
d^{4}{y}
\left(  -\frac {1}{2}\,\mathrm{Tr} \left(F_{mn}F^{mn}\right) -  (D_m Z)_i  (D^m \bar{Z})_i   \right) ,
\]
where $(D_m Z)_i = \left( \delta_{ij}\partial_m  - i T^A_{ij}A_m^A \right) Z_j$. 
With these conventions the free propagator for the scalar field is
\begin{equation}
\langle Z_i( w) \bar{Z}_j(0) \rangle 
= \frac{g_{\mathrm{YM}}^{2}}{4\pi^{2}} \,\frac{\delta_{ij}}{w^{2}+i\epsilon}~,
\label{ScalarProp1}
\end{equation}
when both points are in the same Poincar\'e patches. If $w$ is in the patch ${\cal P}_1$ and the other point is at the origin of  ${\cal P}_3$,
we have
\begin{equation}
\langle Z_i( w) \bar{Z}_j(0) \rangle 
= \frac{g_{\mathrm{YM}}^{2}}{4\pi^{2}} \,\frac{\delta_{ij}}{w^{2}-i\epsilon_w}~.
\label{ScalarProp2}
\end{equation}
The current operator is simply given by
\begin{equation}
j_m(x) = c\,  i \bar{Z}_i \overleftrightarrow{D}_{\hspace{-0.1cm}m} Z_i 
= c\left( i \bar{Z}_i \overleftrightarrow{\partial}_{\hspace{-0.1cm}m}Z_i   + 2A_m^A  T^A_{ij}\bar{Z}_i Z_j  \right)\,,
\label{ScalarCurrent}
\end{equation}
where  the constant $c$ is again given by (\ref{2ptConstant}).
To leading order in perturbation theory we may set  $j_m(x) = c\,\bar{Z}_i  \overleftrightarrow{\partial}_{\hspace{-0.1cm}m} Z_i $.
Finally, the coupling to the gluon field $\eta^{pq}A_q^A$  reads 
\begin{equation}
g_p^A (w)= - \frac{i}{g_{\mathrm{YM}}^{2}}\,\bar{Z}_i \overleftrightarrow{\partial}_{\hspace{-0.1cm}p}Z_j  \,T^A_{ij} \,.
\label{ScalarCoupling}
\end{equation}

We consider first the contribution of  the diagram in figure \ref{vertex}a to the impact factor. Again we start with the case 
$w_1^+>0$ and $w_3^+>0$, so that both points are in the patch ${\cal P}_1$.  The final result is independent of this choice.
A simple computation gives 
\begin{align*}
&\langle j_m(x_1) j_n(x_3) g_-^A(w_1)g_-^A(w_3)\rangle = - c^2 \left(  -\frac{i}{g_{\mathrm{YM}}^{2}}\right)^{2}  T^A_{ij} \,T^B_{ji}
\\
&
\langle Z(x_1) \bar{Z}(w_1)\rangle
\overleftrightarrow{\partial}_{\hspace{-0.1cm}w_1^-}
\langle Z(w_1) \bar{Z}(x_3)\rangle
(-\overleftrightarrow{\partial}_{\hspace{-0.1cm}x_1^m} )
\overleftrightarrow{\partial}_{\hspace{-0.1cm}x_3^n}
\langle Z(x_3) \bar{Z}(w_3)\rangle
\overleftrightarrow{\partial}_{\hspace{-0.1cm}w_3^-}
\langle Z(w_3) \bar{Z}(x_1)\rangle
\,,
\end{align*}
where the propagators have no color indices, already included in the term $T^A_{ij} \,T^B_{ji}=C(R)\,\delta^{AB}$.
We conclude that the  contribution of this diagram to the impact factor is 
\begin{align}
&
-\frac{c^2}{g_{\mathrm{YM}}^{4}}\, C(R) \left(\frac{g_{\mathrm{YM}}^{2}}{4\pi^{2}}\right)^4
\left(-x^2\right)^3 \left(z_{13}\right)^2 \sigma_1^2 \sigma_3^2
\int d\lambda_1 d\lambda_3
\nonumber\\
&
\left(\frac{1}{(x_1-w_1)^2 +i\epsilon}
\overleftrightarrow{\partial}_{\hspace{-0.1cm}w_1^-}
\frac{1}{(w_1-x_3)^2 -i\epsilon_{w_1-x_3}}\right)
\overleftrightarrow{\partial}_{\hspace{-0.1cm}x_1^m} \overleftrightarrow{\partial}_{\hspace{-0.1cm}x_3^n}
\label{CompScalarIP1}\\
&
\left(\frac{1}{\left(x_3-{w}_{3}\right)  ^{2}-i\epsilon_{w_3-x_{3}}}
\overleftrightarrow{\partial}_{\hspace{-0.1cm}w_3^-}
\frac{1}{\left({w}_{3}-x_1\right)  ^{2}+i\epsilon}\right)\,,
\nonumber
\end{align}
Now we look at the integration of the term in curved parenthesis in the  second line of this equation
\[
\sigma_1^2\int d\lambda_1  
\left(\frac{1}{(x_1-\sigma_1 z_1 - \lambda_1 n)^2 +i\epsilon}
\overleftrightarrow{\partial}_{\hspace{-0.1cm}\lambda_1}
\frac{1}{ (x_3 - \sigma_1 z_1 - \lambda_1 n)^2 -i\epsilon_{w_1-x_3}}\right)\,.
\]
This integral has poles located at
\begin{align*}
\lambda_1&\simeq - 2 x_1\cdot z_1 + i\epsilon\,,
\\
 \lambda_1 &\simeq -2 x_3\cdot z_1 -i\epsilon_{w_1-x_3} =-2 x_3\cdot z_1 -  i \epsilon\,,
\end{align*}
where again at the second pole $\epsilon_{w_1-x_3}=\epsilon$. Thus,
deforming the $\lambda_1$ contour of integration in the lower half plane, we obtain
\[
\frac{4\pi i}{ \left(-2(x_1-x_3)\cdot z_1 \right)^2}
\,.
\]
A similar argument can be done to integrate the last line of (\ref{CompScalarIP1}), with a similar result up to a minus sign. 
Recalling that $x\approx x_1-x_3$, we conclude that the
contribution of the diagram in figure \ref{vertex}a to the impact factor is 
\[
k\, c^2\, C(R) \,\frac{g^4_{\mathrm{YM}}}{16\pi^6}
\left(-x^2\right)^3 \left(z_{13}\right)^2 
\left(\frac{1}{ (-2 x\cdot z_1 )^2}
\overleftrightarrow{\partial}_{\hspace{-0.1cm}x^m} \overleftrightarrow{\partial}_{\hspace{-0.1cm}x^n}
\frac{1}{ (-2x\cdot  z_3  )^2}\right)\,,
\]
which can be simplified to
\[
k\, c^2\, C(R) \,\frac{g^4_{\mathrm{YM}}}{\pi^6}
\,u^2 \left[ \frac{3}{2}\left( \frac{z_{1m}z_{1n} (-x^2)}{(-2 x\cdot z_1 )^2}  +  \frac{z_{3m}z_{3n} (-x^2)}{(-2 x\cdot z_3 )^2}  \right)  
- u \, \frac{z_{1m}z_{3n}+z_{3m}z_{1n}}{z_{13}} \right]\,.
\]
Expressing the result in terms of the tensor structures defined in (\ref{ITensors}), we have  then
\[
k\,c^2\, C(R) \,\frac{g^4_{\mathrm{YM}}}{\pi^6}\,
u^2 \left(  \frac{3}{2}\,{\cal I}_4^{mn} - u \,{\cal I}_5^{mn}\right)\,,
\]
which obeys the conservation conditions (\ref{hfunctionsdivergence}).

Again, the contribution from the
diagrams in figures \ref{vertex}b and \ref{vertex}c is proportional to $\delta(u)$ and is presented in appendix \ref{DeltaFunction}. 
Including these extra terms, the final result for the impact factor of a complex scalar field  in a representation $R$ of  the colour  gauge  group is
\begin{align}
V^{mn}_{\rm scalar}  (x,z_1,z_3)&= a\,
\left[
u^2  \left(  \frac{3}{2}\,{\cal I}_4^{mn} - u \,{\cal I}_5^{mn}\right) 
 + \frac{u^2\delta(u)}{2} \left(  \frac{1}{4}\,{\cal I}_1^{mn}  - {\cal I}_2^{mn} + \frac{3}{2}\,{\cal I}_3^{mn}   -3\,{\cal I}_4^{mn} \right)
\right]\,,
\label{ScalarIF}
\end{align}
where $a$ is given in (\ref{aConstant}). A simple computation shows that this impact factor 
correctly enforces the  IR finiteness condition (\ref{IRfinite}).

%%%%%%%%%%%%%%%%%%%%%%%%%%%%%%%%%%%%%%
\subsection{Impact factor in ${\cal N}=4$ SYM}\label{Rcurrent}
%%%%%%%%%%%%%%%%%%%%%%%%%%%%%%%%%%%%%%

Four-dimensional ${\cal N}=4$ SYM can be obtained from dimensional reduction of ten-dimensional 
${\cal N}=1$ SYM. We start by considering a convenient reduction of the ten-dimensional Dirac matrices. We
will denote with $\gamma_a$ and with $\Gamma_\alpha$, respectively, the $SO(3,1)$ and $SO(6)$ Dirac matrices, satisfying
\[
\left[ \gamma_a,\gamma_b \right] = 2\eta_{ab}\,,\ \ \ \ \ \ \ \ \ \ \ 
\left[ \Gamma_\alpha,\Gamma_\beta \right] = 2\delta_{\alpha\beta}\,,
\]
where we use Greek indices $\alpha,\beta,\cdots$ for the internal directions. We will choose a Majorana basis where the 
$\gamma_a$ are $4\times4$ purely real matrices and the $\Gamma_\alpha$ are $8\times 8$ purely imaginary matrices. Moreover we will denote the imaginary chirality matrices
as
\[
\tilde{\gamma} = i \gamma^0\cdots \gamma^3\,,\ \ \ \ \ \ \ \ \ \ \ 
\tilde{\Gamma} = -i \Gamma^1\cdots\Gamma^6\,.
\]
With the above conventions we can define the real $SO(9,1)$ Dirac matrices as 
\[
\gamma_a \otimes 1\,,\ \ \ \ \ \ \ \ \ \ \ \tilde{\gamma}\otimes \Gamma_\alpha\,.
\]
The ten-dimensional chirality matrix is then given by $\tilde{\gamma}\otimes\tilde{\Gamma} \equiv \tilde{\gamma}\tilde{\Gamma}$.

Our starting point is a ten-dimensional Weyl-Majorana spinor $\lambda$,
\[
\tilde{\gamma}\tilde{\Gamma} \lambda = \lambda\,,\ \ \ \ \ \ \ \ \ \ \ \lambda^\star =\lambda\,.
\]
This spinor can be written as a Weyl-Weyl spinor of $SO(3,1)\times SO(6)$. In fact,  defining 
\[
\psi = \frac{1+\tilde{\gamma}}{2}\,\frac{1+\tilde{\Gamma}}{2}\,\lambda\,,
\]
it is clear that $\lambda=\psi + \psi^\star$. It is now an exercise to show that the reduction
of ten-dimensional ${\cal N}=1$ SYM yields the following action for the ${\cal N}=4$  theory
\begin{align*}
\frac{1}{g_{\mathrm{YM}}^{2}}\int 
d^{4}{y} \,\mathrm{Tr}
\Big(  &-\frac {1}{2}\,F_{mn}F^{mn} -  D_m\phi_\alpha D^m\phi^\alpha + \frac{1}{2}\left[ \phi_\alpha,\phi_\beta\right]^2
\\&
-2i\bar{\psi} \gamma\cdot D \psi - \psi \Gamma^\alpha \left[ \phi_\alpha,\psi\right] + \bar{\psi}\Gamma^\alpha\left[ \phi_\alpha,\bar{\psi}\right]\Big)\,,
\end{align*}
where the last two fermionic bilinears are defined by $\psi \Gamma^\alpha \psi = \psi^T \gamma^0 \Gamma^\alpha \psi$ and 
$\bar{\psi} \Gamma^\alpha \bar{\psi} = \psi^\dagger \gamma^0 \Gamma^\alpha \psi^\star$. The covariant derivative is defined by 
$D_m=\partial_m - i [A_m,\;]$.
All fields are in the adjoint representation
of the colour gauge group, for instance $(\psi^A)_{ij}= \psi^A T^A_{ij}$, with $i$ and $j$ fundamental and anti-fundamental 
colour indices. 
It is therefore trivial to make direct contact with the notation of sections \ref{Weyl} and \ref{Scalar}, which considered
an arbitrary representation of the gauge group. In particular, the contribution
of the Weyl fermion and scalar fields to the impact factor of the R-current can be readily  determined from the results of those
sections.

It is simple to verify that the Weyl fermion propagator $\left\langle \psi^A(w) \bar{\psi}^B(0) \right\rangle$ is exactly given by
those in  (\ref{FermionProp1}) and (\ref{FermionProp2}) for a field in the adjoint representation. Also, defining the complex scalar field
\[
Z=\frac{1}{\sqrt{2}}\,\left( \phi_1 + i\phi_2 \right)\,,
\] 
the expressions for the propagator $\left\langle Z^A(w) \bar{Z}^B(0) \right\rangle$ exactly match those in  (\ref{ScalarProp1}) and (\ref{ScalarProp2}).

The above action is invariant under the global $SO(6)$ R-symmetry, with transformation
\[
\delta \phi_\alpha = \omega_{\alpha\beta} \phi^\beta\,,\ \ \ \ \ \ \ \ \ 
\delta\psi = \frac{1}{4}\,\omega_{\alpha\beta} \Gamma^{\alpha\beta}\psi\,,
\]
where
\[
\Gamma^{\alpha\beta}\ = \frac{1}{2}\left[ \Gamma^\alpha, \Gamma^\beta \right]\,.
\]
The corresponding conserved R-current reads
\[
j_m^{\alpha\beta} (x)= c\, \mathrm{Tr} \left(  - \phi^\alpha \overleftrightarrow{D}_{\hspace{-0.1cm}m} \phi^\beta +\frac{i}{2}\,\bar{\psi} \gamma_m \Gamma^{\alpha\beta}\psi \right)\,.
\]
In the following we shall consider the $12$ component of this current, which is given by
\[
j_m(x) = j_m^{12} (x)= c\,\mathrm{Tr} \left(  i \bar{Z}  \overleftrightarrow{D}_{\hspace{-0.1cm}m} Z +\frac{i}{2}\,\bar{\psi} \gamma_m \Gamma^{12Ô}\psi \right)\,.
\]
Both terms have exactly the same form as those in (\ref{FermionCurrent}) and (\ref{ScalarCurrent}), except for the factor of $i/2$ and the generator 
$\Gamma^{12}$ in the fermionic piece. The computation of the R-current two-point function is similar to that of a Weyl fermion and a complex scalar. For the Weyl fermion,
the difference with respect to section \ref{Weyl} is a factor of $(i/2)^2$ times a factor from the generators of the R-current of
\[
\mathrm{Tr} \left(  \Gamma^{12}\Gamma^{12}\, \frac{1+\tilde{\Gamma}}{2}\right) = -4\,,
\]
which gives unit. Taking this fact into account the normalisation of the 
two-point function fixes the constant $c$ to be
\[
c=\frac{2\pi^2}{g^2_{\mathrm{YM}} \sqrt{N^2-1}} \,.
\]

Finally, the coupling to the gluon field $\eta^{pq}A_q^A$  reads 
\[
g_p^A (w)= - \frac{1}{g_{\mathrm{YM}}^{2}}\,\bar{Z}^B \overleftrightarrow{\partial}_{\hspace{-0.1cm}p}Z^C  f^{ABC} + \frac{i}{g_{\mathrm{YM}}^{2}}\,\bar{\psi}_B  \gamma_p \psi_C f^{ABC} \,,
\]
which, recalling that  $f^{ABC}=i T^A_{BC}$ for adjoint fields, exactly reproduces the sum of (\ref{FermionCoupling}) and (\ref{ScalarCoupling}).

It is now trivial to import the results of sections  \ref{Weyl} and \ref{Scalar} to determine the R-current impact factor. Again the only difference is the
computation of the fermionic contribution, with the same factor of $(i/2)^2$ times $-4$ from the trace of the generators of the R-current.
We conclude that the R-current impact factor is  given by
\begin{equation*}
V^{mn}  = V_{\rm scalar}^{mn} +V_{\rm fermion}^{mn}  \,,
\end{equation*}
with each term given by  (\ref{FermionIF}) and (\ref{ScalarIF}). In this case the overall constant 
\[
a=\frac{N^2}{\sqrt{N^2-1}}\,\frac{g^2_{\mathrm{YM}}}{2\pi^3} \simeq \frac{g^2}{2\pi^3}\,,
\]
where $g^2=g^2_{\mathrm{YM}} N$ is the 't Hooft coupling.

%%%%%%%%%%%%%%%%%%%%%%%%%%%%%%%%%%%%%%
\subsection{Spin 0 and Spin 2 components}
%%%%%%%%%%%%%%%%%%%%%%%%%%%%%%%%%%%%%%

In this section we shall decompose the fermion and scalar impact factors (\ref{FermionIF}) and (\ref{ScalarIF}) in their spin
0 and spin 2 components. We follow the general procedure explained in section \ref{Disentangling} to determine
the functions $S_{k}(\mu)$ and $T(\mu)$ in the Fourier decomposition (\ref{Spin0Expansion})  and (\ref{Spin2Expansion}), which 
enter the expression for the amplitude in its Regge form.

First we consider the impact factor for a Weyl fermion given by  (\ref{FermionIF}). Following section  \ref{Disentangling}, the scalar functions 
$S_k(u)$, as given by (\ref{S1S2S3final}) and (\ref{S4}),  are
\begin{align*}
&S_{1}   =  \frac{1}{3}\,u^{2}+\frac{1}{3}\,u^{3}\,,\ \ \ \ \ \ \ \ 
S_{2}   =  -u^{2}+u^{3}\,,\ \ \ \ \ \ \ \ 
S_{3}   =  \frac{u^{2}}{4}\,,\ \ \ \ \ \ \ \ 
\left(\nabla^2-3\right)S_{4}  = -\frac{1}{4}\,u^{2}+\frac{1}{2}\,u^{3}\,,
\end{align*}
where we dropped the overall factor of $a$. It is convenient to normalise 
the transform  $S_k(\nu) $ in the expansion (\ref{Spin0Expansion}) with respect to the transform of $u^2$ as 
 \begin{equation}
S_k(\nu) = \frac{2\pi^{3}}{\mbox{cosh\ensuremath{\left(\frac{\pi\nu}{2}\right)}}} \,s_k(\nu)\,.
\label{normalisation}
\end{equation}
Then we have
\begin{align*}
s_{1}   =  \frac{\nu^{2}+25}{48}\,,\ \ \ \ \ \ \ \    s_{2}   =  \frac{\nu^{2}-7}{16}\,,\ \ \ \ \ \ \ \ \ \  
s_{3}   =  \frac{1}{4}\,,\ \ \ \ \ \ \ \ \ 
s_{4}   =  -\frac{1}{32}\:\frac{\nu^{2}+1}{\nu^{2}+4}\,,
\end{align*}
which satisfy the algebraic current conservation conditions (\ref{nuconservation}). To determine the spin 2
component, first we compute the  Cotton
function $C(u)$ as given by (\ref{Cotton}),
\[
C=18u^{3}(1-u)(1-2u)\,.\\
\]
Then the transform in  (\ref{Spin2Expansion}) is given by 
\[
T=  - \frac{2\pi^{3}}{\mbox{cosh\ensuremath{\left(\frac{\pi\nu}{2}\right)}}} \,\frac{\left(1+\nu^{2}\right)^{2}\left(9+\nu^{2}\right)}{256\left(4+\nu^{2}\right)}\,.
\]

Next we consider the  impact factor for a complex scalar given by  (\ref{FermionIF}). Again dropping the overall factor of $a$, the
scalar functions given by (\ref{S1S2S3final}) and (\ref{S4}) are
\begin{align*}
S_{1}   =  \frac{1}{12}\,u^{2}-\frac{1}{3}\,u^{3}\:,\ \ \ \ \  \ \ \ \ \
S_{2}   =  -\frac{1}{4}\,u^{2}\:,\ \ \ \ \  \ \ \ \ \
S_{3}   =  \frac{u^{2}}{16}\:,\ \ \ \ \  \ \ \ \ \
\left(\nabla^2-3\right)S_{4}   =  -\frac{1}{16}\,u^{2}+ \frac{1}{2}\,u^{3}\:.
\end{align*}
With the normalisation (\ref{normalisation}), the transforms in (\ref{Spin0Expansion}) are given by 
\begin{align*}
s_{1}   =  \frac{\nu^{2}+13}{48}\:,\ \ \ \ \ \ \ \ \ 
s_{2}   =  -\frac{1}{4}\:,\ \ \ \ \ \ \ \ \ 
s_{3}   =  \frac{1}{16}\:,\ \ \ \ \ \ \ \ \ 
s_{4}   =  -\frac{1}{32}\:\frac{\nu^{2}+7}{\nu^{2}+4}\:,
\end{align*}
which satisfy the algebraic current conservation conditions (\ref{nuconservation}). Finally, the Cotton
function $C(u)$ in (\ref{Cotton}) is
\[
C = - 18u^{3}(1-u)(1-2u)\,,
\]
which leads to
\[
T=\frac{2\pi^{3}}{\mbox{cosh\ensuremath{\left(\frac{\pi\nu}{2}\right)}}} \,\frac{\left(1+\nu^{2}\right)^{2}\left(9+\nu^{2}\right)}{256\left(4+\nu^{2}\right)}\:.
\]

%%%%%%%%%%%%%%%%%%%%%%%%%%%%%%%%%%%%%%
\section{A conjecture and concluding remarks}
%%%%%%%%%%%%%%%%%%%%%%%%%%%%%%%%%%%%%%

Looking at the R-current impact factor computed in section \ref{Rcurrent} and at the transverse spin 2 components for a Weyl fermion and a complex scalar field just computed, we conclude
that the spin 2 component of the R-current impact factor vanishes. The R-current impact factor only has overlap with the spin 0 component of the BFKL kernel, which is dual to
the graviton Regge trajectory. Other current impact factors in ${\cal N}=4$ SYM,  that are not half-BPS operators, will in general have a spin 2 component. We believe this cancelation
is related to supersymmetry and should hold to all operators that are dual to the supergravity multiplet. We are then led to the following conjecture:
{\it half-BPS single-trace operators in ${\cal N}=4$ SYM have impact factors with zero transverse conformal spin}.
  We checked this conjecture for the R-current operator and to
 leading order in the coupling constant. At strong coupling, in the gravity approximation, this is also true since in this limit all strings states are decoupled and we are only left with the
 supergravity multiplet. 
 It would be 
  interesting to check this conjecture at one-loop order in the computation of the impact factor.

Next to leading order corrections to the impact factor of scalars operators have been studied recently in \cite{Balitsky}. A scheme to renormalize the impact factors 
was presented, while preserving conformal invariance. In general we expect the amplitude to have the following form, for any value of the coupling, 
\begin{align*}
\mathcal{A}_{mn} (x,\bar{x})
  \simeq&-\frac{1}{N^2}\,\int_{\partial H_{3}} \frac{dz_1 dz_3}{\left(z_{13}\right)^2}\,
   \frac{dz_2 dz_4}{\left(z_{24}\right)^2}~
  V_{mn}(  x,z_1,z_3)  \; F(z_1,z_3,z_2,z_4,\sigma) \; \bar{V}(\bar{x},z_2,z_4)  \,,
\end{align*}
where $\sigma = |x||\bar{x}|$. In this expression the pomeron propagator depends on $\sigma$. Its decomposition in conformal partial waves,
as given by (\ref{Kernel}), will now include a factor of $\sigma^{1-j(\nu,n)}$.
The impact factor $V^{mn}$ depends on functions of the
cross ratio $u$, which also depend on the coupling constant. In the case of  ${\cal N}=4$ SYM, the expression above assumes a renormalization scheme that does
not break conformal invariance of the impact factors, order by order in perturbation theory.

Expression (\ref{ImpactFactor}) gives the weak coupling impact factor in terms of a four-point function of the external currents and the gluonic current that couples to the exchanged gluons.
If a similar form of the impact factor can be generalized for any coupling (possibly by using lightlike Wilson line operators \cite{Balitskyreview,BanksFestuccia}), then we can try to compute the impact factor at higher orders in perturbation theory. The next-to-leading order has been recently obtained in \cite{Balitsky}.
Another important related open question is the nature of the impact factors at strong coupling.
If we can find a definition of the impact factor  compatible with conformal symmetry for all values of the coupling then we can hope that the integrability of ${\cal N}=4$ SYM will be enough to determine it.
The simplicity of the weak coupling result certainly points in this direction.

%%%%%%%%%%%%%%%%%%%%%%%%%%%%%%%%%%%%%%%%%%%%%%%%%%%%%%%%%%%%%%%%%
\section*{Acknowledgments}

We wish to thank I. Balitsky, L. Lipatov, A. Mueller, J. Polchinski and P. Vieira for   discussions. 
LC is funded by the \textit{Museo Storico della Fisica e Centro Studi e
Ricerche "Enrico Fermi"} and is partially funded by INFN, by the MIUR--PRIN
contract 2005--024045--002 and by the EU contracts MRTN--CT--2004--005104. 
MC wishes to thank the theory group at CERN, in particular L. \'Alvarez-Gaum\'e, for the very kind hospitality during the period most of
this work was done. MC is partially supported by the grants CERN/FP/83508/2008 and SFRH/BSAB/870/2008.
JP is funded by  FCT grant SFRH/BPD/34052/2006. 
This research was supported in part by the National Science Foundation under 
Grant No. NSF PHY05-51164.
\emph{Centro de F\'{\i}sica do Porto} is partially funded
by FCT through the POCI program.

%%%%%%%%%%%%%%%%%%%%%%%%%%%%%%%%%%%%%%%%%%%%%%%%%%%%%%%%%%%%%%%%%

\vfill

\eject

\appendix

%%%%%%%%%%%%%%%%%%%%%%%%%%%%%%%%%%%%%%
\section{Some Fourier transforms  \label{Ftrans}}
%%%%%%%%%%%%%%%%%%%%%%%%%%%%%%%%%%%%%%

In this appendix we compute some Fourier transforms used in section \ref{Regge}.
We start by 
\begin{align}
B^{mn}(p,\bar{p})=  \frac{1}{\pi^8} \int dx  d\bar{x} \,  e^{ 2 i p\cdot x +2 i \bar{p} \cdot \bar{x}}
A^{mn}(x,\bar{x}) \ , \label{Bft}
\end{align}
with  $A^{mn}(x,\bar{x}) $ given by the Regge form (\ref{ReggeA}).
It is convenient to rewrite (\ref{ReggeA}) as 
\begin{align*}
A^{mn}(x,\bar{x})  \approx -2\pi i  \int 
  d\nu~\sum_{k=0}^4\alpha_k (  \nu)\,x^2\mathcal{D}_k^{mn}  \frac{ \Omega_{i\nu} ( \rho)}{
( x^2-i\epsilon_{x})^{\xi +\frac{j(\nu)+1}{2}}  ( \bar{x}^2-i\epsilon_{\bar{x}})^{\Delta+\frac{j(\nu)-1}{2}}     }~,
\end{align*}
where in this equation we redefined the operators $\mathcal{D}_k^{mn}$ by 
\begin{align*}
x^2\mathcal{D}_1^{mn}=&\; x^2\eta^{mn}- x^mx^n \,,
\\ x^2\mathcal{D}_2^{mn}=&\; x^mx^n \,,
\\ x^2\mathcal{D}_3^{mn}=&\;x^2\left(x^m\partial^n+x^n\partial^m\right) -2 x^m x^n x\cdot \partial\,,
\\ x^2\mathcal{D}_4^{mn}=&\;x^4\partial^n \partial^m -x^2 x^q \left(x^m\partial^n+x^n\partial^m\right) \partial_q \\
&-\frac{1}{3} \left( x^2\eta^{mn}-x^mx^n\right) x^2 \partial^2+\frac{1}{3} \left(x^2\eta^{mn} +2x^nx^m\right) x^qx^s \partial_q \partial_s\,,
\end{align*}
 so that they commute with any function of $x^2$.
 One can then use integration by parts in (\ref{Bft}) to write
\begin{align}
B^{mn}(p,\bar{p})\approx &\; -2\pi i  \int
d\nu~\sum_{k=0}^4 \alpha_k(  \nu) \tilde{\mathcal{D}}_k^{mn} %
%\\&
 \frac{1}{\pi^8} \int dx d\bar{x} \,
\frac{ e^{ 2i x\cdot p +2i \bar{x} \cdot \bar{p}}\, \Omega_{i\nu} ( \rho )}{
( x^2-i\epsilon_{x})^{\xi +\frac{j(\nu)+1}{2}}  ( \bar{x}^2-i\epsilon_{\bar{x}})^{\Delta+\frac{j(\nu)-1}{2}}    } \,,
 \label{Bint} 
\end{align}
where 
\begin{align*}
-4\tilde{\mathcal{D}}_1^{mn}=&\; \eta^{mn}\hat{\partial}^2- \hat{\partial}^m\hat{\partial}^n \,,
\\ -4\tilde{\mathcal{D}}_2^{mn}=&\;  \hat{\partial}^m\hat{\partial}^n \,,
\\ -4\tilde{\mathcal{D}}_3^{mn}= &\;-\hat{\partial}^2 \left(\hat{\partial}^n p^m+\hat{\partial}^m p^n\right)
 +2 \hat{\partial}^m\hat{\partial}^n \hat{\partial}   \cdot p\,,
\\ -4\tilde{\mathcal{D}}_4^{mn} = &\; \hat{\partial}^4p^np^m  -  
\hat{\partial}^2 \hat{\partial}_s \left(\hat{\partial}^n p^m+\hat{\partial}^mp^n\right) p^s
\\
&-\frac{1}{3} \hat{\partial}^2\left( \eta^{mn}\hat{\partial}^2- \hat{\partial}^m\hat{\partial}^n \right)p^2  
+\frac{1}{3}  \hat{\partial}_s\hat{\partial}_q\left( \eta^{mn}\hat{\partial}^2+2\hat{\partial}^m\hat{\partial}^n\right)  p^s p^q\,,
\end{align*}
and $\displaystyle{\hat{\partial}_n =\frac{\partial\ }{\partial p^n}}$.
The scalar integral in the second line of (\ref{Bint}) can be done explicitly.
First notice that the $i\epsilon$-prescription implies that the integral vanishes if either $p$ or $\bar{p}$ is spacelike or past-directed.
We can then write
\begin{align*}
 \frac{1}{\pi^8} \int dx d\bar{x} \,
\frac{ e^{ 2i x\cdot p +2i \bar{x} \cdot \bar{p}}\,\Omega_{i\nu} ( \rho )}{
( x^2-i\epsilon_{x})^{\xi +\frac{j(\nu)+1}{2}}  ( \bar{x}^2-i\epsilon_{\bar{x}})^{\Delta+\frac{j(\nu)-1}{2}}    }=
\frac{ \theta(p^0) \theta(-p^2) \theta(\bar{p}^0) \theta(-\bar{p}^2) G \left( e\cdot \bar{e} \right)}{
(-p^2)^{2 -\xi-\frac{j(\nu)+1}{2}}  (-\bar{p}^2)^{2-\Delta-\frac{j(\nu)-1}{2}}     }\,,
\end{align*}
just using Lorentz invariance and scaling.
Performing a Fourier transform we have 
\begin{align*}
\frac{  \Omega_{i\nu} ( \rho )}{
( x^2-i\epsilon_{x})^{\xi +\frac{j(\nu)+1}{2}}  ( \bar{x}^2-i\epsilon_{\bar{x}})^{\Delta+\frac{j(\nu)-1}{2}}    }=
\int_{\rm M} dp d\bar{p}  \,
\frac{ e^{ -2i x\cdot p -2i \bar{x} \cdot \bar{p}}\, G \left( e\cdot \bar{e} \right)}{
(-p^2)^{2 -\xi-\frac{j(\nu)+1}{2}}  (-\bar{p}^2)^{2-\Delta-\frac{j(\nu)-1}{2}}     }\,,
\end{align*}
where we denote by ${\rm M}$ the future light-cone or Milne wedge.
To determine the function $G$ it is sufficient to consider future directed $x$ and $\bar{x}$.
In this case, after integrating over $E$ and $\bar{E}$ (recall that $p=E \,e$) we find
\begin{align*}
 \Omega_{i\nu} ( \rho ) 
 =\int_{H_3} de d\bar{e}\,  
\frac{ \Gamma(2\xi+j(\nu)+1) \Gamma(2\Delta+j(\nu)-1)  \,G \left( e\cdot \bar{e} \right)}{
\left(-2 e \cdot x /|x|  \right)^{2\xi+j(\nu)+1}  
\left(-2\bar{e}\cdot \bar{x}/|\bar{x}|\right)^{2\Delta+j(\nu)-1}     }\,.
\end{align*}
Each integral is a convolution of radial functions on $H_3$ that is easily done using the harmonic basis \cite{Mythesis}. This gives
$
G \left( e\cdot \bar{e} \right)= \zeta(\nu) \,\Omega_{i\nu}(L) 
$, with
\begin{align*}
\zeta(\nu)= \frac{4  }
{\pi^2  
\Gamma\left( \frac{2\xi+j(\nu)+i\nu}{2} \right)\Gamma\left( \frac{2\xi+j(\nu)-i\nu}{2} \right)
\Gamma\left( \frac{2\Delta+j(\nu)-2+i\nu}{2} \right)\Gamma\left( \frac{2\Delta+j(\nu)-2-i\nu}{2} \right)
}\,.
\end{align*}

We may now return to (\ref{Bint}) to find 
\begin{align*}
B^{mn}(p,\bar{p})&\approx -2\pi i  \int
d\nu~\sum_{k=0}^4 \alpha_k(  \nu) \tilde{\mathcal{D}}_k^{mn}  
\frac{\zeta(\nu) \,\Omega_{i\nu}(L) }{
(-p^2)^{2 -\xi-\frac{j(\nu)+1}{2}}  (-\bar{p}^2)^{2-\Delta-\frac{j(\nu)-1}{2}}     }\,.
\end{align*}
With long but trivial manipulations we can rewrite the operators $\tilde{D}$ in the following convenient form
\begin{align*}
-4p^2\tilde{\mathcal{D}}_1^{mn}= &\;\eta^{mn}p^2\hat{\partial}^2- p^2\hat{\partial}^m\hat{\partial}^n \,,
\\ -4p^2\tilde{\mathcal{D}}_2^{mn}= &\; p^2\hat{\partial}^m\hat{\partial}^n \,,
\\ -4p^2\tilde{\mathcal{D}}_3^{mn}=&\;- \left(p^m\hat{\partial}^n+p^n\hat{\partial}^m\right) p^2\hat{\partial}^2
-2\left( \eta^{mn}-2\frac{p^np^m}{p^2}\right) p^2\hat{\partial}^2
\\& +2  p^2\hat{\partial}^m\hat{\partial}^n \left(p\cdot \hat{\partial} +2\right)\,,
\\ -4p^2\tilde{\mathcal{D}}_4^{mn} =&\;
\frac{p^np^m}{p^2} \left( p^2\hat{\partial}^2  +4  \right)  p^2\hat{\partial}^2
- \left(p^m\hat{\partial}^n+p^n\hat{\partial}^m\right)  p^2\hat{\partial}^2\left( p\cdot \hat{\partial} +1\right)\\&
 +\frac{1}{3}  p^2 \hat{\partial}^m\hat{\partial}^n 
 \left[  p^2 \hat{\partial}^2+2(p\cdot \hat{\partial})^2  +10 p\cdot \hat{\partial}+12\right]
\\
&-\frac{1}{3}  \eta^{mn}p^2 \hat{\partial}^2
\left[   p^2 \hat{\partial}^2 -  (p\cdot \hat{\partial})^2 +p\cdot \hat{\partial} +6\right] \,,
\end{align*}
so that the commuting operators $p\cdot \hat{\partial}$ and $p^2 \hat{\partial}^2$ can be traded by their eigenvalues,
\begin{align*}
p\cdot \hat{\partial}&\to 2\xi+j(\nu)-3 \ ,\\
p^2 \hat{\partial}^2
&= \left(p\cdot \hat{\partial}\right)^2 +2 p\cdot \hat{\partial} -\nabla^2  
 \to \big(2\xi+j(\nu)-3\big)\big(2\xi+j(\nu)-1\big) +1+\nu^2\ ,
\end{align*}
where $ \nabla^2$ is the laplacian on the three-dimensional hyperboloid $p^2=-1$.
It is then a trivial computation to obtain the form (\ref{calB}) with\footnote{We suppressed the argument $\nu$ of 
the functions $j(\nu)$ and $\alpha_k(\nu)$ on the right-hand-side to reduce the size of the expressions.}  
\begin{align*}
\frac{4\beta_1(\nu)}{\zeta(\nu)}=& -\left(j^2+(4 \xi -5) j+\frac{2}{3} \left(\nu ^2+3 \xi  (2 \xi -5)+10\right)\right)
   \alpha _1-\frac{1}{3} \left(\nu ^2+3 j+6 \xi -8\right) \alpha _2  \\&-\frac{2}{3} (j+2 \xi
   -4) \left(\nu ^2+1\right) \alpha _3
   +\frac{2}{9} \left(\nu ^4+5 \nu ^2+4\right) \alpha _4 \,,
   \\
   \frac{4\beta_2(\nu)}{\zeta(\nu)}=&-\left(\nu ^2+3 j+6 \xi -8\right) \alpha _1
   -\left(j+2 \xi -4\right) \left(j+2 \xi -3\right) \alpha _2\\&+2 (j+2
   \xi -4) \left(\nu ^2+1\right) \alpha _3-\frac{2}{3} \left(\nu ^4+5 \nu ^2+4\right)
   \alpha _4\,,
   \\
   \frac{4\beta_3(\nu)}{\zeta(\nu)}=& -\left(-j-2 \xi +4\right) \alpha _1
   -\left(j+2 \xi -4\right) \alpha _2-\left(j^2+(4 \xi -6) j-\nu ^2+4 (\xi   -3) \xi +4\right) \alpha _3
   \\& +\frac{2}{3} \left(j+2 \xi -1\right) \left(\nu ^2+4\right) \alpha _4\,,
   \\
   \frac{4\beta_4(\nu)}{\zeta(\nu)}=&\alpha _1-\alpha _2-2 \left(j+2 \xi -1\right) \alpha _3-\left(j^2+(4 \xi -2) j+\frac{1}{3}
   \left(\nu ^2+12 (\xi -1) \xi +4\right)\right) \alpha _4\,.
   \end{align*}

The conservation condition $p_mB^{mn}$ implies that $\beta_2=\beta_3=0$. This is equivalent to 
the conditions (recall that $\xi=3$ for a conserved current)
\begin{align*}
0&=3 \alpha _1(\nu)+\left(j(\nu)+2\right) \alpha _2(\nu)-\left(\nu ^2+1\right) \alpha _3(\nu) \,,\\
0&=3 \alpha _1(\nu)-3 \left(j(\nu)+1\right) \alpha _3(\nu)+2 \left(\nu ^2+4\right) \alpha _4(\nu)\,,
\end{align*}
in agreement with (\ref{nuconservation}) for the case $j(\nu)=1$.

%%%%%%%%%%%%%%%%%%%%%%%%%%%%%%%%%%%%%%
\subsection{Wave functions $F_i$}  \label{Ftrans1}
%%%%%%%%%%%%%%%%%%%%%%%%%%%%%%%%%%%%%%

We now turn to the computation of the functions $F_i$ defined in section \ref{relscat}.
Let us consider, for instance, the integral over $y_4$
$$
 \int  dy_4  \,   ( y_4^-  )^{-\Delta}  \, e^{ i  k_4 \cdot y_4+2i \bar{p} \cdot  x_4}\,.
$$  
Using (\ref{ConformalTrans2}) we find
$$
  2 \bar{p} \cdot  x_4 =-\frac{ \bar{E}}{ \bar{r}}\,y_4^+ + \frac{\bar{E}}{\bar{r}y^-_4}
 \left[ \bar{r}^2+ (y_{4\perp}-\bar{e}_\perp)^2\right]\,,
$$
which allows us to do the $y_4$ integral explicitly and obtain\footnote{We restrict the integration region to $y_4^- >0$ because this is the relevant region in the Regge limit.}   
$$
 \int  dy_4  \,   ( y_4^-  )^{-\Delta}  \, e^{ i  k_4 \cdot y_4+2i \bar{p} \cdot  x_4}=   2\sqrt{\pi}  \,
 \delta( \bar{r}\omega_4-\bar{E}) 
 \bar{E}^{1-\Delta}  \,e^{i\bar{q}_\perp\cdot \bar{e}_\perp} F_4(\bar{r})\,,
$$  
where
$$
F_4(\bar{r})=-i\pi^{\frac{3}{2}} i^{\Delta} 2^{\Delta-2}\bar{r}^2 \left(\sqrt{k_4^2} \right)^{\Delta-2}
 K_{2-\Delta}\left(\bar{r}\sqrt{k_4^2} \right)\,,
$$
with $K$ the modified Bessel function of the second kind.
Similarly, the function $F_2$ is defined by
$$
 \int  dy_2  \,   ( -y_2^-  )^{-\Delta}  \, e^{ i  k_2 \cdot y_2-2i \bar{p} \cdot  x_2}=  2\sqrt{\pi} \,
 \delta( \bar{r}\omega_2-\bar{E})   
 \bar{E}^{1-\Delta} F_2(\bar{r})\,,
$$  
which gives  
$$
F_2(\bar{r})= i\pi^{\frac{3}{2}} i^{-\Delta} 2^{\Delta-2}\bar{r}^2 \left(\sqrt{k_2^2} \right)^{\Delta-2}
 K_{2-\Delta}\left(\bar{r}\sqrt{k_2^2} \right)\,.
$$

The functions $F_1$ and $F_3$ are more complicated because of the index structure.
Consider first
\begin{align*}
  \int  dy_3 \, ( y_3^+)^{-1-\xi}  
  \,\frac{\partial y_3^b}{\partial x_3^n}\, e^{i  k_3 \cdot y_3+  2i p\cdot x_3  }
  =  2\sqrt{\pi}  \, \delta(  r \omega_3- E )   
E^{1-\xi} \,e^{-i q_\perp \cdot e_\perp}  F_{3\,n}^{\,b}(r)\,.
\end{align*}
The exponent reads
\begin{align*}
 i  k_3 \cdot y_3+  2i p\cdot x_3  = \,&\,i \omega_3 y_3^- -i \frac{k_3^2-q_\perp^2}{4 \omega_3} \,y_3^+ -iq_\perp \cdot y_{3\perp}
-i\frac{E}{ r}\,y_3^- + i\frac{E}{r y^+_3}
 \left[ r^2+ (y_{3\perp}-e_\perp)^2\right]   \\
  =\,&\,i  y_3^- \left(  \omega_3-\frac{E}{ r}\right)  -iq_\perp \cdot e_{\perp}
  -i \frac{k_3^2 r}{4 E} \,y_3^+  + i\frac{Er}{y^+_3}
 + i\frac{E}{r y^+_3}
  \left(y_{3\perp}-e_\perp- \frac{ry_3^+}{2E} q_\perp \right)^2 \,,
\end{align*}
where we have used $ \omega_3=E/ r$, which is enforced after integrating over $y_3^-$.
The index structure can be written in matrix form
\begin{align*}
\frac{\partial y_3^b}{\partial x_3^n}=
\left(
\begin{array}{ccc}
 (y_3^+)^2 & y_{3\perp}^2 & -y_3^+ y_{3\perp}  \\
 0 & 1 & 0 \\
 0 & -2 y_{3\perp}& y_3^+  
 \end{array}
\right)\,.
\end{align*}
where the indices are ordered as $(+,-,\perp)$ and the index $b$ labels columns and the index $n$ labels rows.
Changing to the integration variable $s=-i E r /y_3^+$ we obtain
%\begin{align*}
%  &\int  dy_3 \, ( y_3^+)^{-1-\xi}  
%  \,\frac{\partial y_3^b}{\partial x_3^n}\, e^{i  k_3 \cdot y_3+  2i p\cdot x_3  }
%  = i^\xi \pi^2   \delta(  r \omega_3- E )   e^{-i q_\perp \cdot e_\perp}  r^3 (Er)^{-\xi}\\
%  &\int_0^\infty ds s^{\xi-2} e^{-s-\frac{k_3^2 r^2}{4s}} 
%  \left(
%\begin{array}{ccc}
% -\left(\frac{Er}{s}\right)^2 &  -\left(\frac{r^2}{2s}q_\perp \right)^2 +\frac{r^2(1-iq_{\perp}\cdot e_\perp)}{s} +e_\perp^2 & \frac{Er^3}{2s^2} q_\perp +i\frac{Er}{s} e_\perp  \\
% 0 & 1 & 0 \\
% 0 & -2 e_{\perp}   +i\frac{r^2}{s} q_\perp  & -i\frac{Er}{s}  
% \end{array}
%\right)
%\end{align*}
\begin{align*}
  F_{3\,n}^{\,b}(r)
  &=\frac{ i^\xi  \pi^{\frac{3}{2}} }{2E }   \,r^{3-\xi} \int_0^\infty ds \,s^{\xi-2} e^{-s-\frac{k_3^2 r^2}{4s}} \ \times
\\
  & \times\    \left(
\begin{array}{ccc}
 -\left(\frac{Er}{s}\right)^2 &  -\left(\frac{r^2}{2s}q_\perp \right)^2 +\frac{r^2\left(1-iq_{\perp}\cdot e_\perp\right)}{s} +e_\perp^2 & \frac{Er^3}{2s^2} q_\perp +i\frac{Er}{s} e_\perp  \\
 0 & 1 & 0 \\
 0 & -2 e_{\perp}   +i\frac{r^2}{s} q_\perp  & -i\frac{Er}{s}  
 \end{array}
\right)\,.
\end{align*}
The next step is to change the indices to the coordinates $p^\tau=(E,r,e_\perp)$, i.e.
$$
F_{3\,\tau}^{\,b}(r) =   \frac{\partial p^n}{\partial p^\tau }F_{3\,n}^{\,b}(r)\ .
$$
This gives
\begin{align*}
  F_{3\,\tau}^{\,b}(r)
  &=\frac{ i^\xi  \pi^{\frac{3}{2}} }{2  }   \,r^{3-\xi} \int_0^\infty ds \,s^{\xi-2} e^{-s-\frac{k_3^2 r^2}{4s}}   
  % \ \times\\  & \times\    
    \left(
\begin{array}{ccc}
 -\frac{Er}{s^2} &  \frac{r}{E}\left( 1+\frac{1}{s} -\left(\frac{r}{2s}q_\perp \right)^2 \right)  & \frac{r^2}{2s^2} q_\perp  \\
 \left( \frac{E}{s}\right)^2 &  1-\frac{1}{s} +\left(\frac{r}{2s}q_\perp \right)^2  & -\frac{Er}{2s^2} q_\perp  \\
  0 &  i\frac{r}{s} q_\perp  & -i\frac{E}{s}  
 \end{array}
\right)\,,\end{align*}
where the index $\tau$ labels rows ordered as $(E,r,e_\perp)$. Similarly, the function $F_1$ is equal to the complex conjugate of $F_3$ with $q_\perp$ set to zero and $k_3^2$ replaced by $k_1^2$.

%%%%%%%%%%%%%%%%%%%%%%%%%%%%%%%%%%%%%%
\subsection{Structure functions} \label{Ftrans2}
%%%%%%%%%%%%%%%%%%%%%%%%%%%%%%%%%%%%%%

In the context of DIS, we are interested in the case 
 \begin{align*}
\xi=3\ ,\ \ \ \ \ \ q_\perp=0\ ,\ \ \ \ \ \ k_1^2=k_3^2=Q^2\ ,\ \ \ \ \ \ 
k_2^2=k_4^2=\bar{Q}^2\ .
 \end{align*}  
In particular, to determine the structure functions using (\ref{Piintdr}), we need 
 \begin{align*}
r^2 \delta^{kl}\, F_{1\,k}^{\,i}(r) F_{3\,l}^{\,j}(r)&=\frac{\pi^{3 }}{4} Q^2 r^4 K_1^2(Qr)\,\delta^{ij} \,,\\
 F_{1\,r}^{\,+}(r) F_{3}^{\,+r}(r)&= \pi^{3 }  \omega_1^2
     r^4 K_0^2(Qr)\,,
 \end{align*}  
and
$$
F_2(\bar{r})F_4(\bar{r})= \pi^{3}   2^{2\Delta-4} \bar{r}^4 \bar{Q}^{2\Delta-4}
 K_{2-\Delta}^2\left(\bar{Q}\bar{r}\right)\,.
$$
This gives 
\begin{align*}
  \gamma_1(\nu)=&\; \left[ \left(\frac{4}{3}-i\nu\right) \,  \beta_4(\nu) -  \beta_1(\nu)\right]  \pi^6 2^{2\Delta-5} \int dr\,  r^{1-i\nu+j(\nu)} K_1^2(r) 
\int d\bar{r}  \,\bar{r}^{1+i\nu+j(\nu)}   K_{2-\Delta}^2\left(\bar{r}\right)\\
= &\;  \pi^7 2^{2\Delta-9}
\left[ \left(\frac{4}{3}-i\nu\right) \,  \beta_4(\nu) -  \beta_1(\nu)\right]
\frac{ \Gamma
   \left(1+\frac{j(\nu)-i\nu}{2}\right) \Gamma \left(  \frac{j(\nu)-i\nu}{2} 
   \right) \Gamma \left( 2+\frac{j(\nu)-i\nu}{2}\right) }{    \Gamma \left(\frac{3+j(\nu)-i\nu}{2} \right)} \\
&\times \frac{  \Gamma   \left(1+\frac{ j(\nu)+i\nu}{2}\right) \Gamma \left(3+\frac{j(\nu)+ i\nu }{2}-\Delta
   \right) \Gamma \left(\frac{j(\nu)+ i\nu }{2}+\Delta-1 \right)}{  \Gamma
   \left(\frac{3+j(\nu)+i\nu}{2}\right)  } \,,
   \\
  \gamma_2(\nu)=&  \left[ \beta_1(\nu) +\left( \frac{2}{3}-2i\nu-\nu^2 \right)   \beta_4(\nu)\right]  \pi^6 2^{2\Delta-5} \int dr  \,r^{1-i\nu+j(\nu)} K_0^2(r)  
\int d\bar{r}  \,\bar{r}^{1+i\nu+j(\nu)}   K_{2-\Delta}^2\left(\bar{r}\right)\\
=& \;  \pi^7 2^{2\Delta-9} \left[ \beta_1(\nu) +\left( \frac{2}{3}-2i\nu-\nu^2 \right)   \beta_4(\nu)\right]\frac{ \Gamma^3
   \left(1+\frac{j(\nu)-i\nu}{2}\right)   }{    \Gamma \left(\frac{3+j(\nu)-i\nu}{2} \right)} \\
&\times \frac{  \Gamma   \left(1+\frac{ j(\nu)+i\nu}{2}\right) \Gamma \left(3+\frac{j(\nu)+ i\nu }{2}-\Delta
   \right) \Gamma \left(\frac{j(\nu)+ i\nu }{2}+\Delta-1 \right)}{  \Gamma
   \left(\frac{3+j(\nu)+i\nu}{2}\right)  } \,,
  \end{align*}
where the integrals over $r$ and $\bar{r}$ were evaluated using the identity\footnote{The integral converges for $\Re(a )>2 |\Re(b) | $ and it is defined by analytic continuation otherwise.}
$$
\int_0^\infty dt\, t^{a -1} K_{b }(t){}^2=\frac{\sqrt{\pi } \, \Gamma
   \left(\frac{a }{2}\right) \Gamma \left(\frac{a}{2}-b
   \right) \Gamma \left(\frac{a }{2}+b \right)}{4\, \Gamma
   \left(\frac{a +1}{2}\right)}
  \,.
  $$

%%%%%%%%%%%%%%%%%%%%%%%%%%%%%%%%%%%%%%
\section{Regge limit of conformal partial waves \label{CPW}}
%%%%%%%%%%%%%%%%%%%%%%%%%%%%%%%%%%%%%%

The elementary partial wave $G_{\nu,J}^{MN} (P_1,\dots,P_4)$ is given by the integrated product of three-point functions
 \cite{CPWSofia,embedding}
\begin{align}
\int_{\mathbb{R}^d} dP_5\, &
E^{MNA_1\dots A_J}\left(P_1,[1,\xi] ;P_3,[1,\xi];P_5,[J,d/2+i\nu] \right)\times \label{intrepCPW} \\
&\times E_{A_1\dots A_J}\left(P_2,[0,\Delta];P_4 ,[0,\Delta];P_5 ,[J,d/2-i\nu] \right) \ , \nonumber
\end{align}
where the first entry in the square brackets denotes the spin of the operator at that point and the second entry gives its conformal dimension.
In the embedding space formalism, the correlation functions of primary operators are Lorentz covariant and homogeneous functions of the $P_i$'s with weights fixed by the conformal dimensions of the operators.
The three-point functions are completely fixed by conformal invariance up to a finite number of undetermined constants. 

Our goal in this appendix is to determine the Regge limit of the conformal partial wave $G_{\nu,J}^{MN} (P_1,\dots,P_4)$.
More precisely, we want to confirm the small $\sigma$ behavior (\ref{CPWRegge}) of the functions $f^k$ defined by
\[
G_{\nu,J}^{MN} (P_1,\dots,P_4) = \frac{1}{P_{13}^\xi P_{24}^\Delta}\sum_k f_{\nu,J}^k(z,\bar{z}) \,T_k^{MN}\ .
\]
Contracting both sides of this equation with $T_l^{MN}$ we obtain
\[
g^l_{\nu,J}(z,\bar{z})
  =\sum_k f_{\nu,J}^k(z,\bar{z})\, \mathcal{M}_{kl}(z,\bar{z})\ ,
\]
where
\begin{align} 
g^l_{\nu,J}(z,\bar{z})&=
 P_{13}^\xi P_{24}^\Delta \,\eta_{MM'}\eta_{NN'} T_l^{M'N'} G_{\nu,J}^{MN} (P_1,\dots,P_4)\,, \label{defg}\\
\mathcal{M}_{kl}(z,\bar{z})&= \eta_{MM'}\eta_{NN'} T_l^{M'N'}T_k^{MN}\,.\nonumber
\end{align}
The matrix $\mathcal{M}_{kl}$ is symmetric and can be inverted. It is a simple and tedious computation to show that 
\begin{align*} 
\left[ \mathcal{M}^{-1} \right] (z,\bar{z}) = \left(
\begin{array}{cccc}
O(\sigma^0) &  O(\sigma^{-2})  & O(\sigma^0) & O(\sigma^{-1})  \\
 O(\sigma^{-2}) &  O(\sigma^{-4})  & O(\sigma^{-2}) & O(\sigma^{-3})  \\
 O(\sigma^0) &  O(\sigma^{-2})  & O(\sigma^0) & O(\sigma^{-1})  \\
 O(\sigma^{-1}) &  O(\sigma^{-3})  & O(\sigma^{-1}) & O(\sigma^{-2})  
 \end{array}
\right)\,,
\end{align*}
in the limit $\sigma \to 0$.
Therefore, if we can show that
\begin{align} 
g^1_{\nu,J}(z,\bar{z})&\sim \sigma^{1-J} \,,
&g^2_{\nu,J}(z,\bar{z})\sim \sigma^{3-J} \,,\label{Reggeg}\\
g^3_{\nu,J}(z,\bar{z})&\sim \sigma^{1-J}\,,
&g^4_{\nu,J}(z,\bar{z})\sim \sigma^{2-J}\,, \nonumber
\end{align}
then (\ref{CPWRegge}) follows.

In order to verify (\ref{Reggeg}) we used the integral representation (\ref{intrepCPW}) and the explicit form of the 
three-point functions in the embedding space formalism which we shall explain in detail in \cite{embedding}.
We implemented the index contractions in Mathematica\footnote{We can provide the code to the interested reader.}
and reduced the expression (\ref{defg}) to a sum of D-functions,
\begin{align} 
g^l_{\nu,J}(z,\bar{z})= \sum_{\{\Delta_i\}} R^l_{\nu,J} (\{\Delta_i\}, P_{ij})\,   D_{\Delta_1\Delta_2\Delta_3\Delta_4} (P_1,\dots,P_4)    \ , \label{sumDfunctions}
\end{align}
where $R^l_{\nu,J} $ are simple rational functions of the scalar products $P_{ij}$ and
\begin{align} 
  D_{\Delta_1\Delta_2\Delta_3\Delta_4} (P_1,\dots,P_4)  =
  \int_{\mathbb{R}^d} dP_5\, \frac{1}{P_{15}^{\Delta_1}P_{25}^{\Delta_2}P_{35}^{\Delta_3}P_{45}^{\Delta_4}}  \ , \label{defDfunction}
\end{align}
with $\sum \Delta_i = d$ as required by conformal invariance.
Since the functions $g^l_{\nu,J}$ are only a function of the cross ratios,  we can use conformal symmetry to fix the form of the external points $P_i$ in  (\ref{sumDfunctions}) to be
\begin{align}
P_1&=(1,0,0,0,0) \ ,\ \ \  
&P_2=(-\bar{z},1,1,\bar{z},0)\ , \label{Reggechoice} \\
P_3&=(-1,z,1,z,0) \ ,\ \ \  
&P_4=(0,-1,0,0,0)\ , \nonumber
\end{align}
where we have split 
\begin{align}
\mathbb{R}^{2,d}=\mathbb{M}^2\times\mathbb{M}^2\times\mathbb{R}^{d-2} \label{split}
\end{align}
and used light-cone coordinates for both two dimensional Minkowski factors.
The Regge limit is attained by writing $z=\sigma e^\rho$ and $\bar{z}=\sigma e^{-\rho}$ and letting $\sigma \to 0$ with fixed $\rho$.
We have considered this limit explicitly for the functions $R^l_{\nu,J}$ up to $J=3$ \footnote{The running time of our Mathematica code  increases very rapidly with $J$.} and found  
\begin{align*}
R^1_{\nu,J}& \sim \sigma^{\frac{d}{2}-J} \ ,\ \ \ \ \ \ 
&R^2_{\nu,J} \sim \sigma^{2+\frac{d}{2}-J} \ ,\\
R^3_{\nu,J}& \sim \sigma^{\frac{d}{2}-J} \ , 
&R^4_{\nu,J} \sim \sigma^{1+\frac{d}{2}-J} \ .
\end{align*}
This is compatible with (\ref{Reggeg}) if 
\begin{align} 
  D_{\Delta_1  \Delta_2\Delta_3 \Delta_4} (P_1,\dots,P_4)   \sim \sigma^{1-\frac{d}{2}}  \ , 
  \label{ReggeDfunction}
\end{align}
for the choice (\ref{Reggechoice}) and for small $\sigma$.
Fortunately, this result can be easily obtained from the definition (\ref{defDfunction}). 
First, we parametrize the integration point by
\begin{align*}
P_5=(y^+,y^-,1,y^2,y_\perp)\ , 
\end{align*}
where the coordinates correspond to the split (\ref{split}) and $y^2=-y^+y^-+y_\perp^2$.
With this parametrization the scalar products in (\ref{defDfunction}) simplify to
\begin{align*}
&P_{15}=y^- \ , 
&P_{25}=(1-y^-)(\bar{z}+y^+) +y_\perp^2\ ,  \\
&P_{45}=-y^+\ ,
&P_{35}=(1+y^+)(z-y^-) +y_\perp^2 \ .
\end{align*}
In the Regge limit, the dominant contribution to the integral in (\ref{defDfunction}) comes from the region $y^+,y^- \ll1$, which 
corresponds to $P_5$ almost null related to the external points $P_1$ and $P_4$. Therefore, we can approximate 
$  D_{\Delta_1 \Delta_2\Delta_3 \Delta_4}  $ by
\begin{align*} 
   \int 
  \frac{dy^+ dy^- dy_\perp }{ 2 (y^-+i\epsilon)^{\Delta_1}(\sigma e^{-\rho}+y^+ +y_\perp^2+i\epsilon)^{\Delta_2}(\sigma e^\rho -y^- +y_\perp^2+i\epsilon)^{\Delta_3}(-y^++i\epsilon)^{\Delta_4}} 
    \,.
\end{align*}
The change of variables $y^\pm \to \sigma\, y^\pm$, $y_\perp \to \sqrt{\sigma}\, y_\perp$ immediately gives the leading behavior (\ref{ReggeDfunction}).

%%%%%%%%%%%%%%%%%%%%%%%%%%%%%%%%%%%%%%
\section{Some results on the $E_\nu$ tensor functions \label{Efunctions}}
%%%%%%%%%%%%%%%%%%%%%%%%%%%%%%%%%%%%%%

In this appendix we show in detail some results of section \ref{SectionBFKL} that use the embedding formalism to make the action 
of the transverse conformal group $SO(3,1)$ manifest . We shall then consider a particular section of the light-cone to make contact 
with results derived by Lipatov \cite{Lipatov}, who used complex coordinates on the transverse space $\mathbb{R}^{2}$. 

The building blocks in the expansion of the BFKL propagator (\ref{Kernel}), and also of  impact factors, are the conformal 
three-point functions  of two scalar fields of zero dimension and one symmetric and traceless spin $n$ field of dimension $1+i\nu$. 
Placing the scalars at $z_1$ and $z_3$, and the spin $n$ field at $z_7$, we have
\[
E^{a_1\cdots a_n}_\nu(z_1,z_3,z_7) = \left( \frac{z_{13}}{z_{17}z_{37}}\right)^{\frac{1+i\nu}{2}} T^{a_1\cdots a_n}(z_1,z_3,z_7)\,,
\]
where $z_i$ are null vectors in $\mathbb{M}^4$ and $z_{ij}=-2z_i\cdot z_j$. The tensor $T$ has weight zero, is symmetric and traceless, satisfies the $z_7\cdot T=0$  orthogonality condition
\[
z_{7a_i} T^{a_1\cdots a_i\cdots a_n}=0\,,
\]
and is normalized such that $T\cdot T=1$.
To find its explicit form it is useful to consider first the  spin 1 case, for which 
\[
T^a (z_1,z_3,z_7)= \left( \frac{z_{17}z_{37}}{z_{13}}\right)^{\frac{1}{2}}\left( \frac{z_1^a}{z_{17}} - \frac{z_3^a}{z_{37}}\right)\,.
\]
It is also useful to  define a symmetric rank two tensor
\[
G^{ab} (z_1,z_3,z_7)= \eta^{ab} + \frac{z_1^a z_7^b + z_1^b  z_7^a}{z_{17}} + \frac{z_3^a z_7^b + z_3^b z_7^a}{z_{37}} \,,
\]
of weight zero and satisfying
\[
z_{7a} G^{ab} =0\,,\ \ \ \ \ \  \ \ \ \ \  \eta_{ab}\, G^{ab} = G^a_{\ a}  = 2\,.
\]
The tensor $G$ acts on $T$ just as the metric $\eta$, since
\[
G^{ab} \,T_b = T^a \,,\ \ \ \ \ \  \ \ \ \ \  G^{ab} \,T_a T_b = 1\,.
\]
We may now define the tensor structure  $T$ for arbitrary  spin $n$. It is given by
\[
T^{a_1\cdots a_n} (z_1,z_3,z_7)= T^{a_1} T^{a_2} \cdots T^{a_n} - {\rm traces\  computed\ with\ }  G\,.
\]
In particular, for $n=2$ and $n=3$ we have
\begin{align}
T^{a_1a_2} &= T^{a_1} T^{a_2} - \frac{1}{2}\,G^{a_1 a_2}\,,
\nonumber\\
T^{a_1a_2a_3} &= T^{a_1} T^{a_2} T^{a_3} - \frac{1}{4} \left( T^{a_1}G^{a_2 a_3} + T^{a_2}G^{a_1 a_3} +T^{a_3}G^{a_1 a_2}\right)
\nonumber\,,
\end{align}
and so on.

The tensor functions $E_\nu$ obey the important orthogonality condition (\ref{Orthogonality}). In the main text we postponed to this appendix the
description of the tensors $U$ and $V$ that appear in this condition.  The tensor 
$U$ is  simply constructed from the embedding metric
\begin{align}
U^{a_1\cdots a_nb_1\cdots b_{n}} = \frac{1}{2^n} \,\Big( &\eta^{a_1\left(b_1\right.} \eta^{|a_2|b_2} \cdots \eta^{|a_n|\left.b_n\right)} - 
\nonumber\\
& a_i\ {\rm traces \ and} \ b_j\ {\rm traces \ both\ computed\ with}\  \eta \Big)\,.
\nonumber
\end{align}
In particular, for $n=1$ and $n=2$ we have
\begin{align}
U^{ab} &= \frac{1}{2}\,\eta^{ab}\,,
\nonumber\\
U^{a_1a_2b_1b_2} &=   \frac{1}{4} \left( \eta^{a_1\left(b_1\right.} \eta^{|a_2|\left.b_2\right)} - \frac{1}{2}\,\eta^{a_1a_2}\eta^{b_1b_2}\right)
\nonumber\,,
\end{align}
and so on.
To define the tensor $V$ it is convenient to define first a new  symmetric rank two tensor
\[
H^{ab} =\eta^{ab} - \frac{ z_7^a z_5^b + z_7^b z_5^a}{z_5\cdot z_7}
\]
of weight zero and satisfying
\[
z_{5a} \,H^{ab} =0\,,\ \ \ \ \ \  \ \ \ \ \  z_{7a} H^{ab} =0\,.
\]
It is important that
\[
\eta_{ab} H^{ab} = H^a_{\ a} = 2\,,\ \ \ \ \ \ \ \ \ H_{ab} \,H^{bc} = H_a^{\ c} = \eta_{ab} \,H^{bc} \,.
\]
We can now define the tensor $V$ by
\begin{align}
V^{\,a_1\cdots a_nb_1\cdots b_{n}}(z_5,z_7) =& \frac{1}{(-2)^n} \,\Big( H^{a_1\left(b_1\right.} H^{|a_2|b_2} \cdots H^{|a_n|\left.b_n\right)} - 
\nonumber\\
&a_i\ {\rm traces \ and} \ b_j\ {\rm traces \ both\ computed\ with}\  H \Big)\,.
\nonumber
\end{align}
In particular, for $n=1$ and $n=2$ we have
\begin{align}
V^{ab} &= - \frac{1}{2}\,H^{ab}\,,
\nonumber\\
V^{a_1a_2b_1b_2} &=   \frac{1}{4} \left( H^{a_1\left(b_1\right.} H^{|a_2|\left.b_2\right)} - \frac{1}{2}\,H^{a_1a_2}H^{b_1b_2}\right)
\label{Vspin2}\,,
\end{align}
and so on.

%%%%%%%%%%%%%%%%%%%%%%%%%%%%%%%%%%%%%%
\subsection{Complex coordinates on $\mathbb{R}^{2}$ light-cone section}
%%%%%%%%%%%%%%%%%%%%%%%%%%%%%%%%%%%%%%

To make contact with the notation used by Lipatov in \cite{Lipatov}, let us consider as light-cone section a Poincar\'e patch and introduce 
complex coordinates $z$ and $\bar{z}$
\[
z^a  = (z^+,z^-,z, \bar{z}) = \left(1,|z|^2 ,z,\bar{z}\right)\,.
\]
The corresponding metric on $\mathbb{R}^{2}$ has non-vanishing component $g_{z\bar{z}}=g_{\bar{z}z} =1/2$.

Now we  show that the tensor functions $E_\mu$ match those defined by Lipatov.
First compute the projection of the tensor $T$ introduced above. It is simple to verify that the only non-vanishing components of this tensor are
\[
t^{z\cdots z} = \frac{\partial z_7^{a_1}}{\partial z_7} \,\cdots \frac{\partial z_7^{a_n}}{\partial z_7} \, T_{a_1\cdots a_n}  = \left(t^z\right)^n
\]
and its complex conjugate. The component $t^z$ is given by
\[
t^z =  \frac{|z_1-z_7||z_3-z_7|}{|z_1-z_3|} \left(  \frac{z_1-z_7}{|z_1-z_7|^2} - \frac{z_3-z_7}{|z_3-z_7|^2}\right)\,.
\]
This component  is precisely the $e$ function of Lipatov.  Next we must project the symmetric rank two tensor $G_{ab}$ at the point $z_7$. The only 
non-vanishing component gives precisely the  $\mathbb{R}^{2}$ metric
\[
 \frac{\partial z_7^a}{\partial z_7} \, \frac{\partial z_7^b}{\partial \bar{z}_7} \, G_{ab} =g_{z\bar{z}}\,.
\]
It is now simple to see that the only non-vanishing components of the projection of the tensor functions $E_\mu$ are
\[
e_\nu^{z\cdots z} = \frac{\partial z_7^{a_1}}{\partial z_7} \,\cdots \frac{\partial z_7^{a_n}}{\partial z_7} \, E_{\nu\,a_1\cdots a_n} 
= \left( \frac{|z_1-z_7||z_3-z_7|}{|z_1-z_3|} \right)^\frac{1+i\nu}{2} \left(t^z\right)^n
\]
and its complex conjugate. This expression coincides with that given by Lipatov for $E_\nu$, provided one replaces in our expressions $\nu$ by $2\nu$.

Now we consider the projection of the tensors $U$ and $V$, so we verify that the 
orthogonality condition (\ref{Orthogonality}) reproduces that of \cite{Lipatov}. 
First we note  that the only non-vanishing components 
of the projection of the tensor $U$ are given by
\[
u^{z\cdots z \bar{z}\cdots \bar{z}}  = \frac{\partial z^{a_1}}{\partial z} \,\cdots \frac{\partial z^{a_n}}{\partial z} \,
\frac{\partial z^{b_1}}{\partial \bar{z}} \,\cdots \frac{\partial z^{b_n}}{\partial \bar{z}}\,  U_{a_1 \cdots a_n b_1 \cdots b_n}= 1 \,,
\]
together with its complex conjugate. Note that in this case the projection is independent of the points $z_i$. Next let us consider the projection of the tensor
$V$. In this case we need to be careful in projecting the symmetric rank two tensor $H$ to  different boundary points $z_5$ and $z_7$, or to the the same boundary 
point, say $z_5$. The projecting  to the boundary points $z_5$ and $z_7$ has non-vanishing components 
 \[
h_{zz} = \frac{\partial z_5^a}{\partial z_5} \, \frac{\partial z_7^b}{\partial z_7} \, H_{ab} = -\frac{1}{2}\, \frac{\bar{z}_5-\bar{z}_7}{z_5-z_7} \,,
\]
and its complex conjugate. On the other hand, when projecting $H$ to the same boundary point, say to $z_5$, one obtains
the metric tensor on the transverse space. In particular,
\[
\frac{\partial z_5^a}{\partial z_5} \, \frac{\partial z_5^b}{\partial \bar{z}_5} \, H_{ab} = g_{z\bar{z}} \,.
\]
Using the above general expression for $V$ we conclude that the only non-vanishing components are
\[
v^{z\cdots z z\cdots z}  = \frac{\partial z_5^{a_1}}{\partial z} \,\cdots \frac{\partial z_5^{a_n}}{\partial z} \,
\frac{\partial z_7^{b_1}}{\partial z} \,\cdots \frac{\partial z_7^{b_n}}{\partial z}\,  V_{a_1 \cdots a_n b_1 \cdots b_n}= 
 \left(\frac{z_5-z_7}{\bar{z}_5-\bar{z}_7}\right)^n \,,
\]
and its complex conjugate.

%%%%%%%%%%%%%%%%%%%%%%%%%%%%%%%%%%%%%%
\section{Spin 2 conformal integral  \label{ConfInt}}
%%%%%%%%%%%%%%%%%%%%%%%%%%%%%%%%%%%%%%

The purpose of this appendix is to determine the conformal integral
\begin{align}
I^{mn a_1a_2}(x,z_7)=\int_{\mathbb{R}^2} dz_5\,  \Pi_{-\nu}^{mn  b_1b_2} (x,z_5)\,
 \frac{V_{b_1b_2}^{\ \ \ a_1a_2}(z_5,z_7)}{(-2 z_5\cdot z_7)^{1+i\nu}} \ , \label{Idef}
\end{align}
where $V$ is given in (\ref{Vspin2}) and $\Pi$ is the spin 2 bulk to boundary propagator given in (\ref{Pispin2}).
The properties of the integrand guarantee that $I^{mn a_1a_2}(x,z_7)$ is traceless and symmetric in both pairs of indices $mn$ and $a_1a_2$. Moreover, it obeys the transversality conditions
\begin{align*}
x_m I^{mn a_1a_2}(x,z_7)= z_{5\,a_1} I^{mn a_1a_2} (x,z_7)=0\ 
\end{align*}
and is a homogeneous function of weight 0 and $1+i\nu$ in $x$ and $z_7$, respectively.
These features imply that 
\begin{align*}
 I^{mn a_1a_2}(x,z_7)=  \iota(\nu) \Pi_{\nu}^{mn  a_1a_2} (x,z_7) + \Sigma^{mn (a_1 }z_7^{ a_2)}\ ,
\end{align*}
for some function $\Sigma$. We are not interested in the function $\Sigma$ because it  does not contribute to the contraction 
$ I^{mn a_1a_2}(x,z_7)I^{\bar{m}\bar{n}}_{\ \ \ \  a_1a_2}(\bar{x},z_7)$ 
that we need in section \ref{gammagamma}.

To determine the function $\iota(\nu)$ we contract equation (\ref{Idef}) 
with $\eta_{m a_1} \eta_{n a_2}$.
The left hand side gives  
\begin{align*}
2 \iota(\nu)\left(\frac{|x|}{-2x\cdot z_7}\right)^{1+i\nu}  \,,
\end{align*}
because the $\Sigma$ term drops out due to the orthogonality 
$\Sigma^{mn a}z_{7\,m}=0$.
Using the contraction
\begin{align*}
 \pi^{a_1a_2  b_1b_2} (x,z_5)V_{b_1b_2 a_1a_2}(z_5,z_7)=1\ .
\end{align*}
the right hand side becomes
\begin{align*}
\int_{\mathbb{R}^2} dz_5\,
  \frac{ |x|^{1-i\nu} }
  {(-2x\cdot z_5)^{1-i\nu}(-2 z_5\cdot z_7)^{1+i\nu}} =
\frac{i\pi}{\nu}  \left(\frac{|x|}{-2x\cdot z_7}\right)^{1+i\nu} \ ,
\end{align*}
where the last integral was computed in appendix A.2 of \cite{BFKLpaper}.
We conclude that
\begin{align*}
 \iota(\nu)=\frac{i\pi}{2\nu}\ .
\end{align*}

%%%%%%%%%%%%%%%%%%%%%%%%%%%%%%%%%%%%%%
\section{Diagrams in figures \ref{vertex}b and \ref{vertex}c  \label{DeltaFunction}}
%%%%%%%%%%%%%%%%%%%%%%%%%%%%%%%%%%%%%%

In this appendix we complete the computation of the impact factors for a Weyl fermion and a 
complex scalar presented in sections \ref{Weyl} and \ref{Scalar}. We shall compute the contribution 
of the diagrams in figures \ref{vertex}b and \ref{vertex}c, which give the extra delta function terms 
in (\ref{FermionIF}) and (\ref{ScalarIF}).

%%%%%%%%%%%%%%%%%%%%%%%%%%%%%%%%%%%%%%
\subsection{Weyl fermion}
%%%%%%%%%%%%%%%%%%%%%%%%%%%%%%%%%%%%%%

We consider first the contribution of  the diagram in figure \ref{vertex}b to the  impact factor (\ref{ImpactFactor}) and set 
$\sigma_3=w_3^+ >\sigma_1=w_1^+>0$, so that both points are in the patch 
${\cal P}_1$.  We may also set $x=x_1$ and  $x_3=0$. The four-point function in (\ref{ImpactFactor}) is given by
\begin{align*}
&\langle j_m(x) j_n(0) g_-^A(w_1)g_-^B(w_3)\rangle = -  c^2 \left(  -\frac{1}{g_{\mathrm{YM}}^{2}}\right)^{2}  T^A_{ij} \,T^B_{jk} \delta_{ki}
\\
&
\mathrm{Tr}\,\Big\{ \gamma_m\, \langle \psi ({x})\, \bar{\psi} ({w}_1) \rangle 
\,\gamma_- \,\langle \psi ({w}_1)\, \bar{\psi} (w_3) \rangle 
\,\gamma_-\, \langle \psi (w_3)\, \bar{\psi} (0) \rangle
\,\gamma_n \,\langle \psi (0)\, \bar{\psi} ({x}) \rangle\Big\}  \,,
\end{align*}
where the trace in this expression acts on the Dirac indices and all the color dependence is included in the factor $T^A_{ij} \,T^B_{jk} \,\delta_{ki}=C(R)\,\delta^{AB}$.
Dropping the color delta function $\delta^{AB}$, already included in the  BFKL propagator,  the contribution of the diagram in figure \ref{vertex}b to the
impact factor is 
\begin{align*}
&
-\frac{c^2}{g_{\mathrm{YM}}^{4}}\, C(R) \left(\frac{g_{\mathrm{YM}}^{2}}{4\pi^{2}}\right)^4
\left(-x^2\right)^3 \left(z_{13}\right)^2 \sigma_1^2\sigma_3^2
\int  d\lambda_1d\lambda_3
\\
&
\mathrm{Tr}\left\{
\gamma_m\, \slashed{\partial}_{x} \frac{1}{(x-w_1)^2 +i\epsilon}
\,\gamma_-\,\slashed{\partial}_{{w}_1} \frac{1}{(w_1-w_3)^2 + i\epsilon}\right.
\\
&
\ \ \ \ \ \left.\,\gamma_-\, \slashed{\partial}_{{w}_{3}} \,\frac{1}{({w}_{3})^{2}-i\epsilon_{w_3}}
\,\gamma_n \, \left( - \slashed{\partial}_{x}\right) \,\frac{1}{ {x}^{2}-i\epsilon_x}
\frac{1-\tilde{\gamma}}{2}
\right\}\,,
\end{align*}
where we recall that the $w_i$ are parametrized as in (\ref{wParameterization}). We focus  on the integration of the second and third line of the previous equation
\begin{align}
&
 \sigma_1^2\sigma_3^2
\int  d\lambda_1d\lambda_3
\frac{-2(x-w_1)^a}{\big((x-w_1)^2 +i\epsilon\big)^2} \, \frac{-2(w_1-w_3)^b}{\big((w_1-w_3)^2 +i\epsilon\big)^2}\,
\frac{-2w_3^c}{\big((w_3)^2 -i\epsilon_{w_3}\big)^2}
\nonumber
\\
&
\frac{2x^d}{\big( x^2 - i\epsilon_x\big)^2} \, \mathrm{Tr}\left\{
\gamma_m \gamma_a \gamma_- \gamma_b \gamma_- \gamma_c \gamma_n \gamma_d\,
\frac{1-\tilde{\gamma}}{2}
\right\}\,.
\label{IntegralFermion}
\end{align}
These integrals have poles at
\begin{align*}
& \lambda_1\simeq  - 2 x\cdot z_1 + i\epsilon\,,
\\
& \lambda_3= -i\epsilon_{w_3} =-  i \epsilon\,,
\\
& \lambda_1\simeq  \lambda_3  - \frac{\sigma_1\sigma_3 (-2z_1\cdot z_3)}{\sigma_3-\sigma_1}  - i\epsilon\,,
\end{align*}
where  we set  $\epsilon_{w_3}=\epsilon$ because at  $\lambda_3=0$ we have $w_3^-=w_{3\perp}^2/w^+>0$ and therefore $w_3^0>0$. 
Deforming the $\lambda_1$ integral in the upper half plane and the $\lambda_3$ integral in the lower half plane, the first line of (\ref{IntegralFermion}) becomes
\[
-32\pi^2 \sigma_1 \sigma_3 \,
\partial_{\lambda_1} \partial_{\lambda_3}
\frac{ (x-\sigma_1 z_1 - \lambda_1 n)^a   (\sigma_1 z_1 + \lambda_1 n-\sigma_3 z_3 - \lambda_3 n)^b  (\sigma_3 z_3 + \lambda_3 n)^c}
{ \big((x-\sigma_1 z_1 - \lambda_1 n)^2\big)^2}\,,
\]
computed at $\lambda_1=-2x \cdot z_1$ and $\lambda_3=0$. We can drop the derivative of the three terms in the numerator because
these are contracted with the $\gamma$ matrices in (\ref{IntegralFermion}): the first and third terms do not contribute because 
$(\gamma_-)^2=0$ and the second term because we must have $b=+$. We may then replace the previous equation by
\[
-192 \pi^2 \frac{z_1^az_3^c}{(-2x\cdot z_1)^3}\,\frac{\alpha^3}{\left(  -2z_1\cdot z_3 +\alpha \right)^4}\,,
\]
where 
\[
\alpha = \frac{\sigma_3 - \sigma_1}{\sigma_3 \sigma_1}\,(-2x\cdot z_1) >0\,.
\]
and we recall that $-2z_1\cdot z_3 = (z_{1\perp} - z_{3\perp})^2$.
Using the representation of the delta function
\begin{equation}
\lim_{\alpha\rightarrow 0}\frac{\alpha^n}{\big((z_{1\perp} - z_{3\perp})^2+ \alpha\big)^{n+1}} =  \frac{\pi}{n}\,\delta^{(2)}\left( z_{1\perp} - z_{3\perp} \right)\,,
\label{DeltaFunctionRep}
\end{equation}
we conclude that, in the Regge limit of small $x$,  we can replace the first line of (\ref{IntegralFermion}) with
\[
-64 \pi^3\frac{z_1^az_1^c}{(-2x\cdot z_1)^3}\,\delta^{(2)}\left( z_{1\perp} - z_{3\perp} \right)\,,
\]
and set $b=+$. Equation  (\ref{IntegralFermion}) becomes then
\[
-64 \pi^3\frac{z_1^az_1^c}{(-2x\cdot z_1)^3}\,\delta^{(2)}\left( z_{1\perp} - z_{3\perp} \right)\, 
\frac{2x^d}{ x^4 } \, \mathrm{Tr}\left\{\gamma_m \gamma_a (- \gamma_-)  \gamma_c \gamma_n \gamma_d\,
\frac{1-\tilde{\gamma}}{2}
\right\}\,.
\]
Defining the radial delta function
\[
\delta^{(2)}(z_{\perp}) = \frac{1}{\pi}\, \delta(z_\perp^2)\,,\ \ \ \ \ \ \ \int_0^\infty d(z_\perp^2) \delta(z_\perp^2)=1\,,  
\]
a simple computation shows that  (\ref{IntegralFermion})  is given by 
\[
-128\pi^2 \frac{\delta(u)}{-x^2 (-2x\cdot z_1)^4} \,\Big(  \eta_{mn} + 2\, \frac{z_{1m} x_n + z_{1n} x_m}{(-2x\cdot z_1)}\Big)\,.
\]

We can repeat the same computation but with $\sigma_1=w_1^+ >\sigma_3=w_3^+>0$. In  this case the pole structure is such that one obtains
a vanishing result, so that the diagram in figure \ref{vertex}b orders the interaction points in light-cone time. If one of the interaction points, or both, are in the other
Poincar\'e patch ${\cal P}_3$, the non-vanishing contributions are given by the same result,  still ordered in light-cone  time. The other diagram in figure 
 \ref{vertex}c gives exactly the same result but will the opposite ordering in light-cone time. Thus, both diagrams in figures 
 \ref{vertex}b and \ref{vertex}c contribute to the impact factor with
\[
k \,c^2\, C(R) \,\frac{g^4_{\mathrm{YM}}}{\pi^6} 
\frac{u^2\delta(u)}{2} \Big(  {\cal I}_1^{mn} + {\cal I}_3^{mn}\Big)\,.
\]

%%%%%%%%%%%%%%%%%%%%%%%%%%%%%%%%%%%%%%
\subsection{Complex scalar}
%%%%%%%%%%%%%%%%%%%%%%%%%%%%%%%%%%%%%%

We start again with the diagram in figure \ref{vertex}b and consider  the case with 
$w_1^+>0$ and $w_3^+>0$, so that both points are in the patch ${\cal P}_1$. 
A simple computation gives 
\begin{align*}
&\langle j_m(x_1) j_n(x_3) g_-^A(w_1)g_-^A(w_3)\rangle = - c^2 \left(  -\frac{i}{g_{\mathrm{YM}}^{2}}\right)^{2}  T^A_{ij} \,T^B_{jk} \delta_{ki}
\\
&
\langle Z(x_1) \bar{Z}(w_1)\rangle
\overleftrightarrow{\partial}_{\hspace{-0.1cm}w_1^-}
\langle Z(w_1) \bar{Z}(w_3)\rangle
\overleftrightarrow{\partial}_{\hspace{-0.1cm}w_3^-}
\langle Z(w_3) \bar{Z}(x_3)\rangle
(- \overleftrightarrow{\partial}_{\hspace{-0.1cm}x_1^m})
\overleftrightarrow{\partial}_{\hspace{-0.1cm}x_3^n}
\langle Z(x_3) \bar{Z}(x_1)\rangle
\,,
\end{align*}
where the propagators have no color indices. The contribution of this diagram to the impact factor is then
\begin{align}
&
-\frac{c^2}{g_{\mathrm{YM}}^{4}}\, C(R) \left(\frac{g_{\mathrm{YM}}^{2}}{4\pi^{2}}\right)^4
\left(-x^2\right)^3 \left(z_{13}\right)^2 \sigma_1^2 \sigma_3^2
\int d\lambda_1 d\lambda_3
\nonumber\\
&
\left(\frac{1}{(x_1-w_1)^2 +i\epsilon}
\overleftrightarrow{\partial}_{\hspace{-0.1cm}w_1^-}
\frac{1}{(w_1-w_3)^2 +i\epsilon}
\overleftrightarrow{\partial}_{\hspace{-0.1cm}w_3^-}
\frac{1}{\left(w_3-x_{3}\right)  ^{2}-i\epsilon_{w_3-{x}_{3}}}\right) \label{Last}
\\
&
\overleftrightarrow{\partial}_{\hspace{-0.1cm}x_1^m}
\overleftrightarrow{\partial}_{\hspace{-0.1cm}x_3^n}\,
\frac{1}{\left(x_1-x_3\right)  ^{2} - i\epsilon_{x_1-x_3}}\,,
\nonumber
\end{align}
Now we focus on the integral of the second line of this equation
\begin{equation*}
\sigma_1^2 \sigma_3^2 \int d\lambda_1\lambda_3  
\frac{1}{(x_1-w_1)^2 +i\epsilon}
\overleftrightarrow{\partial}_{\hspace{-0.1cm}w_1^-}
\frac{1}{(w_1-w_3)^2 +i\epsilon}
\overleftrightarrow{\partial}_{\hspace{-0.1cm}w_3^-}
\frac{1}{\left(w_3-x_{3}\right)  ^{2}-i\epsilon_{w_3-{x}_{3}}}\,.
\end{equation*}
The poles are located at
\begin{align*}
& \lambda_1\simeq  - 2 x_1\cdot z_1 + i\epsilon\,,
\\
& \lambda_3 \simeq - 2 x_3\cdot z_3 -i\epsilon_{w_3} = - 2 x_3\cdot z_3 -  i \epsilon\,,
\\
& \lambda_1\simeq  \lambda_3  - \frac{\sigma_1\sigma_3 (-2z_1\cdot z_3)}{\sigma_3-\sigma_1}  - i\epsilon\,,
\end{align*}
where at the second pole $\epsilon_{w_1-x_3}=\epsilon$. 
Deforming the $\lambda_1$ integral in the upper half plane and the $\lambda_3$ integral in the lower half plane, and using the 
representation of the delta function (\ref{DeltaFunctionRep}), in the Regge limit the previous integral is given by
\[
16\pi^3 \delta^{(2)} \left( z_{1\perp} - z_{3\perp} \right)\frac{1}{(-2 (x_1 - x_3) \cdot z_1)^2}\,.
\]
Setting $x\approx x_1-x_3$, a simple computation shows that (\ref{Last}) is given by
\[
k \,c^2\, C(R) \,\frac{g^4_{\mathrm{YM}}}{\pi^6} \frac{u^2\delta(u)}{2} \left(  \frac{1}{4}\,{\cal I}_1^{mn}  - {\cal I}_2^{mn} + \frac{3}{2}\,{\cal I}_3^{mn}   -3\,{\cal I}_4^{mn} \right)\,.
\]
As for the fermionic field, after considering the other possibilities for the signs of $\sigma_1$ and $\sigma_3$, as well as their ordering, we conclude 
the previous result gives the contribution of diagrams in figures \ref{vertex}b and \ref{vertex}c to the impact factor of a complex scalar field.

\end{document}